%**************************** SACMACE.TEX %%%%%%***********************************
%********************************************************************%%%*******
% Macros for articles using a modified version of macros of
%``Quantum field theory  and critical phenomena" combined with harvmac
% Version with 2 type of fonts: 10 pts, 10 pts magnified
% 10 pts uses double column format with "landscape" 
% J. Zinn-Justin 9/02/1996
% bug corrected 17/02/99, 15/05/99, 21/10/99
% modif 03/03/99
% addition of \figlbl and \label protection on 25/05/99
% modification of sslbl concerning appendix 29/05/99
% bug hypertex corrected 03/07/99
% table numbering 16/02/2000
% addition \dd,\ddi 24/03/02
% date bug 2000 corrected, page number centred in draft mode 12/04/02
% 11/09/02 Reference added in table of contents
% 06/12/02 \sslbl modified . Index added
% 27/12/02 ARTICLE MACROS
% 28/07/2007 lfont replaced by lfontnew
% 23/02/09 supression of \def\n{\noindent}
% 08/03/09 bug corrigé dans \ssubsection
% 25/03/09 bug corrigé dans \ssubsection
% 26/03/09 introduction de \firstpage
% 10/02/2010 correction in \nrf correction de \fpage en \firstpage dans %\title
% 16/01/2013 change after jobname.equ
% *************************** output macros %%%%********************************
\input hyperbasics
\catcode`\@=11
%%% saclay A4 paper:
\def\unredoffs{\voffset=13mm \hoffset=6.5truemm} 
\def\redoffs{\voffset=-12.truemm\hoffset=-3truemm} 

\newif\ifbookmode
\bookmodefalse
%
% ****************************** INDEX
\newwrite\inx
\def\book{\bookmodetrue\immediate\openout\inx=\jobname.inx%
 \hsize=120mm\vsize=195mm} 
\def\@#1{\noindent\ifbookmode\write\inx{#1,\space
\number\pageno.\par}\fi}

%---------------------------------------------------------------------%
\newbox\leftpage \newdimen\fullhsize \newdimen\hstitle \newdimen\hsbody
\newdimen\hdim
%\tolerance=1000
\hfuzz=1pt
\ifx\hyperdef\UNd@FiNeD\def\hyperdef#1#2#3#4{#4}\def\hyperref#1#2#3#4{#4}\fi
%%%%%%%%%%%%%%%%%%%%%%%%%%%%%%%%%%%%%%%%%%%%%%%%%
\def\bigans{b }
%\message{ big or little (b/l)? }\read-1 to\answ
\def\answ{b }
\ifx\answ\bigans\message{(Format simple colonne 12pts.}
\magnification=1200 \unredoffs%\ifbookmode \hsize=144truemm\vsize=234truemm%
%\else
\hsize=122.5mm\vsize=182.5mm
%\fi   
\hsbody=\hsize \hstitle=\hsize %take default values for unreduced format
\else\message{(Format simple colonne, 10pts.} \let\l@r=L
\magnification=1000 
\redoffs%
\hsize=122.5mm\vsize=182.5mm
\hsbody=\hsize \hstitle=\hsize %take default values for unreduced format
\fi
% 
%%%%%%%%%%%%%%%%%%%%%% fonts, Dirac slash %%%%%%%%%%%%%%%
% 5/12/02  twbfx added,
% 5/12/02   \chapfnt=cmbx10 scaled 1440 \font\headbf=cmbx9
% 30/04/07 adddition of chapfonts
% 25/02/09 MACRO \Bfg for bold lowercase greek letters 
% 22/04/09 \skewchar for \Bfg

\def\sla#1{\mkern-1.5mu\raise0.4pt\hbox{$\not$}\mkern1.2mu #1\mkern 0.7mu}
\def\Dbar{\mkern-1.5mu\raise0.4pt\hbox{$\not$}\mkern-.1mu {\rm D}\mkern.1mu}
\def\Abar{\mkern1.mu\raise0.4pt\hbox{$\not$}\mkern-1.3mu A\mkern.1mu}
\def\Bbar{\mkern-0.mu\raise0.4pt\hbox{$\not$}\mkern-.3mu B\mkern 0.6mu}
\newskip\tableskipamount \tableskipamount=8pt plus 3pt minus 3pt

%****************************
%%%% chapters
\newdimen\chapskip
\chapskip=17.5mm
\font\twbfx=cmbx10 scaled 1200
\font\chapfnt=cmbx10 scaled 1440
\font\chapbxten=cmbx10
\font\chapbxseven=cmbx7

\font\ssbx=cmssbx10  

\font\chaprm=cmr10 scaled 1440
\font\chaprmscript=cmr10
\font\chaprmseven=cmr7
\font\chapssfnt=cmssbx10 scaled 1440

\font\chapibfnt=cmmib10 scaled 1440
\font\chapmifnt=cmmi10 scaled 1440
\font\chapsyfnt=cmsy10 scaled 1440
\font\chapexfnt=cmex10 scaled 1440
\font\chapibten=cmmib10
\font\chapibseven=cmmib7
\font\chapmiscript=cmmi10  
\font\chapsyscript=cmsy10  
\font\chapexscript=cmex10
\font\chapmiseven=cmmi7
\font\chapsyseven=cmsy7
\font\chapexseven=cmex7  
%%%%%%%%%%%%%%%%%%%%%%%%%%
\def\chapfont{
%\textfont0=\chapssfnt \scriptfont0=\chapssten \scriptscriptfont0=\chapssseven
%\def\rm{\fam0\chapssfnt}
\textfont0=\chaprm\scriptfont0=\chaprmscript\scriptscriptfont0=\chaprmseven
\textfont1=\chapmifnt \scriptfont1=\chapmiscript  \scriptscriptfont1=\chapmiseven
\textfont2=\chapsyfnt \scriptfont2=\chapsyscript\scriptscriptfont2=\chapsyseven
\textfont3=\chapexfnt \scriptfont3=\chapexscript \scriptscriptfont3=\chapexseven
%\textfont\itfam=\capit \def\it{\fam\itfam\capit} % \it is family 4
%\textfont\slfam=\capsl  \def\sl{\fam\slfam\capsl} % \sl is family 5
\textfont\bffam=\chapfnt \scriptfont\bffam=\chapbxten
\scriptscriptfont\bffam=\chapbxseven
\def\bf{\fam\bffam\chapfnt} % \bf is family 6
\textfont4=\chapibfnt \scriptfont4=\chapibten \scriptscriptfont4=\chapibseven
\abovedisplayskip=17pt plus 5pt minus 13pt
\belowdisplayskip=\abovedisplayskip
%\smallskipamount=2.7pt plus 1pt minus 1pt
%\medskipamount=5.4pt plus 2pt minus 2pt
%\bigskipamount=10.8pt plus 3.6pt minus 3.6pt
\normalbaselineskip=17pt
\setbox\strutbox=\hbox{\vrule height12.2pt depth5.0pt width0pt}
\normalbaselines \chapssfnt}

\font\caprm=cmr9
\font\capit=cmti9
\font\capbf=cmbx9
\font\capsl=cmsl9
\font\capmi=cmmi9
\font\capex=cmex9
\font\capsy=cmsy9

\def\makeheadline{\vbox to 0pt{\vskip-22.5pt
\line{\vbox to8.5pt{}\the\headline}\vss}\nointerlineskip}
%%%%%%%%%%%%%%%%%%%%%%%%%%%%%%%%%%%%%%%%%%%%%%%%%%%%%%%%%%%%%%%%%%%%%%%%
\font\headrm=cmr10

%****************************
\font\sixrm=cmr6
\font\fiverm=cmr5
\font\sixmi=cmmi6
\font\fivemi=cmmi5
\font\sixsy=cmsy6
\font\fivesy=cmsy5
\font\sixbf=cmbx6
\font\fivebf=cmbx5
\skewchar\capmi='177 \skewchar\sixmi='177 \skewchar\fivemi='177
\skewchar\capsy='60 \skewchar\sixsy='60 \skewchar\fivesy='60

% *************************************************************************

% ****************************************************************************
%		*****	  MSSYMB.TeX	*****		       20 Sept 91
%
%	This file contains the definitions for the symbols in the two
%	"extra symbols" fonts created at the American Math. Society.
%
%       The old fonts msxm et msym have been replaced by msam et msbm. 

\catcode`\@=11
%%%%%%%%%%%%%%%%%%%%%%%%%%%%%%%%%%%%%%%
%***************************************************
  %obsolete??
\font\tenbi=cmmib10 
\font\ninebi=cmmib9
\font\sevenbi=cmmib7 
\font\fivebi=cmmib5
\textfont4=\tenbi \scriptfont4=\sevenbi \scriptscriptfont4=\fivebi
\newfam\mibfam
\textfont\mibfam=\tenbi \scriptfont\mibfam=\sevenbi \scriptscriptfont\mibfam=\fivebi
\def\Bfg{\ifmmode\let\next\Bfg@\else
 \def\next{\errmessage{Use \string\Bfg\space only in math mode}}\fi\next}
\def\Bfg@#1{{\Bfg@@{#1}}}
\def\Bfg@@#1{\fam\mibfam#1}
\skewchar\tenbi='177\skewchar\sevenbi='177
%%%%%%%%%%%%%%%%%%%%%%%%%%%%%%%%%%%%%%%%%%%%%%%%%%%%%%%%%%%%%%
\mathchardef\alpha="710B
\mathchardef\beta="710C 
\mathchardef\gamma="710D
\mathchardef\delta="710E
\mathchardef\epsilon="710F  
\mathchardef\zeta="7110  
\mathchardef\eta="7111 
\mathchardef\theta="7112 
\mathchardef\iota="7113 
\mathchardef\kappa="7114 
\mathchardef\lambda="7115 
\mathchardef\mu="7116 
\mathchardef\nu="7117 
\mathchardef\xi="7118 
\mathchardef\pi="7119 
\mathchardef\rho="711A  
\mathchardef\sigma="711B 
\mathchardef\tau="711C 
\mathchardef\upsilon="711D 
\mathchardef\phi="711E 
\mathchardef\chi="711F 
\mathchardef\psi="7120 
\mathchardef\omega="7121 
\mathchardef\varepsilon="7122 
\mathchardef\vartheta="7123
\mathchardef\varpi="7124 
\mathchardef\varrho="7125 
\mathchardef\varsigma="7126 
\mathchardef\varphi="7127
%%%%%%%%%%%%%%%%%%%%%%%% OBSOLETE except for titles %%%%%%%%%%%%%%%%%%%%% %%%%%%%%%%%%%%%%%%%%%%%%%%%%%%%%%%%%%%%%%%%%%%%%%%%%%%%%%%%%%%%%%%%%%%%%%
\mathchardef\alphab="040B
\mathchardef\betab="040C 
\mathchardef\gammab="040D
\mathchardef\deltab="040E
\mathchardef\epsilonb="040F  
\mathchardef\zetab="0410  
\mathchardef\etab="0411 
\mathchardef\thetab="0412 
\mathchardef\iotab="0413 
\mathchardef\kappab="0414 
\mathchardef\lambdab="0415 
\mathchardef\mub="0416 
\mathchardef\nub="0417 
\mathchardef\xib="0418 
\mathchardef\pib="0419 
\mathchardef\rhob="041A  
\mathchardef\sigmab="041B 
\mathchardef\taub="041C 
\mathchardef\upsilonb="041D 
\mathchardef\phib="041E 
\mathchardef\chib="041F 
\mathchardef\psib="0420 
\mathchardef\omegab="0421 
\mathchardef\varepsilonb="0422 
\mathchardef\varthetab="0423
\mathchardef\varpib="0424 
\mathchardef\varrhob="0425 
\mathchardef\varsigmab="0426 
\mathchardef\varphib="0427 
 %%%%%%%%%%%%%%%%%%%%%%%%%%%%%%%%%%%%%%%%%%%%%%%%%%%%%%%%%%%%
\font\tenmsa=msam10
\font\sevenmsa=msam7
\font\fivemsa=msam5
\font\tenmsb=msbm10
\font\sevenmsb=msbm7
\font\fivemsb=msbm5
\newfam\msafam
\newfam\msbfam
\textfont\msafam=\tenmsa  \scriptfont\msafam=\sevenmsa
  \scriptscriptfont\msafam=\fivemsa
\textfont\msbfam=\tenmsb  \scriptfont\msbfam=\sevenmsb
  \scriptscriptfont\msbfam=\fivemsb

\def\hexnumber@#1{\ifcase#1 0\or1\or2\or3\or4\or5\or6\or7\or8\or9\or
	A\or B\or C\or D\or E\or F\fi }
%  The following 13 lines establish the use of the Euler Fraktur font.
%  To use this font, remove % from beginning of these lines.
\font\teneuf=eufm10
\font\seveneuf=eufm7
\font\fiveeuf=eufm5
\newfam\euffam
\textfont\euffam=\teneuf
\scriptfont\euffam=\seveneuf
\scriptscriptfont\euffam=\fiveeuf
\def\frak{\ifmmode\let\next\frak@\else
 \def\next{\Err@{Use \string\frak\space only in math mode}}\fi\next}
\def\goth{\ifmmode\let\next\frak@\else
 \def\next{\Err@{Use \string\goth\space only in math mode}}\fi\next}
\def\frak@#1{{\frak@@{#1}}}
\def\frak@@#1{\fam\euffam#1}
%  End definition of Euler Fraktur font.

\edef\msa@{\hexnumber@\msafam}
\edef\msb@{\hexnumber@\msbfam}

\mathchardef\boxdot="2\msa@00
\mathchardef\boxplus="2\msa@01
\mathchardef\boxtimes="2\msa@02
\mathchardef\square="0\msa@03
\mathchardef\blacksquare="0\msa@04
\mathchardef\centerdot="2\msa@05
\mathchardef\lozenge="0\msa@06
\mathchardef\blacklozenge="0\msa@07
\mathchardef\circlearrowright="3\msa@08
\mathchardef\circlearrowleft="3\msa@09
\mathchardef\rightleftharpoons="3\msa@0A
\mathchardef\leftrightharpoons="3\msa@0B
\mathchardef\boxminus="2\msa@0C
\mathchardef\Vdash="3\msa@0D
\mathchardef\Vvdash="3\msa@0E
\mathchardef\vDash="3\msa@0F
\mathchardef\twoheadrightarrow="3\msa@10
\mathchardef\twoheadleftarrow="3\msa@11
\mathchardef\leftleftarrows="3\msa@12
\mathchardef\rightrightarrows="3\msa@13
\mathchardef\upuparrows="3\msa@14
\mathchardef\downdownarrows="3\msa@15
\mathchardef\upharpoonright="3\msa@16

\mathchardef\downharpoonright="3\msa@17
\mathchardef\upharpoonleft="3\msa@18
\mathchardef\downharpoonleft="3\msa@19
\mathchardef\rightarrowtail="3\msa@1A
\mathchardef\leftarrowtail="3\msa@1B
\mathchardef\leftrightarrows="3\msa@1C
\mathchardef\rightleftarrows="3\msa@1D
\mathchardef\Lsh="3\msa@1E
\mathchardef\Rsh="3\msa@1F
\mathchardef\rightsquigarrow="3\msa@20
\mathchardef\leftrightsquigarrow="3\msa@21
\mathchardef\looparrowleft="3\msa@22
\mathchardef\looparrowright="3\msa@23
\mathchardef\circeq="3\msa@24
\mathchardef\succsim="3\msa@25
\mathchardef\gtrsim="3\msa@26
\mathchardef\gtrapprox="3\msa@27
\mathchardef\multimap="3\msa@28
\mathchardef\therefore="3\msa@29
\mathchardef\because="3\msa@2A
\mathchardef\doteqdot="3\msa@2B

\mathchardef\triangleq="3\msa@2C
\mathchardef\precsim="3\msa@2D
\mathchardef\lesssim="3\msa@2E
\mathchardef\lessapprox="3\msa@2F
\mathchardef\eqslantless="3\msa@30
\mathchardef\eqslantgtr="3\msa@31
\mathchardef\curlyeqprec="3\msa@32
\mathchardef\curlyeqsucc="3\msa@33
\mathchardef\preccurlyeq="3\msa@34
\mathchardef\leqq="3\msa@35
\mathchardef\leqslant="3\msa@36
\mathchardef\lessgtr="3\msa@37
\mathchardef\backprime="0\msa@38
\mathchardef\risingdotseq="3\msa@3A
\mathchardef\fallingdotseq="3\msa@3B
\mathchardef\succcurlyeq="3\msa@3C
\mathchardef\geqq="3\msa@3D
\mathchardef\geqslant="3\msa@3E
\mathchardef\gtrless="3\msa@3F
\mathchardef\sqsubset="3\msa@40
\mathchardef\sqsupset="3\msa@41
\mathchardef\vartriangleright="3\msa@42
\mathchardef\vartriangleleft="3\msa@43
\mathchardef\trianglerighteq="3\msa@44
\mathchardef\trianglelefteq="3\msa@45
\mathchardef\bigstar="0\msa@46
\mathchardef\between="3\msa@47
\mathchardef\blacktriangledown="0\msa@48
\mathchardef\blacktriangleright="3\msa@49
\mathchardef\blacktriangleleft="3\msa@4A
\mathchardef\vartriangle="0\msa@4D
\mathchardef\blacktriangle="0\msa@4E
\mathchardef\triangledown="0\msa@4F
\mathchardef\eqcirc="3\msa@50
\mathchardef\lesseqgtr="3\msa@51
\mathchardef\gtreqless="3\msa@52
\mathchardef\lesseqqgtr="3\msa@53
\mathchardef\gtreqqless="3\msa@54
\mathchardef\Rrightarrow="3\msa@56
\mathchardef\Lleftarrow="3\msa@57
\mathchardef\veebar="2\msa@59
\mathchardef\barwedge="2\msa@5A
\mathchardef\doublebarwedge="2\msa@5B
\mathchardef\angle="0\msa@5C
\mathchardef\measuredangle="0\msa@5D
\mathchardef\sphericalangle="0\msa@5E
\mathchardef\varpropto="3\msa@5F
\mathchardef\smallsmile="3\msa@60
\mathchardef\smallfrown="3\msa@61
\mathchardef\Subset="3\msa@62
\mathchardef\Supset="3\msa@63
\mathchardef\Cup="2\msa@64

\mathchardef\Cap="2\msa@65

\mathchardef\curlywedge="2\msa@66
\mathchardef\curlyvee="2\msa@67
\mathchardef\leftthreetimes="2\msa@68
\mathchardef\rightthreetimes="2\msa@69
\mathchardef\subseteqq="3\msa@6A
\mathchardef\supseteqq="3\msa@6B
\mathchardef\bumpeq="3\msa@6C
\mathchardef\Bumpeq="3\msa@6D
\mathchardef\lll="3\msa@6E

\mathchardef\ggg="3\msa@6F

\mathchardef\circledS="0\msa@73
\mathchardef\pitchfork="3\msa@74
\mathchardef\dotplus="2\msa@75
\mathchardef\backsim="3\msa@76
\mathchardef\backsimeq="3\msa@77
\mathchardef\complement="0\msa@7B
\mathchardef\intercal="2\msa@7C
\mathchardef\circledcirc="2\msa@7D
\mathchardef\circledast="2\msa@7E
\mathchardef\circleddash="2\msa@7F
\def\ulcorner{\delimiter"4\msa@70\msa@70 }
\def\urcorner{\delimiter"5\msa@71\msa@71 }
\def\llcorner{\delimiter"4\msa@78\msa@78 }
\def\lrcorner{\delimiter"5\msa@79\msa@79 }
\def\yen{\mathhexbox\msa@55 }
\def\checkmark{\mathhexbox\msa@58 }
\def\circledR{\mathhexbox\msa@72 }
\def\maltese{\mathhexbox\msa@7A }
\mathchardef\lvertneqq="3\msb@00
\mathchardef\gvertneqq="3\msb@01
\mathchardef\nleq="3\msb@02
\mathchardef\ngeq="3\msb@03
\mathchardef\nless="3\msb@04
\mathchardef\ngtr="3\msb@05
\mathchardef\nprec="3\msb@06
\mathchardef\nsucc="3\msb@07
\mathchardef\lneqq="3\msb@08
\mathchardef\gneqq="3\msb@09
\mathchardef\nleqslant="3\msb@0A
\mathchardef\ngeqslant="3\msb@0B
\mathchardef\lneq="3\msb@0C
\mathchardef\gneq="3\msb@0D
\mathchardef\npreceq="3\msb@0E
\mathchardef\nsucceq="3\msb@0F
\mathchardef\precnsim="3\msb@10
\mathchardef\succnsim="3\msb@11
\mathchardef\lnsim="3\msb@12
\mathchardef\gnsim="3\msb@13
\mathchardef\nleqq="3\msb@14
\mathchardef\ngeqq="3\msb@15
\mathchardef\precneqq="3\msb@16
\mathchardef\succneqq="3\msb@17
\mathchardef\precnapprox="3\msb@18
\mathchardef\succnapprox="3\msb@19
\mathchardef\lnapprox="3\msb@1A
\mathchardef\gnapprox="3\msb@1B
\mathchardef\nsim="3\msb@1C
%\mathchardef\napprox="3\msb@1D
\mathchardef\ncong="3\msb@1D

\mathchardef\varsubsetneq="3\msb@20
\mathchardef\varsupsetneq="3\msb@21
\mathchardef\nsubseteqq="3\msb@22
\mathchardef\nsupseteqq="3\msb@23
\mathchardef\subsetneqq="3\msb@24
\mathchardef\supsetneqq="3\msb@25
\mathchardef\varsubsetneqq="3\msb@26
\mathchardef\varsupsetneqq="3\msb@27
\mathchardef\subsetneq="3\msb@28
\mathchardef\supsetneq="3\msb@29
\mathchardef\nsubseteq="3\msb@2A
\mathchardef\nsupseteq="3\msb@2B
\mathchardef\nparallel="3\msb@2C
\mathchardef\nmid="3\msb@2D
\mathchardef\nshortmid="3\msb@2E
\mathchardef\nshortparallel="3\msb@2F
\mathchardef\nvdash="3\msb@30
\mathchardef\nVdash="3\msb@31
\mathchardef\nvDash="3\msb@32
\mathchardef\nVDash="3\msb@33
\mathchardef\ntrianglerighteq="3\msb@34
\mathchardef\ntrianglelefteq="3\msb@35
\mathchardef\ntriangleleft="3\msb@36
\mathchardef\ntriangleright="3\msb@37
\mathchardef\nleftarrow="3\msb@38
\mathchardef\nrightarrow="3\msb@39
\mathchardef\nLeftarrow="3\msb@3A
\mathchardef\nRightarrow="3\msb@3B
\mathchardef\nLeftrightarrow="3\msb@3C
\mathchardef\nleftrightarrow="3\msb@3D
\mathchardef\divideontimes="2\msb@3E
\mathchardef\varnothing="0\msb@3F
\mathchardef\nexists="0\msb@40
\mathchardef\mho="0\msb@66
\mathchardef\eth="0\msb@67
\mathchardef\eqsim="3\msb@68
\mathchardef\beth="0\msb@69
\mathchardef\gimel="0\msb@6A
\mathchardef\daleth="0\msb@6B
\mathchardef\lessdot="3\msb@6C
\mathchardef\gtrdot="3\msb@6D
\mathchardef\ltimes="2\msb@6E
\mathchardef\rtimes="2\msb@6F
\mathchardef\shortmid="3\msb@70
\mathchardef\shortparallel="3\msb@71
\mathchardef\smallsetminus="2\msb@72
\mathchardef\thicksim="3\msb@73
\mathchardef\thickapprox="3\msb@74
\mathchardef\approxeq="3\msb@75
\mathchardef\succapprox="3\msb@76
\mathchardef\precapprox="3\msb@77
\mathchardef\curvearrowleft="3\msb@78
\mathchardef\curvearrowright="3\msb@79
\mathchardef\digamma="0\msb@7A
\mathchardef\varkappa="0\msb@7B
\mathchardef\hslash="0\msb@7D
\mathchardef\hbar="0\msb@7E
\mathchardef\backepsilon="3\msb@7F
\def\Bbb{\ifmmode\let\next\Bbb@\else
 \def\next{\errmessage{Use \string\Bbb\space only in math mode}}\fi\next}
\def\Bbb@#1{{\Bbb@@{#1}}}
\def\Bbb@@#1{\fam\msbfam#1}
%%%%%%%%%%%%%%%%%%%%%%%%%%%%%%%%%%%%%%%%%%%%%%%%%%%%%%%%%%%%%

\def\elevenpoint{
\textfont0=\caprm \scriptfont0=\sixrm \scriptscriptfont0=\fiverm
\def\rm{\fam0\caprm}
\textfont1=\capmi \scriptfont1=\sixmi \scriptscriptfont1=\fivemi
\textfont2=\capsy \scriptfont2=\sixsy \scriptscriptfont2=\fivesy
\textfont3=\capex \scriptfont3=\capex \scriptscriptfont3=\capex
\textfont\itfam=\capit \def\it{\fam\itfam\capit} % \it is family 4
\textfont\slfam=\capsl  \def\sl{\fam\slfam\capsl} % \sl is family 5
\textfont\bffam=\capbf \scriptfont\bffam=\sixbf
\scriptscriptfont\bffam=\fivebf
\def\bf{\fam\bffam\capbf} % \bf is family 6
\textfont4=\ninebi \scriptfont4=\sevenbi \scriptscriptfont4=\fivebi
\abovedisplayskip=11pt plus 3pt minus 8pt
\belowdisplayskip=\abovedisplayskip
\smallskipamount=2.7pt plus 1pt minus 1pt
\medskipamount=5.4pt plus 2pt minus 2pt
\bigskipamount=10.8pt plus 3.6pt minus 3.6pt
\normalbaselineskip=11pt
\setbox\strutbox=\hbox{\vrule height7.8pt depth3.2pt width0pt}
\normalbaselines \rm}
\catcode`\@=12
\def\sla#1{\mkern-1.5mu\raise0.4pt\hbox{$\not$}\mkern1.2mu #1\mkern 0.7mu}
\def\Dbar{\mkern-1.5mu\raise0.4pt\hbox{$\not$}\mkern-.1mu {\rm D}\mkern.1mu}
\def\Abar{\mkern1.mu\raise0.4pt\hbox{$\not$}\mkern-1.3mu A\mkern.1mu}
\nopagenumbers
%%%%%%%%%%%%%%%%%% headline %%%%%%%%%%%%%%%%%%%%%%%% 
\def\makeheadline{\vbox to 0pt{\vskip-27pt
\line{\vbox to8.5pt{}\the\headline}\vss}\nointerlineskip}
\def\subsectionname{}
\def\sectionname{}
\headline={\ifnum\pageno=1\hfill\else\draftdate\hfil{\headrm\folio}%
\hfil\hphantom{\draftdate}\fi}	 

%%%%%%%%%%%%%%%%%%%%%%%%%%%%%%%%%%%%%%%%%%%%%%%% %********* end ouput macros
%%%%%%%%%%%%%%%%%%%%%%%%%%%%%%%%%%%%%%%%%%%%%%%% 
% ****** extrait de definit.tex (obsolete ?)

% **************************************************************
\newcount\yearltd\yearltd=\year\advance\yearltd by -2000
\newif\ifdraftmode
\draftmodefalse
\def\draft{\draftmodetrue{\count255=\time\divide\count255 by 60
\xdef\hourmin{\number\count255} 
  \multiply\count255 by-60\advance\count255 by\time
  \xdef\hourmin{\hourmin:\ifnum\count255<10 0\fi\the\count255}}}
\def\draftdate{\ifdraftmode{\headrm\quad (\jobname,\ le
\number\day/\number\month/\number\yearltd\ \ \hourmin)}\else{}\fi} 
\newif\iffrancmode
\francmodefalse
% ********* A few math symbols
\def\e{\mathop{\rm e}\nolimits}
\def\sgn{\mathop{\rm sgn}\nolimits}

\def\d{{\rm d}}
\def\ud{{\textstyle{1\over 2}}}
\def\half{\ud}
\def\tr{\mathop{\rm tr}\nolimits}
\def\det{\mathop{\rm det}\nolimits}
\def\del{\partial}

\chardef\sigmat=27
\def\rf{\par\item{}}

\def\frac#1#2{{\textstyle{#1\over#2}}}

\def\leaderfill{\leaders\hbox to 1em{\hss.\hss}\hfill}
%%%%%%%%%%%%%%%%%%%%%%%%%%%%%%%%%%%%%%%%%%%%%%%% 
\catcode`\@=11

% ************** double alignment in eqalignno style **********************
\def\deqalignno#1{\displ@y\tabskip\centering \halign to
\displaywidth{\hfil$\displaystyle{##}$\tabskip0pt&$\displaystyle{{}##}$
\hfil\tabskip0pt &\quad
\hfil$\displaystyle{##}$\tabskip0pt&$\displaystyle{{}##}$ 
\hfil\tabskip\centering& \llap{$##$}\tabskip0pt \crcr #1 \crcr}}
%%%%%%%%%%%%%%% double eqalign %%%%%%%%%%%%%%%%%%%%%%%
\def\deqalign#1{\null\,\vcenter{\openup\jot\m@th\ialign{
\strut\hfil$\displaystyle{##}$&$\displaystyle{{}##}$\hfil
&&\quad\strut\hfil$\displaystyle{##}$&$\displaystyle{{}##}$
\hfil\crcr#1\crcr}}\,}
%***************************************************************************
% protection macro for undefined macros
\def\xlabel#1{\expandafter\xl@bel#1}\def\xl@bel#1{#1}
\def\label#1{\l@bel #1{\hbox{}}}
\def\l@bel#1{\ifx\UNd@FiNeD#1\message{label \string#1 is undefined.}%
\xdef#1{?.? }\fi{\let\hyperref=\relax\xdef\next{#1}}%
\ifx\next\em@rk\def\next{}%
%\else\ifx\next#1\xlabel#1\fi\fi\next
\else\def\next{#1}\fi\next}
%***************************************************************************
%********* titlepage, headline, chapter section, subsection, sub, appendix *********
%***************************************************************************
%**************** input with check of file existence ***********************
% Warning macro
\def\DefWarn#1{\ifx\UNd@FiNeD#1\else
\immediate\write16{*** WARNING: the label \string#1 is already defined%
***}\fi}% 
%NOW WORK syntax \cinput{filename} 
\newread\ch@ckfile
\def\cinput#1{\def\filen@me{#1 }% space mandatory after #1 !!
\immediate\openin\ch@ckfile=\filen@me
\ifeof\ch@ckfile\message{<< (\filen@me\ DOES NOT EXIST in this pass)>>}\else% 
\closein \ch@ckfile\input\filen@me\fi}
%********* introduce equation number file: for non-causal quotation
\ifx\UNd@FiNeD\prefix\def\prefix{}\fi % correction added
\newread\ch@ckfile
\immediate\openin\ch@ckfile=\jobname.def
\ifeof\ch@ckfile\message{<< (\jobname.def DOES NOT EXIST in this pass) >>}
\else
\def\DefWarn#1{}%
\closein \ch@ckfile% 
\input\jobname.def\fi
%**********
% Autre utilitaire
\def\listcontent{%\immediate\closeout\tfile%
\immediate\openin\ch@ckfile=\jobname.tab % space mandatory after tab!!
\ifeof\ch@ckfile\message{no file \jobname.tab, no table of contents this
pass}%
\else\closein\ch@ckfile%
\def\sectionname{\iffrancmode Table des
mati\`eres \else Contents\fi}
\centerline{\twbfx\iffrancmode Table des
mati\`eres \else Contents\fi}\nobreak\medskip% 
{\baselineskip=12pt\parskip=0pt\catcode`\@=11\input\jobname.tab
\catcode`\@=12\bigbreak\bigskip}\fi}
%**************************************************************************
\newcount\nochapter
\newcount\nosection
\newcount\nosubsection
\newcount\neqno
\newcount\notenumber
\newcount\nofigure
\newcount\notable
\newcount\noexerc
\newcount\fpage
\newcount\firstpage
\newif\ifappmode
\newwrite\equa
%\newwrite\tab 
%\newwrite\eqdf
% ******************* titlepage **********************************

\newdimen\hulp
\def\maketitle#1{
\edef\oneliner##1{\centerline{##1}}
\edef\twoliner##1{\vbox{\parindent=0pt\leftskip=0pt plus 1fill\rightskip=0pt
plus 1fill 
                     \parfillskip=0pt\relax##1}} 
\setbox0=\vbox{#1}\hulp=0.5\hsize
                 \ifdim\wd0<\hulp\oneliner{#1}\else
                 \twoliner{#1}\fi}
\def\preprint#1{\ifdraftmode\gdef\prepname{\jobname/#1}\else%
\gdef\prepname{#1}\fi\hfill{%\sacfont
\expandafter{\prepname}}\vskip20mm} 
% **************** beginning
\def\title#1\par{\gdef\titlename{#1}
%\global\fpage=\pageno
\global\firstpage=\pageno
\nosection=0
\mark{\the\nosection}
\maketitle{\ssbx\uppercase\expandafter{\titlename}}
\vskip17mm
\nochapter=0
\neqno=0
\notenumber=0
\nofigure=0
\notable=0
\def\prefix{}
\appmodefalse
\mark{\the\nochapter}
\message{#1}
%\immediate\openout\tab=\table
%%%%%%%%%%%%%%%%%%%%%\immediate\openout\equa=\equation%
\immediate\openout\equa=\jobname.equ %
%\immediate\openout\eqdf=\labeldefs
%\edef\ecrire{\write\tab{\par\noindent{\ssbx\ \titlename} 
%\string\leaderfill{\noexpand\number\pageno}}}\ecrire
%\edef\ecrire{\write\equa{{\titlename},
%{\noexpand\number\pageno}, Date \today}\write\equa{}}\ecrire
}
\ifbookmode% 
\headline={\ifnum\pageno=\firstpage\hfill\else\ifodd\pageno\rightheadline
\else\leftheadline\fi\fi}
\else
\headline={\ifnum\pageno=\firstpage\hfill\else\draftdate\hfil{\headrm\folio}%
\hfil\hphantom{\draftdate}\fi}\fi 
%}
\def\abstract{\vskip8mm\iffrancmode\centerline{R\'ESUM\'E}\else%
\centerline{ABSTRACT}\fi \smallskip \begingroup\narrower
\elevenpoint\baselineskip10pt} 
\def\endabstract{\par\endgroup \bigskip}
% ***************** input table of contents
%\def\content{\vfill\eject% A un bug dans le format double colonne
%\begingroup\centerline{\uppercase\expandafter{Table of contents}}% 
%\bigskip\elevenpoint\noindent% 
%\parindent=2em
%\openin 1=\jobname.tab
%\ifeof 1\closein1\message{<< (\jobname.tab DOES NOT EXIST. TeX again)>>}%
%\else\input\jobname.tab\closein 1\fi 
%\endgroup\vfill\eject}
%******************************* SECTION ****************************
%******************************************************************
\def\section#1\par{\vskip0pt plus.1\vsize\penalty-100\vskip0pt plus-.1
\vsize\bigskip\vglue\parskip\par%
\global\fpage=\pageno
\ifnum\nochapter=0\ifappmode\relax\else\writetoc
\fi\fi% ajout
\advance\nochapter by 1\global\nosection=0\global\neqno=0
\gdef\sectionname{#1}
%%%%%%%%%%%%%%%%%%%%%%%%%%%%%%%%%%%%%%%%%%%%%%
\vbox{\noindent\twbfx{\hyperdef\hypernoname{section}{\prefix\the\nochapter}%
{\prefix\the\nochapter}\ #1}}%
%%%%%%%%%%%%%%%%%%%%%%%%%%%%%%%%%%%%%%%%%
%\vbox{\noindent\twbfx\ifappmode\iffrancmode{Appendice\ }\else {Appendix\ }\fi
%\else\iffrancmode{Chapitre\ }\else {Chapter\ }\fi\fi%
\message{\prefix\the\nochapter\ \sectionname}%
%{\hyperdef\hypernoname{chapter}{\prefix\the\nochapter}{\prefix\the\nochapter}%
%\vskip11.5mm
%\vbox{\noindent\twbfx{#1}}}}\vskip9mm%
\writetoca{{\string\hyperref{}{section}{\prefix\the\nochapter}%
{\prefix\the\nochapter}} {#1}}%
%\mark{\the\nosection
\bigskip\noindent%
}

% appendix
\def\appendix#1#2\par{\bigbreak\nochapter=0
\appmodetrue
\notenumber=0
\neqno=0
\def\prefix{A}
\mark{\the\nochapter}
\message{APPENDICES}
%\leftline{APPENDICES}
{\hyperdef\hypernoname{section}{\prefix}{ 
\leftline{\uppercase\expandafter{#1}}
\leftline{\uppercase\expandafter{#2}}}}
\noindent\nonfrenchspacing
\writetoca{\string\hyperref{}{section}{\prefix}{Appendices}.\ #1 \ #2}%
}
% **************************** SUBSECTION *************************
\def\subsection#1\par{\vskip0pt plus.05\vsize\penalty-100\vskip0pt
plus-.05\vsize\bigskip\vskip1mm\advance\nosection by 1
\global\nosubsection=0
\def\subsectionname{#1}
\mark{#1}
\message{\the\nochapter.\the\nosection\ #1}
\vbox{\noindent\bf{\hyperdef\hypernoname{section}{\prefix\the\nochapter.%
\the\nosection}{\prefix\the\nochapter.\the\nosection\ #1}}}\vskip1mm%
%\mark{\the\nosection}
\smallskip\noindent% 
\writetoca{{\string\hyperref{}{section}{\prefix\the\nochapter.%
\the\nosection}{\prefix\the\nochapter.\the\nosection}} {#1}}%
} 
%%%%%%%%%%%%%%%%%% SUBSUBSECTION %%%%%%%%%%%%%%%%%%%%
\def\ssubsection#1\par{\vskip0pt plus.05\vsize\penalty-100\vskip0pt%
plus-.05\vsize\medskip\vskip\parskip\advance\nosubsection by 1%
\message{\the\nochapter.\the\nosection.\the\nosubsection\ #1}%
\vbox{\noindent\bf{\hyperdef\hypernoname{section}{\prefix\the\nochapter.%
\the\nosection.\the\nosubsection}{\prefix\the\nochapter.\the\nosection.%
\the\nosubsection\ \bf #1}}}%
%\vbox{\noindent\it\prefix\the\nochapter.\the\nosection.\the\nosubsection\
%\it #1}
\smallskip\noindent% 
\writetoca{{\string\hyperref{}{section}{\prefix\the\nochapter.%
\the\nosection.\the\nosubsection}{\prefix\the\nochapter.\the\nosection.%
\the\nosubsection}} {#1}}%
}   
%%%%%%%%%%%%%%%%%%%%%%%%%%%%%%%%%%%%%%%%%%%%%%%
%
\def\note #1{\advance\notenumber by 1
\footnote{$^{\the\notenumber}$}{\sevenrm #1}} 
% ?????

%\def\enchapter{\end}
\parindent=1em 
\newinsert\margin
\dimen\margin=\maxdimen
\count\margin=0 \skip\margin=0pt
%*****************************************************************
% IMPORTANT, new version demands chapter be defined before any section,
% section be defined before any subsection
\def\sslbl#1{\DefWarn#1%
\ifdraftmode{\leavevmode\vadjust{\smash%
{\line{{\escapechar=` \hfill\rlap{\sevenrm\hskip1mm\string#1}}}}}}%
\fi 
\ifnum\nochapter=0%
\if\prefix{}\xdef#1{}%
\edef\ewrite{\write\equa{{\string#1}}%
\write\equa{}}\ewrite%
\else
%%%%%%%%%%%%%%%%%%%%%%%%%%%%%%%%%%%%%%
\xdef#1{\noexpand\hyperref{}{section}{\prefix}{\prefix}}%
\edef\ewrite{\write\equa{{\string#1},\prefix}%
\write\equa{}}\ewrite%
\writedef{#1\leftbracket#1}
%%%%%%%%%%%%%%%%%%%%%%%%%%%%%%%%%%%%%%
\fi
\else%
\ifnum\nosection=0%
\xdef#1{\noexpand\hyperref{}{section}{\prefix\the\nochapter}%
{\prefix\the\nochapter}}%
\edef\ewrite{\write\equa{{\string#1},\prefix\the\nochapter}%
\write\equa{}}\ewrite%
\writedef{#1\leftbracket#1}
\else%
\ifnum\nosubsection=0%
\xdef#1{\noexpand\hyperref{}{section}{\prefix\the\nochapter.%
\the\nosection}{\prefix\the\nochapter.\the\nosection}}%
\writedef{#1\leftbracket#1}
\edef\ewrite{\write\equa{{\string#1},\prefix\the\nochapter%
.\the\nosection}\write\equa{}}\ewrite%
\else%
\xdef#1{\noexpand\hyperref{}{section}%
{\prefix\the\nochapter.\the\nosection.\the\nosubsection}%
{\prefix\the\nochapter.\the\nosection.\the\nosubsection}}%
\writedef{#1\leftbracket#1}
%\xdef#1{\prefix\the\nochapter.\the\nosection.\the\nosubsection}%
%\edef\ewrite{\write\eqdf{\string\def\string#1{\prefix\the\nochapter.%
%\the\nosection.\the\nosubsection}}\write\eqdf{}}\ewrite%
\edef\ewrite{\write\equa{{\string#1},\prefix\the\nochapter.\the\nosection%
.\the\nosubsection}\write\equa{}}\ewrite%
\fi\fi\fi}%
%**********************************************************************

% *************
\newwrite\tfile \def\writetoca#1{}
%       use this to write file with table of contents
\def\writetoc{\immediate\openout\tfile=\jobname.tab
\def\writetoca##1{{\edef\next{\write\tfile{\noindent ##1 \string\leaderfill%
%{\string\hyperref{}{page}{\noexpand\number\pageno}{\noexpand\number\pageno}}
\noexpand\number\pageno\par}}\next}}}

% ********************* references harvmac style ***********************
%     \ref\label{text}
% generates a number, assigns it to \label, generates an entry.
% To list the refs on a separate page,  \listrefs
%
\def\nolabels{\def\wrlabeL##1{}\def\eqlabeL##1{}\def\reflabeL##1{}}
\def\writelabels{\def\wrlabeL##1{\leavevmode\vadjust{\rlap{\smash%
{\line{{\escapechar=` \hfill\rlap{\sevenrm\hskip.03in\string##1}}}}}}}%
\def\eqlabeL##1{{\escapechar-1\rlap{\sevenrm\hskip.05in\string##1}}}%
\def\reflabeL##1{\noexpand\llap{\noexpand\sevenrm\string\string\string##1}}}
\ifdraftmode\writelabels\else\nolabels\fi

\global\newcount\refno \global\refno=1
\newwrite\rfile
\def\ref{[\hyperref{}{reference}{\the\refno}{\the\refno}]\nref}
\def\nref#1{\DefWarn#1%
\xdef#1{[\noexpand\hyperref{}{reference}{\the\refno}{\the\refno}]}%
\writedef{#1\leftbracket#1}%
\ifnum\refno=1\immediate\openout\rfile=\jobname.ref\fi
\chardef\wfile=\rfile\immediate\write\rfile{\noexpand\item{[\noexpand\hyperdef%
\noexpand\hypernoname{reference}{\the\refno}{\the\refno}]\ }%
\reflabeL{#1\hskip.31in}\pctsign}\global\advance\refno by1\findarg}
%       horrible hack to sidestep tex \write limitation%
\def\findarg#1#{\begingroup\obeylines\newlinechar=`\^^M\pass@rg}
{\obeylines\gdef\pass@rg#1{\writ@line\relax #1^^M\hbox{}^^M}%
\gdef\writ@line#1^^M{\expandafter\toks0\expandafter{\striprel@x #1}%
\edef\next{\the\toks0}\ifx\next\em@rk\let\next=\endgroup\else\ifx\next\empty%
\else\immediate\write\wfile{\the\toks0}\fi\let\next=\writ@line\fi\next\relax}}
\def\striprel@x#1{} \def\em@rk{\hbox{}}
\def\lref{\begingroup\obeylines\lr@f}
\def\lr@f#1#2{\DefWarn#1\gdef#1{\let#1=\UNd@FiNeD\ref#1{#2}}\endgroup\unskip}

\def\addref#1{\immediate\write\rfile{\noexpand\item{}#1}} %now unnecessary
\def\listrefs{{}\vfill\supereject\immediate\closeout\rfile\writestoppt
\baselineskip=14pt
\gdef\reference{\iffrancmode  R\'eferences \else References\fi}
\mark{\reference}  
\nochapter=0\def\sectionname{\reference} 
%%%%%%%%%%%%%%%%%%%%%%%%%%%%%%%%%
{\centerline{\bf \hyperdef\hypernoname{refer}{refer}{\bf \reference}}}
\nobreak\bigskip\noindent 
\writetoca{\string\hyperref{}{refer}{refer}{\reference}}
%%%%%%%%%%%%%%%%%%%%%%%%%%%%%%%%%
{\parindent=20pt%
\frenchspacing\escapechar=` \input \jobname.ref\vfill\eject}\nonfrenchspacing}
\def\startrefs#1{\immediate\openout\rfile=\jobname.ref\refno=#1}
\def\xref{\expandafter\xr@f}\def\xr@f[#1]{#1}
\def\refs#1{\count255=1[\r@fs #1{\hbox{}}]}
\def\r@fs#1{\ifx\UNd@FiNeD#1\message{reflabel \string#1 is undefined.}%
\nref#1{need to supply reference \string#1.}\fi%
\vphantom{\hphantom{#1}}{\let\hyperref=\relax\xdef\next{#1}}%
\ifx\next\em@rk\def\next{}%
\else\ifx\next#1\ifodd\count255\relax\xref#1\count255=0\fi%
\else#1\count255=1\fi\let\next=\r@fs\fi\next}
%************************
%*******
%
\newwrite\lfile
{\escapechar-1\xdef\pctsign{\string\%}\xdef\leftbracket{\string\{}
\xdef\rightbracket{\string\}}\xdef\numbersign{\string\#}}
\def\writedefs{\immediate\openout\lfile=\jobname.def \def\writedef##1{%
{\let\hyperref=\relax\let\hyperdef=\relax\let\hypernoname=\relax
 \immediate\write\lfile{\string\def\string##1\rightbracket}}}}%
\def\writestop{\def\writestoppt{\immediate\write\lfile{\string\pageno%
\the\pageno\string\startrefs\leftbracket\the\refno\rightbracket%
\string\def\string\secsym\leftbracket\secsym\rightbracket%
\string\secno\the\secno\string\meqno\the\meqno}\immediate\closeout\lfile}}
\def\writestoppt{}\def\writedef#1{}
\writedefs
% ******
% bibliography: not very satisfactory
\def\biblio\par{\vskip0pt plus.1\vsize\penalty-100\vskip0pt plus-.1
\vsize\bigskip\vskip\parskip
\message{Bibliographie}
{\leftline{\bf \hyperdef\hypernoname{biblio}{bib}{Bibliographical Notes}}}
\nobreak\medskip\noindent\frenchspacing
\writetoca{\string\hyperref{}{biblio}{bib}{Bibliographical Notes}}}%

%**************** autre version si plusieurs biblio ************************
\def\biblionote{\iffrancmode Notes Bibliographiques\else Bibliographical Notes
\fi}
\def\beginbib\par{\vskip0pt plus.1\vsize\penalty-100\vskip0pt plus-.1
\vsize\bigskip\vskip\parskip
\message{Bibliographie}
{\leftline{\bf \hyperdef\hypernoname{biblio}{\prefix\the\nochapter}%
{\biblionote}}}
\nobreak\medskip\noindent\frenchspacing
\writetoca{\string\hyperref{}{biblio}{\prefix\the\nochapter}%
{\biblionote}}}%

% *************** exercises: same comment
\def\Exercises{\iffrancmode Exercices\else Exercises
\fi}
\def\exerc\par{\vskip0pt plus.1\vsize\penalty-100\vskip0pt plus-.1
\vsize\bigskip\vskip\parskip\global\noexerc=0
\message{Exercises}
\iffrancmode\mark{Exercices}\else\mark{Exercises}\fi
{\leftline{\bf\hyperdef\hypernoname{exercise}{\the\nochapter}{\Exercises}}}
\nobreak\medskip\noindent\frenchspacing
\writetoca{\string\hyperref{}{exercise}{\the\nochapter}{\Exercises}}
}
\def\esubsec{\ifnum\noexerc=0\vskip-12pt\else\vskip0pt plus.05\vsize%
\penalty-100\vskip0pt plus-.05\vsize\bigskip\vskip\parskip\fi%
\global\advance\noexerc by 1
\hyperdef\hypernoname{exercise}{\the\nochapter.\the\noexerc}{}%
\vbox{\noindent\it \iffrancmode Exercice\else Exercise\fi\ \the\nochapter.\the\noexerc}\smallskip\noindent}
%%%%%%%%%%%%%%%%%%%%%%%%%%%
\def\exelbl#1{\ifdraftmode{\hfill\escapechar-1{\rlap{\hskip-1mm%
\sevenrm\string#1}}}\fi%
{\xdef#1{\noexpand\hyperref{}{exercise}{\the\nochapter.\the\noexerc}%
{\the\nochapter.\the\noexerc}}}%
\edef\ewrite{\write\equa{{\string#1}\the\nochapter.\the\noexerc}%
\write\equa{}}\ewrite%
\writedef{#1\leftbracket#1}}
%****************************

%*************************************************************************
%Macro de numerotation automatique
%*************************************************************************
% numbering without naming
\def\eqnn{\global\advance\neqno by 1 \ifinner\relax\else%
\eqno\fi(\prefix\the\nochapter.\the\neqno)}
%
% numbering and attaching a name: \eqnd{\ename}
\def\eqnd#1{\DefWarn#1%
\global\advance\neqno by 1 
{\xdef#1{($\noexpand\hyperref{}{equation}{\prefix\the\nochapter.\the\neqno}%
{\prefix\the\nochapter.\the\neqno}$)}}%???
\ifinner\relax\else\eqno\fi(\hyperdef\hypernoname{equation}{\prefix\the%
\nochapter.\the\neqno}{\prefix\the\nochapter.\the\neqno})
\writedef{#1\leftbracket#1}
\ifdraftmode{\escapechar-1{\rlap{\hskip.2mm\sevenrm\string#1}}}\fi
\edef\ewrite{\write\equa{{\string#1},(\prefix\the\nochapter.\the\neqno)
{\noexpand\number\pageno}}\write\equa{}}\ewrite}
%
% for eqalignno, allows (1a) (1b)...
\def\checkm@de#1#2{\ifmmode{\def\f@rst##1{##1}\hyperdef\hypernoname{equation}%
{#1}{#2}}\else\hyperref{}{equation}{#1}{#2}\fi}
\def\f@rst#1{\c@t#1a\em@ark}\def\c@t#1#2\em@ark{#1}
\def\eqna#1{\DefWarn#1%
\global\advance\neqno by1\ifdraftmode{\hfill%
\escapechar-1{\rlap{\sevenrm\string#1}}}\fi%
\xdef #1##1{(\noexpand\relax\noexpand%
\checkm@de{\prefix\the\nochapter.\the\neqno\noexpand\f@rst{##1}1}%
{\hbox{$\prefix\the\nochapter.\the\neqno##1$}})}
\writedef{#1\numbersign1\leftbracket#1{\numbersign1}}%
} 
\def\em@rk{\hbox{}} 
\def\xeqn{\expandafter\xe@n}\def\xe@n(#1){#1}
\def\xeqna#1{\expandafter\xe@na#1}\def\xe@na\hbox#1{\xe@nap #1}
\def\xe@nap$(#1)${\hbox{$#1$}}
% \eqns allows to quote several equations, suppressing unnecessary ()
\def\eqns#1{(\e@ns #1{\hbox{}})}
\def\e@ns#1{\ifx\UNd@FiNeD#1\message{eqnlabel \string#1 is undefined.}%
\xdef#1{(?.?)}\fi{\let\hyperref=\relax\xdef\next{#1}}%
\ifx\next\em@rk\def\next{}%
\else\ifx\next#1\xeqn#1\else\def\n@xt{#1}\ifx\n@xt\next#1\else\xeqna#1\fi
\fi\let\next=\e@ns\fi\next}
%**********************************************************************
%*************************** figure macros ****************************
% Pour centrer ajouter 16mm a la taille de la boite
\def\figure#1#2{\global\advance\nofigure by 1 \vglue#1%
\hyperdef\hypernoname{figure}{\the\nofigure}{}%
{\elevenpoint
\setbox1=\hbox{#2}
\ifdim\wd1=0pt\centerline{Fig.\ \the\nofigure\hskip0.5mm}%
\else\def\caption{Fig.\ \the\nofigure\quad#2\hskip0mm}
\setbox0=\hbox{\caption}
\ifdim\wd0>\hsize\noindent Fig.\ \the\nofigure\quad#2\else
                 \centerline{\caption}\fi\fi}\par}
% le bigskip a la fin a ete enleve!
 % obsolete, for compatibility
%***************
%figure alignee a gauche
\def\lfigure#1#2{\global\advance\nofigure by
1\vglue#1%
\hyperdef\hypernoname{figure}{\the\nofigure}{}%
\leftline{\elevenpoint\hskip10truemm  Fig.\
\the\nofigure\quad #2}} 
%***************
\def\figlbl#1{\ifdraftmode{\hfill\escapechar-1{\rlap{\hskip-1mm%
\sevenrm\string#1}}}\fi%
{\xdef#1{\noexpand\hyperref{}{figure}{\the\nofigure}%
{\the\nofigure}}}%
\edef\ewrite{\write\equa{{\string#1}\the\nofigure}%
\write\equa{}}\ewrite%
\writedef{#1\leftbracket#1}}
%****************************
\def\tablbl#1{\global\advance\notable by
1\ifdraftmode{\hfill\escapechar-1{\rlap{\hskip-1mm%
\sevenrm\string#1}}}\fi%
\hyperdef\hypernoname{table}{\the\notable}{}
{\xdef#1{\noexpand\hyperref{}{table}{\the\notable}%
{\the\notable}}}%
\edef\ewrite{\write\equa{{\string#1}\the\notable}%
\write\equa{}}\ewrite%
\writedef{#1\leftbracket#1}}

%***********************************************************************
%***********************************************************************
\catcode`@=12

\input epsf
\catcode`\À=\active\defÀ{\ifmmode\grave A\else\`A\fi}
\catcode`\É=\active\defÉ{\ifmmode\acute E\else\'E\fi}
\catcode`\à=\active\defà{\ifmmode\grave a\else\`a\fi}
\catcode`\â=\active\defâ{\ifmmode\hat a\else\^a\fi}
\catcode`\ä=\active\defä{\ifmmode\ddot a\else\"a\fi}
\catcode`\ç=\active\defç{\c c}
\catcode`\é=\active\defé{\ifmmode\acute e\else\'e\fi}
\catcode`\è=\active\defè{\ifmmode\grave e\else\`e\fi}
\catcode`\ê=\active\defê{\ifmmode\hat e\else\^e\fi}
\catcode`\ë=\active\defë{\ifmmode\ddot e\else\"e\fi}
\catcode`\ï=\active\defï{\ifmmode\ddot i\else\"\i\fi}
\catcode`\î=\active\defî{\ifmmode\hat i\else\^\i\fi}
\catcode`\ö=\active\defö{\ifmmode\ddot o\else\"o\fi}
\catcode`\ô=\active\defô{\ifmmode\hat o\else\^o\fi}
\catcode`\ù=\active\defù{\ifmmode\grave u\else\`u\fi}
\catcode`\ü=\active\defü{\ifmmode\ddot u\else\"u\fi}
\catcode`\û=\active\defû{\ifmmode\hat u\else\^u\fi}

\input colordvi
%\draft % PROBLEME AVEC LA NUMEROTATION DE LA PREMIERE PAGE
%%% U(\omega) or U_\r(\omega) ?
%%%%%%%%%%%%%%% subsection ??? CORRECT
\def\CS{{\rm CS}}

\def\g{g}
\def\Abbar{\mkern0.5mu\raise0.5pt\hbox{$\not$}\mkern-0.8mu{\bf A}\mkern.1mu}
\def\Ds{\mkern-0.5mu\raise0.5pt\hbox{$\not$}\mkern-.6mu {\bf D}\mkern.1mu}
\def\M{M}

\preprint{}

\title{3D field theories with Chern--Simons term for large $N$ in the Weyl gauge}

\centerline{Moshe Moshe${}^{*}$  and Jean Zinn-Justin${}^{**}$}

\medskip
{\baselineskip14pt

\centerline{${}^{*}$\it Department of Physics, Technion - Israel
Institute of Technology,} \centerline{\it Haifa, 32000 Israel}
\centerline{${}^{**}$ \it CEA/IRFU,  Centre de Saclay,}
%\centerline{and Institut de Physique
%Th\'eorique}
\centerline{\it 91191 Gif-sur-Yvette Cedex, France}
\footnote{}{${}^{*}$ ~E-mail: moshe@technion.ac.il}
\footnote{}{${}^{**}$ E-mail: jean.zinn-justin@cea.fr}}

\abstract
Three dimensional, $U(N)$ symmetric, field theory with fermion matter coupled to a topological Chern--Simons term, in the large $N$ limit is analyzed in details.  We determine the conditions for the existence of a massless conformal invariant ground state as well as the conditions for a massive phase. We analyze the phase structure and calculate gauge invariant corelators comparing them in several cases to existing results.  In addition to the non-critical explicitly broken scale invariance massive case we consider also a massive ground state where the scale symmetry is spontaneously broken. We show that such a phase appears only in the presence of a marginal deformation that is introduced by adding a certain scalar auxiliary field  and discuss the fermion-boson dual mapping. The ground state contains in this case a massless $U(N)$ singlet bound state goldstone boson- the dilaton whose properties are  determined.  We employ here the temporal gauge which is at variance with respect to past calculations using the light-cone gauge and thus, a check (though limited) of gauge independence is at hand. The large $N$ properties are determined by using a field integral formalism and the steepest descent method. The saddle point equations, which take here the form of integral equations for non-local fields, determine the mass gap and the dressed fermion propagator. Vertex functions are calculated at leading order in $1/N$ as exact solutions of integral equations. From the vertex functions, we infer gauge invariant two-point correlation functions for scalar operators and a current. Indications about the consistency of the method are obtained by verifying that gauge-invariant quantities have a natural $O(3)$ covariant form. As a further verification, in several occasions, a few terms of the perturbative expansion are calculated and compared with the exact results in the appropriate order.
\endabstract

\vfill\eject
\listcontent
\vfill\eject

%%%%%%%%%%%%%%%%%%   REFERENCES  HERE  %%%%%%%%%%%%%%%%%%%%
%%%%%%%%%%%%%%%%%%%%%  LISTREF at the end of the paper
\nref\Klebanov{
  I.R.~Klebanov and A.M.~Polyakov,
  ``AdS dual of the critical $O(N)$ vector model,''
{\it Phys.~Lett.~B} {\bf 550}  (2002) 213;
[hep-th/0210114].}

\nref\Sezgin{
  E.~Sezgin and P.~Sundell,
  ``Massless higher spins and holography,''\rf
{\it Nucl.~Phys.~} {\bf B644} (2002) 303; [{\it Erratum-ibid.} {\bf B 660} (2003) 403]; [hep-th/0205131] and
{\it JHEP} {\bf 0507}  (2005) 044;
[hep-th/0305040].
%%CITATION = hep-th/0205131%%
  R.~G.~Leigh and A.~C.~Petkou,
``Holography of the N=1 higher spin theory on AdS(4),''
JHEP {\bf 0306} (2003) 011;
[hep-th/0304217].
%%CITATION = hep-th/0304217%%
}

\nref\Fradkin{
  E.S.~Fradkin  and  M.A.~Vasiliev
  ``On the Gravitational Interaction of Massless Higher Spin Fields,''
{\it Phys.~Lett.~B} {\bf 189} (1987) 89  and  M.A.~Vasiliev
  ``Higher spin gauge theories: Star product and AdS space,''
in *Shifman, M.A. (ed.): The many faces of the superworld* 533-610.
[hep-th/9910096].
}
\nref\GiombiA{
  S.~Giombi and X.~Yin,
  ``Higher Spin Gauge Theory and Holography: The Three-Point Functions,''
{\it JHEP} {\bf 1009} (2010) 115;
[arXiv:0912.3462 [hep-th]];
%%CITATION = arXiv:0912.3462%%
  ``Higher Spins in AdS and Twistorial Holography,''
{\it JHEP} {\bf 1104}  (2011) 086;
[arXiv:1004.3736 [hep-th]];
%%CITATION = arXiv:1004.3736%%
  ``On Higher Spin Gauge Theory and the Critical $O(N)$ Model,''
{\it Phys.~Rev.~D} {\bf 85} (2012) 086005;
[arXiv:1105.4011 [hep-th]].
%%CITATION = arXiv:1105.4011%%
}
\nref\GiombiB{
  S.~Giombi, S.~Prakash and X.~Yin,
  ``A Note on CFT Correlators in Three Dimensions,''
{\it JHEP} {\bf 1307} (2013)  105;
[arXiv:1104.4317 [hep-th]].
%%CITATION = arXiv:1104.4317%%
}

\nref\rWadia{
  S.~Giombi, S.~Minwalla, S.~Prakash, S.P.~Trivedi, S.R.~Wadia and X.~Yin,
  ``Chern--Simons Theory with Vector Fermion Matter,''
{\it Eur.~Phys.~J.~C} {\bf 72} (2012)  2112;
[arXiv:1110.4386 [hep-th]].
%%CITATION = arXiv:1110.4386%%
}

\nref\AharonyI{
  O.~Aharony, G.~Gur-Ari and R.~Yacoby,
  ``$d=3$ Bosonic Vector Models Coupled to Chern--Simons Gauge Theories,''
{\it JHEP} {\bf 1203} (2012) 037;\rf
[arXiv:1110.4382 [hep-th]].
%%CITATION = arXiv:1110.4382%%
}

\nref\Jain{
  S.~Jain, S.P.~Trivedi, S.R.~Wadia and S.~Yokoyama,
  ``Supersymmetric Chern--Simons Theories with Vector Matter,''
{\it JHEP} {\bf 1210} (2012) 194;\rf
[arXiv:1207.4750 [hep-th]].
%%CITATION = arXiv:1207.4750%%
}

\nref\AharonyII{
  O.~Aharony, G.~Gur-Ari and R.~Yacoby,
  ``Correlation Functions of Large $N$ Chern--Simons-Matter Theories and Bosonization in Three Dimensions,''
{\it JHEP} {\bf 1212} (2012) 028;
[arXiv:1207.4593 [hep-th]]}
\nref\GurAri{
  G.~Gur-Ari and R.~Yacoby,
  ``Correlators of Large $N$ Fermionic Chern--Simons Vector Models,''
  {\it JHEP} {\bf 1302}, 150 (2013)
  [arXiv:1211.1866 [hep-th]].}

\nref\AharonyIII{
  O.~Aharony, S.~Giombi, G.~Gur-Ari, J.~Maldacena and R.~Yacoby,
  ``The Thermal Free Energy in Large $N$ Chern--Simons-Matter Theories,''
{\it JHEP} {\bf 1303} (2013) 121;
[arXiv:1211.4843 [hep-th]].
%%CITATION = arXiv:1211.4843%%
}

\nref\rMMZJ{M.~Moshe and J.~Zinn-Justin,
``Quantum field theory in the large $N$ limit: A Review,''
{\it Phys. Repts.} 385 (2003) 69-228; arXiv:hep-th/0306133v1.}
\nref\rBMB{
  W.A.~Bardeen, M.~Moshe and M.~Bander,
  ``Spontaneous Breaking of Scale Invariance and the Ultraviolet
  Fixed Point in $O(N)$ Symmetric $\Phi^{6}$ in Three-Dimensions Theory,''
{\it Phys.~Rev.~Lett.}  {\bf 52} (1984) 1188;
%%CITATION = FERMILAB-PUB-83-053-THY%%
D.J.~Amit and E.~Rabinovici,
  ``Breaking of Scale Invariance in $\phi^6$ Theory: Tricriticality and Critical End Points,''
{\it Nucl.~Phys.~B} {\bf 257} (1985)  371.
}
\nref\BHM{
  W.A.~Bardeen, K.~Higashijima and M.~Moshe,
  ``Spontaneous Breaking of Scale Invariance in a Supersymmetric Model,''
{\it Nucl.~Phys.~B} {\bf 250} (1985) 437;
 J.~Feinberg, M.~Moshe, M.~Smolkin and J.~Zinn-Justin,
  ``Spontaneous breaking of scale invariance and supersymmetric models at finite temperature,''
{\it Int.~J.~Mod.~Phys.~A}  (2005) {\bf 20}  4475.
}

\nref\rBM{
  W.A.~Bardeen and M.~Moshe,
  ``Spontaneous breaking of scale invariance in a $D=3$ $U(N)$ model with Chern--Simons gauge fields,''
{\it JHEP} {\bf 1406} (2014)  113;
[arXiv:1402.4196 [hep-th]].
%%CITATION = CERN-PH-TH-2013-299-%%
}

\nref\BardeenFermions{
  W.A.~Bardeen,
  ``The Massive Fermion Phase for the $U(N)$ Chern--Simons Gauge Theory in $D=3$ at Large $N$'',
%Submitted to: JHEP.
[arXiv:1404.7477 [hep-th]], to be published in JHEP.
%%CITATION = FERMILAB-PUB-14-113-T%%
}
%%%%%%%%%%%%%%%%%%%%%   end of added references    %%%%%

\nref\rZJrot{For an introduction see, for example,\rf
L. Alvarez-Gaum�, in   {\it Fundamental problems of gauge theory}, Erice 1985
G. Velo and A.S. Wightman eds. (Plenum Press, New-York 1986);\rf
J. Zinn-Justin, Chiral Anomalies and Topology School: Topology and Geometry in Physics,
Rot an der Rot, Germany, September 24-28 2001, Lecture Notes in Physics 659 (2005) 167-236, arXiv:hep-th/0201220v1.}

%\nref\ranomalies{The problem of chiral anomalies is discussed in\rf
%J.S. Bell and R. Jackiw, {\it Nuovo Cimento} A60 (1969) 47;
%S.L. Adler, {\it Phys. Rev.} 177 (1969) 2426;
%W.A. Bardeen, {\it Phys. Rev.} 184 (1969) 1848;
%D.J. Gross and R. Jackiw, {\it Phys. Rev.} D6 (1972) 477;
%H. Georgi and S.L. Glashow, {\it Phys. Rev.} D6 (1972) 429;
%C. Bouchiat, J. Iliopoulos and Ph. Meyer, {\it Phys. Lett.} 38B (1972) 519.
%\nrf See also the lectures\rf
%S.L. Adler, in {\it Lectures on Elementary Particles and Quantum Field Theory},
%S. Deser {\it et al}\/ eds. (MIT Press, Cambridge 1970);
%M.E.~Peskin, in {\it Recent Advances in Field Theory and Statistical Mechanics},
%Les Houches 1982, R. Stora and J.-B. Zuber eds. (North-Holland, Amsterdam
%1984);
%L. Alvarez-Gaum�, in   {\it Fundamental problems of gauge theory}, Erice 1985
%G. Velo and A.S. Wightman eds. (Plenum Press, New-York 1986).}

\nref\rtHooft{G. 't Hooft, ``A planar diagram theory for strong interactions'', {\it Nucl. Phys.} B72 (1974) 461-473.}
\nref\rPhDPGZJ{For a review on matrix models see
 P.~Di Francesco, P.~Ginsparg, J.~Zinn-Justin, ``2D
gravity and random matrices'', {\it Phys. Repts.} 254 (1995) 1-133, arXiv:hep-th/9306153v2.}
%\nref\rRLASIF{The axial gauge was first proposed for the non-Abelian gauge theory in
%R.L. Arnowitt and S.I. Fickler, {\it Phys. Rev.} 127 (1962) 1821.}

\section {Introduction}

It has been conjectured \refs{\Klebanov,\ \Sezgin}
  that  AdS/CFT  correspondence  implies a close relation between the singlet
sector of  $O(N)$ and $U(N)$
bosons vector theory  in $d=3$ space-time dimensions at large $N$ and Vasiliev's higher spin gravity theory \refs{\Fradkin}
on  ${\rm AdS}_4$. \sslbl\sIntrod
Computation of
correlation functions of the higher spin gauge theory \refs{\GiombiA,\ \GiombiB} strengthen this conjecture.
Since the quantum completion of the gravitational side of this duality is not known, only tree level could be considered.
The AdS/CFT correspondence in this case  provides a testing ground
of AdS/CFT ideas.
The large $N$ limit of
$O(N)_\kappa$ and $ U(N)_\kappa$ level $\kappa$ Chern--Simons gauge
theories, at $N,\kappa \to \infty$ with a fixed 't Hooft coupling
$\lambda= N/\kappa$, coupled to scalar and fermion matter fields
in the fundamental representation in d=3 dimensions were studied recently~ \refs{\rWadia{--}\AharonyIII}. Since pure
Chern--Simons gauge theories has no propagating degrees of
freedom, only the matter fields in the fundamental
representation provide the true independent canonical degrees of freedom in
these $d=3$ theories.
\par
$O(N)$ and  $U(N)$ symmetric
field theories at large $N$ are reasonably well understood \refs{\rMMZJ}.
 In the case of Chern--Simons matter theories in $d=3$ dimensions, the explicit calculations that were performed \refs{\rWadia{--}\AharonyIII}
shed extra light on the  AdS/CFT  correspondence. The original
$O(N)$ model of $g({\Bfg\phi}^2)^2$ in \refs{\Klebanov} was also deformed
to include a marginal
interaction term $\thicksim\lambda_6({\Bfg\phi}^2)^3$.
The massless conformal invariant phase
was analysed either by a generalized Hubbard--Stratanovich
method \refs{\rWadia} or in perturbation theory \refs{\AharonyI}.
It was
conjectured that the gravity dual of these 3D theories on the
boundary  is a parity broken version of Vasiliev's higher spin
theory \refs{\Fradkin} on  ${\rm AdS}_4$  in the bulk with a parity
breaking parameter $\theta$ which depends on the 't Hooft coupling
$\lambda= N/\kappa$. Supersymmetric extension of this idea was
introduced in \refs{\Jain}.
It has also
been conjectured that there is a duality
between the boson and fermion theories \refs{\AharonyII} since the theory of bosons and fermions in the fundamental representation of $U(N)$ coupled
to Chern--Simons gauge fields
are dual to the same bulk gravity theory on  ${\rm AdS}_4$.  Calculation of
the thermal free energy in large $N$ Chern--Simons field coupled to fermions and bosons
has further strengthened this duality \refs{\AharonyIII}. The thermal free energy has been calculated also
in \refs{\rWadia}.

\par

It was stated in Ref.~\refs{\AharonyIII} that the introduction of masses through
spontaneous breaking of scale invariance \refs{\rBMB,\ \BHM} is an alternative
at large $N$. Though at finite $N$ the breakdown of the scale symmetry
is explicit, it is of order ${\cal O}(N^{-1})$, namely the same order by which the  AdS/CFT  correspondence
is approximated in the above mentioned conjectures.
Therefore it was found appealing to further analyse the Chern--Simons matter theory with bosons \refs{\rBM} and fermions \refs{\BardeenFermions} with masses introduced through spontaneous breaking of scale invariance,
which ensures the existence of a massive ground state
but leaves the boundary $d=3$ theory
conformal to order ${\cal O}(N^{-1})$. The calculations mentioned above were done using the light cone gauge fixing for the Chern--Simons gauge theory.

\par

In this article we analyse the large $N$ limit of three dimensional, $U(N)$ symmetric, field theory with fermion matter coupled by the topological Chern--Simons gauge field.  Our calculational framework  differs from  the above mentioned  calculations as we are employing the $A_3=0$ gauge rather than the light-cone gauge. In the temporal gauge, as in the light-cone gauge,  the Chern--Simons action is reduced to a quadratic form and allows integrating out the gauge field. The large $N$ properties are determined by using then a path integral formalism and the steepest descent method in Euclidean space-time. In section \label{\sAnonab}
%and \label{\sTopoCS}
we start with a short discussion of the origin and properties of the topological Chern--Simons action.
In section  \label{\sCSfermions},  the detailed description is presented of the Chern--Simons gauge field in the $A_3=0$ ~gauge coupled to  $U(N)$ fermions in the fundamental representation. The large $N$ action is discussed and the saddle point equations are determined. The saddle point equations, which take here the form of integral equations for non-local fields are then solved. In particular, the fermion dressed two point function $\tilde\Gamma^{(2)}(p)$, the gap equation and the ground state energy in the limit $N\to \infty$ are calculated in details. in section
\label{\sCSfermions} we also derive
the gauge-independent correlators of $\bar\psi_\alpha(x)\psi_\alpha(x)$ two- and three-point functions at zero momentum. As mentioned, these results are obtained by calculating the relevant path integral at large $N$ and thus they are further checked in section \label{\sPerturbative} by calculating in perturbation theory the fermion propagator and the ground state energy up to two-loops and three-loops, respectively.
In section
 \label{\ssCSNexpand}
the vertex function ${\tilde \Gamma}^{1,2}(k;p_1,p_2)$ is calculated at leading order as solutions of integral equations.
\par

Using the exact expression of the vertex function given in section %\label{\svertex}
\label{\ssCSNexpand}, we compute in section \label{\ssCSvertextoRR} the exact expression for the two-point function of
 $R (x)= {1\over N}\bar{ \psi}  (x)\cdot{ \psi} (x)$.
 Several correlation functions involving the $J_\mu$ current are calculated in section
 \label{\sJJcorrel}.  Clearly, all these quantities are gauge-independent and can thus be compared to results obtained in other gauges.
Indications about the consistency of the method are obtained by verifying that gauge-invariant quantities have a natural $O(3)$ covariant form. As a verification, we calculate a few terms of the perturbative expansion of the same correlation functions.  In section \label{\sSigmaTerm},
 we introduce  a scalar field $\sigma(x)$ coupled to the fermions. This deformation is analogous to the triple-trace deformation $\lambda_6\phi^6(x) $ added to the Chern--Simons boson theory \refs{\AharonyIII,\ \rBM}. We show  that when a certain relation between the couplings constants is satisfied, the model exhibits spontaneous breaking of scale invariance, again in clear analogy to the Cherns--Simons boson case~\refs{\AharonyIII,\ \rBM}.
  The dilaton massless Goldstone boson excitation is a $U(N)$  singlet bound state  represented by an effective field  which, in the large $N$ limit, is proportional to to $\sigma(x)-\langle\sigma \rangle \propto \bar{ \psi}  (x)\cdot{ \psi} (x) $.
Summary and conclusions are found in section \label{\sconclusions}. Several technical points and useful relations are included in
 several appendices.
%%%%%%%%%%%\label{\appCSzero{--}\appCSJthree}.
\vfill\eject

\section 3D topological Chern--Simons action

We first shortly recall the origin of the 3D Chern--Simons action and its main properties.\sslbl\sAnonab

\subsection  Chiral anomaly, topology and instantons

We consider an Abelian axial current  in the framework of a non-Abelian
gauge theory in four Euclidean dimensions. \sslbl\ssCSrot The fermion fields transform non-trivially under a unitary gauge group
$G$ and ${\bf A}_{\mu}$ is the corresponding gauge field. The fermion action
reads
$$ {\cal S} (\bar \psi ,\psi;A ) = - \int \d  ^{4}x\, \bar
\psi(x) \Ds \psi (x) \eqnn $$
with the convention
$${\bf D}_\mu ={\bf 1}\partial _\mu+ {\bf A}_\mu(x)\,,\ \Ds = \sla{\partial} + \Abbar\,,  \eqnd\enonabcov   $$
where ${\bf A}_\mu$ is an anti-hermitian matrix, and  the curvature tensor
$${\bf F}_{\mu \nu}=[{\bf D}_\mu,{\bf D}_\nu]= \partial_\mu {\bf A}_\nu -\partial _\nu {\bf A}_\mu +[{\bf A}_\mu ,{\bf A}_\nu ].\eqnd\enonabcur $$
In a gauge transformation represented by a unitary matrix ${\bf g}(x)$, the gauge field ${\bf A}_\mu$ and the  Dirac operator become
$${\bf A}_\mu(x) \mapsto  {\bf g}  (x )\partial_ \mu {\bf g}
^{-1}  (x)+  {\bf g}  (x ){\bf A}_\mu(x) {\bf g}
^{-1}  (x)  \ \Rightarrow \Ds\mapsto {\bf g}^{-1}(x)\Ds{\bf g} (x)\  .\eqnd\eAnogaugDir $$
The axial current
$$ J_{\mu}^5(x)=i \bar\psi(x)\gamma_5 \gamma_{\mu}\psi(x) $$
is  gauge invariant. The axial anomaly \refs{\rZJrot} leads to a non-vanishing of the divergence of the axial current given by
$$\left<\partial_{\lambda}J^5_{\lambda}(x)\right>=-{i \over
16\pi^2}\epsilon_{\mu\nu\rho\sigma}\tr{\bf F}_{\mu\nu}{\bf F}_{\rho\sigma}\,.
\eqnd \enabanom  $$
From general arguments, one knows that the expression \eqns{\enabanom} must be a total derivative. Indeed, one verifies that
$$\epsilon_{\mu\nu\rho\sigma}\tr{\bf F}_{\mu\nu}{\bf F}_{\rho\sigma}=4\,
 \epsilon_{\mu\nu\rho\sigma}\partial_{\mu}\tr( {\bf A}_ \nu \partial_ \rho  {\bf A}_ \sigma + \frac{2}{3}  {\bf
A}_\nu {\bf A}_ \rho{\bf A}_\sigma). \eqnd \etotder  $$
\medskip
{\it Anomaly, topology and instantons.}
Gauge field configurations can be found that contribute to the chiral anomaly.   An especially interesting example is provided  by instantons, that is finite action solutions of Euclidean field equations.
%\sslbl{\ssCStopology}
\par
General instanton solutions can be exhibited by considering only  pure gauge theories and the gauge group $SU(2)$, since for a Lie group containing $SU(2)$ as a
subgroup the instantons are those of the $ SU(2) $
subgroup.\par
Instantons can be classified by a topological charge
$$Q({\bf A}_\mu)= {1\over 32\pi^2}  \int \d^{4}x\, {\bf F}_{\mu \nu}\cdot {\bf \tilde F}_{\mu \nu}\,  . \eqnd{\eineqFFd} $$
The expression $Q({\bf A}_\mu)$ is proportional to the integral of the chiral anomaly \eqns{\enabanom},  here written in $SO(3)$ notation. \par
We know that $ {\bf F}_{\mu \nu}\cdot {\bf \tilde F}_{\mu \nu}
$  is a pure divergence (Eq.~\eqns{\etotder}). In $SO(3)$ notation,
$$ {\bf F}_{\mu \nu}\cdot {\bf \tilde F}_{\mu \nu}= \partial_{\mu}  V_{\mu} $$
with
$$  V_\mu =   2\epsilon_{\mu \nu \rho \sigma} \left[ {\bf A}
_{\nu}\cdot \partial_{\rho} {\bf A}_{\sigma}+{\textstyle{1 \over 3}}
{\bf A}_{\nu}\cdot \left( {\bf A}_{\rho}\times {\bf A}_{\sigma}
\right) \right] .   \eqnd\etotderii    $$

The integral thus depends only on the behaviour of the
gauge field at large distances and its values are quantized.
\par
Stokes theorem implies
$$\int_{\cal D}\d^4 x\,\partial _\mu V_\mu =\int_{\partial {\cal D}}\d  \Omega\ \hat  n_\mu   V _{\mu}\,, $$
where $ \d  \Omega $ is the measure  on the boundary $\partial  {\cal D} $  of the four-volume $\cal D$ and $\hat n_\mu $ the unit vector normal to $\partial  {\cal D} $. We take for $\cal D$  a sphere of large radius $R$ and find for the topological charge
$$Q({\bf A}_\mu)=  {1\over 32\pi^ 2 }\int \d^{4}x\,\tr {\bf F}_{\mu \nu} \cdot  \tilde{\bf F}_{\mu
\nu}=  {1\over 32\pi^ 2 }R^3\int_{r=R} \d  \Omega\  \hat   n_\mu   V _{\mu}\,,
\eqnd\eSUiitop $$
The finiteness of the gauge action  implies that  classical solutions
must asymptotically become  pure gauges.
Since $SU(2)$ is topologically equivalent
to the sphere $S_ 3  $,  the pure gauge configurations on a sphere of large radius $|x|=R$
define a mapping from $  S_ 3  $ to $ S_ 3  $. Such mappings belong to different homotopy classes that are characterized by an integer
called the {\it winding number}.
Here, we identify the homotopy group $\pi_3(S_3)$, which again is isomorphic to the additive group of integers ${\Bbb Z}$.\par
Without explicit calculation,  one knows   from the analysis of the index of the Dirac operator, that the topological charge is an integer:
$$Q({\bf A}_\mu)={1\over 32\pi^2} \int \d^{4}x\, {\bf F}_{\mu \nu}\cdot {\bf \tilde F}_{\mu
\nu}= n\in {\Bbb Z}\,.\eqnn $$
%%%%%%%%%%%%%%%%%%%%%%%%%%%%%%%%%
\subsection Topological Chern--Simons action in 3D field theory

We now study some applications  to 3D field theory of the existence of topological terms  involving gauge fields in 4D quantum field theory, as recalled in   section \label{\ssCSrot}. We first discuss the pure Chern--Simons action and, then, the  Chern--Simons action coupled to fermion matter in the large $N$ limit.\sslbl\sTopoCS
%%%%%%%%%%%%%%%%%%%%%%%%%

%\subsection The  action

We first consider the 3D Euclidean action
in the form of a Chern--Simons (CS) term
$$\eqalignno{{\cal S}_{\rm CS}({\bf A})&=-{i\theta\over 8\pi^2}  \CS_3({\bf A})\ {\rm with}\cr
 \CS_3({\bf A})&=\epsilon_{\mu\nu\rho}\int\d^3 x\, \tr\left[ {\bf A}_ \mu(x) \partial_ \nu  {\bf A}_ \rho(x) + \frac{2}{3}  {\bf
A}_\mu(x) {\bf A}_ \nu(x){\bf A}_\rho(x)\right],&\eqnd{\eiiiDtop}\cr}$$
where ${\bf A}_\mu$ is a gauge field associated with $U(N)$ gauge transformations as defined by equation \eqns{\eAnogaugDir}:
$$ {\bf A}_\mu(x) \mapsto {\bf A}_\mu^g\equiv {\bf g}  (x )\partial_ \mu {\bf g}
^{-1}  (x)+  {\bf g}  (x ){\bf A}_\mu(x) {\bf g}
^{-1}  (x) .$$
The CS action is locally gauge invariant but globally  gauge invariant only up to a constant.
\par
Indeed, starting from four dimensions, using the relation \eqns{\etotder} and Stokes theorem, one infers that the expression
$$\int\d \sigma_3\,{\hat n}_\mu V_\mu\,,$$
where the ${\hat n}_\mu$ is the normal to a 3-surface,   $\d \sigma_3$ the corresponding surface element and
$$V_\mu=-4\epsilon_{\mu\nu\rho\sigma} \tr\left( {\bf A}_ \nu \partial_ \rho  {\bf A}_ \sigma + \frac{2}{3}  {\bf
A}_\nu {\bf A}_ \rho{\bf A}_\sigma\right),$$
is gauge invariant   up to a constant that depends only on a topological index,  a property explained in section \label{\ssCSrot}. The general result can then be applied to three-dimensional flat space.\par
As a consequence, strict gauge invariance of the integral over gauge fields then implies
$$\theta=2\pi \kappa\,,\eqnd{\eCSqunatized} $$
where $\kappa$ is an integer.
\medskip
{\it Chern--Simons term  and gauge transformations.}
As an exercise, we directly verify how $ \CS_3({\bf A}_\mu)$ transforms in a gauge transformation.
We first rewrite $ \CS_3({\bf A}_\mu)$ as
$$ \CS_3({\bf A}_\mu)=\epsilon_{\mu\nu\rho}\int\d^3 x\, \tr\left[\ud {\bf A}_ \mu(x)  {\bf F}_ {\nu\rho}(x) - \frac{1}{3}  {\bf
A}_\mu(x) {\bf A}_ \nu(x){\bf A}_\rho(x)\right].$$
It is convenient to  define
$${\bf B}_\mu(x)=-\partial_ \mu {\bf g}^{-1}  (x){\bf g}(x)=  {\bf g} ^{-1}  (x)\partial_ \mu {\bf g}  (x), $$
in such a way that gauge transformations read
$${\bf A}_\mu^g=    {\bf g}  (x )\left[{\bf A}_\mu(x)-{\bf B}_\mu(x) \right] {\bf g} ^{-1}  (x) .\eqnd{\eAnogaugDirb} $$
Then,
$$\partial_\nu{\bf B}_\rho(x)=-{\bf B}_\nu(x){\bf B}_\rho(x)+ {\bf g} ^{-1}  (x)\partial_ \nu\partial_\rho {\bf g}  (x) .$$
For a pure gauge, $\CS_3$ becomes
$$\CS_3({\bf g} )=\frac{1}{3}\epsilon_{\mu\nu\rho}\int\d^3 x\, \tr{\bf B}_\mu(x) {\bf B}_\nu(x){\bf B}_\rho(x). $$
The quantity,
$${\cal T}=\frac{1}{3}\epsilon_{\mu\nu\rho}\int\d^3 x\, \tr{\bf B}_\mu(x) {\bf B}_\nu(x){\bf B}_\rho(x).\eqnd\eCSiiiadd$$
 is  a topological term. Let us verify this statement directly. We set
$${\bf g}\mapsto {\bf g}\left({\bf 1}+{\bf t}+O\|{\bf t}^2\|\right).$$
Then,
$$\delta{\bf B}_\mu(x)=\partial_\mu{\bf t}(x)+ [{\bf B}_\mu(x), {\bf t}(x)]O\|{\bf t}^2\|.$$
The variation of expression \eqns{\eCSiiiadd} is
$$\delta {\cal T} \sim \epsilon_{\mu\nu\rho}\int\d^3 x\, \tr\partial_\mu{\bf t} (x) {\bf B}_\nu(x){\bf B}_\rho(x).$$
We integrate by parts. The integrated term vanishes because the manifold has no  boundaries. Thus,
$$\delta {\cal T}\sim \epsilon_{\mu\nu\rho}\int\d^3 x\, \tr{\bf t}(x)\left[{\bf B}_\mu(x){\bf B}_\nu(x){\bf B}_\rho(x)+{\bf B}_\nu(x){\bf B}_\mu(x){\bf B}_\rho(x)\right]=0\,.$$
Therefore, the quantity ${\cal T}$ does not depend on local changes of ${\bf g}(x)$ but only on its global properties.\par
Using gauge transformations in the form \eqns{\eAnogaugDirb}, we find
$$\eqalign{\CS_3({\bf A}^g_\mu)&=\epsilon_{\mu\nu\rho}\int\d^3 x\, \tr\left\{\ud\left[{\bf A}_\mu(x)-{\bf B}_\mu(x) \right]{\bf F}_ {\nu\rho}(x)  \right.\cr&\left.\quad-\frac{1}{3}\left[{\bf A}_\mu(x)-{\bf B}_\mu(x) \right]\left[{\bf A}_\nu(x)-{\bf B}_\nu(x) \right]  \left[{\bf A}_\rho(x)-{\bf B}_\rho(x) \right]         \right\}\cr
&= \CS_3({\bf A}_\mu)+\epsilon_{\mu\nu\rho}\int\d^3 x\, \tr\left\{ -{\bf B}_\mu(x)  \left[\partial_ \nu{\bf A}_\rho(x) +{\bf A}_\nu(x){\bf A}_\rho(x)\right] \right.\cr&\left.\quad+{\bf A}_\mu(x){\bf A}_\nu(x){\bf B}_\rho(x)-{\bf A}_\mu(x){\bf B}_\nu(x){\bf B}_\rho(x)        \right\}\cr&\quad+\frac{1}{3}\tr{\bf B}_\mu(x) {\bf B}_\nu(x){\bf B}_\rho(x)  .\cr}$$
We integrate by parts,
$$\eqalign{&-\int\d^3x\,\tr{\bf B}_\mu(x)\partial_ \nu  {\bf A}_ \rho (x)=-\int\d x_\mu\wedge\d x_\rho \tr{\bf B}_\mu(x){\bf A}_ \rho (x)\cr&\quad+\int\d^3 x\,\tr
 \left[-{\bf B}_\nu(x){\bf B}_\mu(x)+ {\bf g} ^{-1}  (x)\partial_ \nu\partial_\mu {\bf g}  (x) \right]{\bf A}_ \rho (x).\cr}$$
The vanishing of the boundary term implies that ${\bf B}_\mu$ should vanish and, thus, asymptotically ${\bf g}(x)$ should go to a constant for $|x|$ large. Therefore, ${\Bbb R}^3$ should be given the topology of $S_3$, as in the discussion of  section \label{\ssCSrot}. \par
Assuming this condition satisfied and expanding, we conclude
$$ \CS_3({\bf A}^g_\mu)=\CS_3({\bf A} _\mu) +\frac{1}{3}\epsilon_{\mu\nu\rho}\int\d^3 x\, \tr{\bf B}_\mu(x) {\bf B}_\nu(x){\bf B}_\rho(x). $$
The Chern--Simons changes by a quantity that depends on global properties of the gauge transformation. Its values have indeed been calculated in section  \label{\ssCSrot}.
\smallskip
{\it The classical equation of motion in an external source.} We add a source to the action and consider  the new action
$${\cal S}({\bf A},{\bf J})=-i{\kappa\over 4\pi}\CS_3({\bf A})+\int\d^3 x\tr{\bf J}_\mu(x){\bf A}_\mu(x).$$
The equation of motion obtained by varying ${\bf A}_\mu$ in $\CS_3$  yields
$${\delta \CS_3\over \delta {\bf A}_\mu}=\epsilon_{\mu\nu\rho}{\bf F}_{\nu\rho} $$
and, thus,
$${\bf F}_{\mu\nu}=-i{2\pi\over\kappa}\epsilon_{\mu\nu\rho}{\bf J}_\rho\ \Rightarrow\ {\bf D}_\mu{\bf J}_\mu=0\,.$$
For ${\bf J}=0$, the stationary solutions are pure gauges. In general, the gauge field has to be coupled to a gauge-covariant conserved source.
%%%%%%%%%%%%%%%%%%%%%%%

%%%%%%%%%%%%%%%%%%%%%%%%%%%%%%%%%%%%%
\subsection The Chern--Simons action in the temporal gauge

In general, to be able to define   integrals over gauge fields,   gauge fixing is required. In a general gauge, this leads to the introduction of Faddeev--Popov ghosts and  the solution of the quantum field theory in the large $N$ limit involves the summation of planar diagrams \refs{\rtHooft}, a problem whose solution is only known in lower dimensions \refs{\rPhDPGZJ}. However, in the case of the Chern--Simons action, in some special gauges like the   light-cone or axial gauges,
 the cubic interaction term in the CS action vanishes, and the integral over gauge fields can be performed explicitly. \par
In contrast to previous work that  mainly uses the light-cone gauge, we
  choose the gauge ${\bf A}_3=0$. Since we use an Euclidean formalism, we may identify $x_3$ with the Euclidean time (and below occasionally use the notation $t\equiv x_3 $ and $\omega\equiv p_3$ in the Fourier representation). The CS action reduces to the quadratic form
$$ \CS_3({\bf A}) =\int\d^3 x\, \tr\left[{\bf A}_2(x)\partial_3{\bf A}_1(x)-{\bf A}_1(x)\partial_3{\bf A}_2(x)\right].\eqnd{\eCStempaction}$$
Gauge invariance restricted to gauge functions ${\bf g}(x_1,x_2)$ should be maintained. Assuming some boundary conditions at Euclidean times $t_i$ and $t_f$
and integrating over time between $t_i$ and $t_f$, one finds
$$\int\d^2 x\tr\left[ \partial_2{\bf  g}\,{\bf g}^{-1}\bigl({\bf A}_1(t_f) - {\bf A}_1(t_i)\bigr)- \partial_1{\bf  g}\,{\bf g}^{-1}\bigl({\bf A}_2(t_f) - {\bf A}_2(t_i)\bigr)\right]=0\,,$$
where $\d^2 x\equiv \d x_1\,\d x_2$.
This condition is automatically satisfied for the quantum partition function at finite temperature. Here, we work at zero temperature and set topological issues aside.
\medskip
{\it The gauge propagator.} For convenience we now set,
$$  {\g\over 4\pi}={N\over \kappa}\,.\eqnd{\eCScouplingdef}$$
The topological action then reads
$${\cal S}_{\rm CS}({\bf A})={N\over  i\g}\CS_3({\bf A})={N\over i\g}\int\d^3 x \,  \tr\left[{\bf A}_2(x)\partial_3{\bf A}_1(x) -{\bf A}_1(x)\partial_3{\bf A}_2(x)\right].\eqnn   $$

The propagator can be calculated by adding  external sources. Then,
$${\cal S}_{\rm CS}({\bf A},{\bf J}) ={\cal S}_{\rm CS}({\bf A})-\int\d^3 x\,  \tr\left[{\bf J}_1(x){\bf A}_1(x)+{\bf J}_2(x){\bf A}_2(x)\right]  .\eqnd\eCSAsources   $$

The solutions of the corresponding equations of motion are
$$ {\bf A}_1(x)={ i\g\over 2N}\int^{x_3}\d x'_3\, {\bf J}_2(x_1,x_2,x_3') ,\quad  {\bf A}_2(x)=-{ i\g\over 2N}\int^{x_3}\d x'_3\, {\bf J}_1(x_1,x_2,x_3')\,.$$
Note that, in the absence of explicit boundary conditions, the inverse of $\partial_3$ is only defined up to the addition of a function of $x_1,x_2$. Independence of boundary terms implies
$$\int\d x_3\,{\bf J}_{1,2}(x)=0\,.$$
These conditions enforce the remaining gauge invariance  of the action corresponding to group elements ${\bf g}(x_1,x_2)$. They can only be used if the time Fourier components are quantized, for example, in the case of the partition function where fermions satisfy antiperiodic boundary conditions.\par
The action calculated for the solution   then is
$${\cal S}_{\rm CS}= {i\g\over4N}\int\d^3 x\, \d^3 x'\,\delta^{(2)}(x-x')\sgn(x_3 -x_3')\tr{\bf J}_2(x ){\bf J}_1(x') . $$
Using
$$\sgn(t)\mathop{=}_{\varepsilon\to0}{1\over 2 \pi}\int\d   \omega\,\e^{i\omega t}{-2i\omega\over\omega^2+\varepsilon^2}\Leftrightarrow -{2\omega i\over \omega^2+\varepsilon^2}\mathop{=}_{\varepsilon\to0}\int\d t\e^{-i\omega t}\sgn(t) \,,\eqnd{\esgnFourier}$$
we find that the ${\bf A}_1{\bf A}_2$ propagator, which is proportional to
$(i\g/4N) \sgn(t -t')\delta^{(2)}(x-x')$, has the Fourier representation
$$\Delta_{\bf A}(p)={\g\over2 N}\lim_{\varepsilon\to0} { p_3\over p_3^2+ \varepsilon^2} = { g\over 2N}{\rm PP}{1\over p_3}\,\eqnd{\ePropA} .$$
%%%%%%%%%%%%%%%%%%%%%%%%%%%%%%%%%
%%%%%%%%%%%%%%%%%%%%%%
\section Chern--Simons gauge fields coupled to $U(N)$ fermions

We first describe a $U(N)$ symmetric fermion  theory with  Chern--Simons term and then study its large $N$ limit.\sslbl\sCSfermions

\subsection Conventions

 In this section, we assume that ${\bf A}_\mu$ are  antihermitian matrices belonging to the adjoint representation of the group $U(N)$.
As a basis of the Lie algebra, one can take $N^2$ antihermitian matrices ${\bf t}^a$ orthogonal by the trace
$$\tr{\bf t}^a{\bf t}^b=-\delta_{ab}\,.\eqnd{\eorthogonalgen}$$
With this convention,
$$[{\bf t}^a,{\bf t}^b]=f_{abc}{\bf t}^c, $$
where the structure constants $f_{abc}$ are real and totally antisymmetric.
Moreover, the orthogonality relations \eqns{\eorthogonalgen} imply
$$t^a_{ij}t^a_{\ell k}=- \delta_{ik}\delta_{j\ell}\ \Rightarrow\  {\bf t}^a{\bf t}^a=-N{\bf 1}\,.\eqnd{\ecomplete} $$
In terms of the ${\bf t}^a$ matrices, the gauge field then can be parametrized as
$${\bf A}_\mu(x)=A_\mu^a(x){\bf t}^a.$$
The Chern--Simons action then reads
$${\cal S}({\bf A})={i N\over \g}\int\d^3 x\left[A_2^a(x)\partial_3 A_1^a(x)-A_1^a(x)\partial_3 A_2^a(x) \right].\eqnn $$
If we add to the action the source terms
$$-\int\d^3 x\,\left[J_1^a(x)A_1^a(x)+J_2^a(x)A_2^a(x)\right] $$
and integrate over the gauge field, we obtain
$${\cal S}={\g \over 4i N}\int\d^3 x\, \d^3 x'\,\delta^{(2)}(x-x')\sgn(x_3 -x_3')J_1
^a(x)J_2^a(x').\eqnd{\eCSintegrated}$$
The   non-vanishing component  of the gauge propagator then is
$$\Delta^{ab}_{12}(x-x')\equiv\left<A_1^a(x)A_2^b(x')\right>={i \g\over 4N}\delta_{ab}\delta^{(2)}(x-x')\sgn(x_3 -x_3').$$
In the Fourier representation,
$$A^a_\mu(x)=\int\d^3 p\,\e^{ipx}\tilde A^a_\mu(p),\quad J^a_\mu(x)=\int\d^3 p\,\e^{ipx}\tilde J^a_\mu(p),$$
the CS action takes the form
$${\cal S}({\bf A})=-{  2N\over \g}(2\pi)^3\int\d^3 p\, \tilde A_1^a(p) p_3\tilde A_2^a(-p)  \eqnd{\eCSactionFourier}$$
and the source terms become
$$ -(2\pi)^3\int\d^3 p\,\left[\tilde J_1^a(-p)\tilde A_1(p)+\tilde J_2^a(-p)\tilde A_2(p)\right]. $$
An integration over the gauge field yields the free energy
$$\ln{\cal Z}=   (2\pi)^3 {\g \over 2N}\int\d^3 p\,\tilde J_2^a(- p){1\over p_3}\tilde J_1^a(p). $$
 The Fourier representation of the gauge field propagator is
$$\Delta^{ab}_{\alpha\beta}(x )={1\over(2\pi)^3}\int\d^3 p\,\e^{ipx}\tilde\Delta^{ab}_{\alpha\beta}(p) $$
  where ${\alpha,\beta}=\{1,2\}$ with
$$\tilde\Delta^{ab}_{\alpha\beta}(p) =\epsilon_{\alpha\beta}\delta^{ab}{\g \over 2N} {1\over p_3}\,.\eqnd{\eApropFourier}$$
 where $\epsilon_{12}=1=-\epsilon_{21}$.
Later we  use a simple pole notation, but the pole term always stands for a principal part: PP($1/p_3$) as depicted in Eqs.~\eqns{\esgnFourier,\ \ePropA}.
%%%%%%%%%%%%%%%%%%%%%%%%%%%%%%%%%%%%%
\subsection The CS action coupled to a  gauge-invariant fermion action

We now add to the Chern--Simons action, quantized in the ${\bf A}_3=0$ gauge, a $U(N)$ gauge-invariant action for an $N$-component spinor field ${\Bfg\psi}$,
$${\cal S}(\psi,\bar\psi,{\bf A})={\cal S}_{\rm CS}({\bf A})+{\cal S}_{\rm F}(\psi,\bar\psi,{\bf A})\eqnd{\eCSUNApsi}$$
with
$${\cal S}_{\rm F}(\psi,\bar\psi,{\bf A})=-\int\d^3 x\,\bar{\Bfg\psi}(x)(\Ds+M_0){\Bfg\psi}(x) \,,\eqnd{\eCSpsiaction} $$
where $\Ds$ is defined in \eqns{\enonabcov} and we have added a mass term, which  violates parity, like the CS action. The $\gamma$-matrices here reduce  to the three Pauli matrices:
$$   \sigma_ 1    = \pmatrix{ 0 &  1 \cr 1 &
0 \cr} ,\ \sigma_2   = \pmatrix{
0 &  -i \cr i & 0 \cr},\ \sigma_3   = \pmatrix{ 1 &
0 \cr 0 &  -1 \cr}.\eqnd {\esigmadef}  $$
\smallskip
{\it Regularization.}
Power counting shows that the coupling constant is dimensionless and, thus, UV divergences are expected. Moreover, the gauge field propagator is not isotropic, which leads to additional difficulties. In a first step, we write formal expressions and postpone the regularization problem, a non-trivial issue in a non-Abelian gauge theory with a Chern--Simons term.
Fortunately, we will discover that here the number of independent divergent contributions is small and we will thus deal with the problem in a rather empirical way.
%%%%%%%%%%%%%%%%%%%%%%%%%%%%%%%%%%%
\subsection Large $N$ expansion: field integral formalism

Solving non-Abelian gauge theories in the large $N$ limit is in general a highly non-trivial problem \refs{\rtHooft}, but the problem drastically simplifies when the gauge action reduces to a Chern--Simons term, at least in some specific gauges.\par
To solve the field theory with the action \eqns{\eCSUNApsi} in the large $N$ limit we use a field integral formalism, a standard method to generate large $N$ expansions \refs{\rMMZJ}, even if here the context is different and new features can be expected \refs{\rWadia}.\sslbl\sCSNpath\par
We thus consider the field integral
$${\cal Z}=\int[\d\psi][\d\bar\psi][\d{\bf A}]\e^{-{\cal S}(\bar\psi,\psi,{\bf A})},$$
where ${\cal S}$ is the Euclidean action \eqns{\eCSUNApsi}.\par
In the ${\bf A}_3=0$ gauge, the gauge action is quadratic and the integration over the gauge field can be performed explicitly. One then obtains an effective  quartic interaction for the fermions, non-local in Euclidean time. The components of the currents coupled to the gauge field are
$$  J_\mu^a(t,x)= \bar\psi_\alpha ^i(x)[\gamma_\mu]_{\alpha\beta}t^a_{ij}\psi_ \beta^j(x)  , $$
where the lower index is the spinor index and the upper index the $U(N)$ vector index.\par
Using the result \eqns{\eCSintegrated}, one finds
$$\eqalign{ {\cal S}& =-\int \d^3 x\,\bar{\Bfg\psi}( x)(\sla{\partial}+M_0)\cdot{\Bfg\psi}( x)+{\g\over4i N} \int\d^3 x\,\d^3 x'\,\sgn(x_3-x'_3)\delta^{(2)}(x-x')\cr&\quad\times\bar\psi_\alpha^i( x)[\sigma_1]_{\alpha\beta} t^a_{ij}\psi_\beta^j(x) \bar\psi_\gamma^k(x')[\sigma_2]_{\gamma\delta}  t^a_{kl}\psi_ \delta^l( x') .\cr}$$
With the help of the identity \eqns{\ecomplete}, the sum over the group index $a$ can be performed and yields
$$\eqalignno{&{\cal S}=-\int \d^3 x\,\bar{\Bfg\psi}( x)\cdot(\sla{\partial}+M_0){\Bfg\psi}( x)
\cr&\quad+{\g \over 2N}\int\d^3 x\,\d^3 x' \,\sgn(x_3-x'_3)\delta^{(2)}_{\rm T}(x-x')\bar{\Bfg\psi} _1( x)\cdot{\Bfg\psi}_1(  x')   \bar{\Bfg\psi}_2(x' )\cdot{\Bfg\psi}_2 ( x )   ,\hskip8mm& \eqnd{\eCSUNpsi}\cr}$$
where
$$\delta^{(2)}_{\rm T}(x-x')\equiv \delta(x_1-x'_1)\delta(x_2-x'_2).$$
In terms of the Fourier components
$$\psi(x)=\int\d^3 p\,\e^{ipx}\psi(p),\ \bar\psi(x)=\int\d^3 p\,\e^{-ipx}\bar\psi(p),$$
the action \eqns{\eCSUNpsi} becomes
$$\eqalignno{{\cal S}&=-(2\pi)^3\int\d^3 p\,\bar{\Bfg\psi}(p)\cdot(i\sla{p}+M_0){\Bfg\psi}(p)
\cr&\quad-{ i\g \over N}(2\pi)^3\int\d^3 p\,\d^3 p'\d^3 q \,\d^3 q'\,\delta^{(3)}(p+q-p'-q')\cr&\quad\times\bar{\Bfg\psi} _1(p)\cdot{\Bfg\psi}_1(p') {\rm PP}{ 1\over  q_3-p'_3 }  \bar{\Bfg\psi}_2(q )\cdot{\Bfg\psi}_2 (q')   .\hskip7mm & \eqnd{\eCSUNtildepsi}\cr}$$
%%%%%%%%%%%%%%%%%%%%%%%%%%%%%%%%%%%%%%%%%%%%%%%%%%%%%%%%%%%%%%%%%
\subsection{The large $N$ action}

To render the $N$-dependence explicit and be able to study the large $N$ limit, we introduce
additional bilocal (in Euclidean  time) composite  fields $\{\rho_\alpha(t',t ,x) \}$
 and $\{\lambda_\alpha(t,t',x)\}$, $\alpha=1,2$, with $x\in{\Bbb R}^2$, to implement the relation \sslbl\sslargeN
$$\rho_\alpha(t',t ,x)={1\over N}\bar{\Bfg\psi} _ \alpha(t,x)\cdot{\Bfg\psi}_ \alpha(t',x).\eqnd\eCSthodef $$
Note that only
$R_\alpha(t,x)\equiv\rho_\alpha(t,t,x)$
is gauge invariant.\par
%%%%%%%%%%%%%%%%%%%%%%%%%%%%%%%%%
We then add to the action \eqns{\eCSUNtildepsi},
$${\cal S}(\lambda,\rho)=  \int\d^2 x\,\d t \,\d t'\sum_{\alpha=1}^2 \lambda_\alpha(t,t',x)\left[  N\rho_\alpha(t',t ,x)- \bar{\Bfg\psi} _ \alpha(t,x)\cdot{\Bfg\psi}_ \alpha(t',x) \right].\eqnn$$
This is a
generalization  (see also Ref.~\refs{\rWadia} and references therein)
of the standard method \refs{\rMMZJ} and a reflection of the required gauge field integration.
%%%%%%%%%%%%%%%%%%%%%%%%%%%%%%%%%%%%%%%
The total action can  be written as
$$\eqalign{{\cal S}&=-\int\d t\,\d^2 x\,\bar{\Bfg\psi}(t,x)\cdot(\sla{\partial}+M_0){\Bfg\psi}(t,x)\cr&\quad+\ud\g N \int\d^2 x\,\d t \,\d t'\sgn(t'-t)\rho_1(t',t ,x)\rho_2(t ,t',x)\cr
&\quad+ \int\d^2 x\,\d t \,\d t'\sum_{\alpha=1}^2 \lambda_\alpha(t,t',x)\left[ N\rho_\alpha(t',t ,x) - \bar{\Bfg\psi} _ \alpha(t,x)\cdot{\Bfg\psi}_ \alpha(t',x)\right].\cr}  $$
The integration over $\bar\psi$ and $\psi$ can  be performed and generates the factor
$$ (\det {\bf K})^N =\e^{N\tr\ln{\bf K}},$$
where the operator $\bf K$ is represented by the kernel
$$K_{\alpha\beta}(t,x;t',x')= \left(\sla{\partial}_{\alpha\beta}+ \delta_{\alpha\beta}M_0\right)\delta(t-t')\delta^{(2)}(x-x')+ \delta_{\alpha\beta}\lambda_\alpha(t,t',x)\delta^{(2)}(x-x') .\eqnd{\eCSferiiiKop}$$
The large $N$ action then reads
$$\eqalignno{&{\cal S}_N/N =  -\tr\ln{\bf K} \cr&+\int\d^2 x\,\d t \,\d t'\left[\sum_{\alpha=1}^2 \lambda_\alpha(t,t',x)   \rho_\alpha(t',t ,x)+ \ud\g\sgn(t'-t)\rho_1(t',t ,x)\rho_2(t ,t',x)\right]   .\cr&&\eqnd{\eCSferSN}\cr}  $$
The integration over the $\rho$-field is also Gaussian and can be performed. It amounts  to replacing $\rho$ by the solution of the field equation.
%%%%%%%%%%
\subsection{Saddle point equations}

A non-trivial large $N$ limit is obtained by taking the   limit with $\g $ fixed \refs{\rtHooft}. The integral is then dominated by saddle points, solution of equations   obtained by varying the $\rho$- and $\lambda $-fields.
A variation of the $\rho$-fields yields
\eqna{\erhoirohii}
$$\eqalignno{ \lambda_1(t ,t',x)&=\ud\g\sgn(t  -t')\rho_2(t  ,t' ,x) , & \erhoirohii{a}\cr \lambda_2(t ,t' ,x)&= -\ud \g\sgn(t -t') \rho_1(t ,t',x).& \erhoirohii{b} \cr}$$
Varying the $\lambda$-fields, one obtains
$$ - \rho_\alpha(t,t',x)+[{\bf K}^{-1}]_{\alpha\alpha} (t,x;t',x)=0\,,\ \alpha=1,2\,.\eqnd{\eCSrhoalphaK}  $$
One looks for solutions that do not break time and space translation invariance and, thus,
$$\lambda_\alpha(t,t';x)=\lambda_\alpha(t-t'),\ \rho_\alpha(t,t';x)=\rho_\alpha(t-t').$$
Equations \erhoirohii{} then reduce to
\eqna{\erhoirohiib}
$$\eqalignno{\lambda_1( t)&= \ud\g\sgn(t)\rho_2(t) , & \erhoirohiib{a}\cr   \lambda_2( t )&=-\ud\g\sgn(t) \rho_1(t ).& \erhoirohiib{b} \cr}$$
We define
$$\tilde \lambda_\alpha(\omega)={1\over2\pi}\int\d t\,\e^{-i\omega t}\lambda_\alpha(t), \ \tilde\rho_\alpha(\omega)={1\over2\pi}\int\d t\,\e^{-i\omega t}\rho_\alpha(t).$$
The operator \eqns{\eCSferiiiKop} in the Fourier representation  takes a diagonal form in $\omega,p$ space ($p\in{\Bbb R}^2$), with elements the $2\times2$ matrices
$$\tilde{\bf K}(\omega,p)=i\omega\sigma_3+i \sla{p}+M_0 +\pi\left(\tilde\lambda_1(\omega)+\tilde\lambda_2(\omega)\right)+\pi\left(\tilde\lambda_1(\omega)-\tilde\lambda_2(\omega)\right)\sigma_3 \,.$$
%%%%%%%%%%%%%%%%%%%%%%%%%%%%%%%%%%%%%%%%%%%%%%%%%%%%%%%%%%%%%%%%%%%%%%%%%%%
%%%%%%%%%%%%%%%%%%%%%%%%%%%%%%%%%%%%%%%%%%%%%%%%%%%%%%%%%%%%%%%%%%%%%%%%%
The action density then  reads
$$\eqalignno{  {\cal S}_N/N/{\rm volume}  &=  2\pi\int \d \omega\,\tilde\rho_\alpha( \omega)\tilde\lambda_\alpha(\omega)+ i\g\int{\d\omega\,\d\omega' \over  \omega-\omega'   }\tilde  \rho_1(\omega)\tilde\rho_2(\omega')\cr&\quad
-{1\over(2\pi)^3}\int\d \omega\,\d^2 p\,\tr\ln\tilde{\bf K}(\omega,p).&\eqnd{\eCSNactiondensity}\cr}
  $$
%%%%%%%%%%%%%%%%%%%%%%%%%%%%%%%%%%%%%%%%%%%%%%%%%%%%%%%%%%%%%%%%%%%%
It is convenient to set
$$\mu_ 1(\omega)=M_0+i\omega+2\pi\tilde \lambda_1(\omega)\,,\quad\mu_ 2(\omega)=M_0-i\omega+2\pi\tilde \lambda_2(\omega).\eqnn $$
Then, $\tilde{\bf K}$ reduces to
$$\tilde{\bf K}= i p_1\sigma_1+i p_2\sigma_2  +\ud\bigl(\mu_1(\omega)+\mu_2(\omega)\bigr)+\ud \bigl(\mu_1(\omega)-\mu_2(\omega)\bigr)  \sigma_3 \,.$$
Its inverse can be obtained in the form
$${\cal D}\tilde{\bf K}^{-1}=\ud\bigl(\mu_1(\omega)+\mu_2(\omega)\bigr)- \ud \bigl(\mu_1(\omega)-\mu_2(\omega)\bigr)  \sigma_3 -i p_1\sigma_1-i p_2\sigma_2 \eqnd{\eCSferNiipt}     $$
with
$${\cal D}=p_1^2+p_2^2+\mu_1(\omega)\mu_2(\omega).$$
Symmetries imply
$$ \tilde\rho_2( \omega)=\tilde\rho_1(- \omega)=\rho^*_1(\omega) , \quad\mu_2( \omega)=\mu_1(- \omega)=\mu_1^*(\omega)  . $$
After Fourier transformation, Eqs.~\erhoirohiib{} become
%%%%%%%%%%%%%%%%%%%%%%%%%%%%%%%%%%%%%%%%%%%%%%%%%%%%%%%%%%%%%%%
%%%%%%%%%%%%%%%%%%%%%%%%%%%%%%%%%%%%%%%%%%%%%%%%%%%%%%%%%%%%%%
\eqna\eCSmusaddleb
$$\eqalignno{\mu_1(\omega)&=M_0+i\omega-i\g\int{\d \omega'\over\omega -\omega'}\tilde\rho_2(\omega'),&\eCSmusaddleb{a} \cr
\mu_2(\omega)&=M_0-i\omega+i\g\int{\d \omega'\over\omega -\omega'}\tilde\rho_1(\omega').& \eCSmusaddleb{b}\cr}$$
Similarly, Eqs.~\eCSrhoalphaK{} become
\eqna\eCSrhosaddleb
 $$\eqalignno{ \tilde\rho_1( \omega  ) & ={\mu_2(\omega)\over(2\pi)^3}\int {\d^2 k \over k^2+   \mu_1(\omega) \mu_2(\omega)},& \eCSrhosaddleb{a}\cr  \tilde\rho_2( \omega  )  &={ \mu_1(\omega)\over(2\pi)^3}\int {\d^2 k \over k^2+   \mu_1(\omega) \mu_2(\omega)}\,.&\eCSrhosaddleb{b} \cr}$$
The four equations can be summarized by the unique pair of equations
\eqna\eCSNfersaddle
$$\eqalignno{ \mu_2(\omega)&=M_0-i\omega+i\g\int{\d \omega'\over\omega -\omega'}\tilde\rho_1 (\omega')\,,&\eCSNfersaddle{a} \cr
\tilde\rho_1( \omega  ) & ={\mu_2(\omega)\over(2\pi)^3}\int {\d^2 k \over k^2+   |\mu_2(\omega)|^2}\, .& \eCSNfersaddle{b}\cr }$$
%%%%%%%%%%%%%%%%%%%%%%%%%%%%%%%%%%%%%%%%%%%
%%%%%%%%%%%%%%%%%%%%%%%%%%%%%%%%%%%%%%%%%%%%%%%%%%%%%%%%%%%%%%%%%%%%%%%%%%%%%%%%
\subsection{Solution}

It is useful to introduce the function
$$ \Theta (\omega)  ={1\over(2\pi)^3}\int{\d^3 p\over \left(\omega-p_3\right)\left(\M ^2+p^2\right)} .  \eqnd{\eCSTheta} $$
Here, and later, $1/(\omega-p_3 )$ stands for  $PP[1/(\omega-p_3 )]$.
The function $\Theta$ requires some regularization  but, with a suitable space symmetric regularization, has a regularization-independent limit because the integrand is globally odd:
it is most easily calculated in real space (see appendix
\label{\ssThetacalcul}) and one finds
$$\Theta(\omega)= {1\over 4\pi}\arctan \left(\omega\over M  \right).  \eqnd{\eCSThetaexplicit}$$
\par
Then, initially motivated by perturbative calculations (see section \label{\ssCSpertiipt}), we set
$$\tilde\rho_1(\omega)=\tilde\rho_2^*(\omega)=\left(\M-i \omega\right)  \exp\left[i \g\Theta(\omega)\right]{ 1 \over(2\pi)^3}\int{\d^2 p\over p^2+\omega^2+\M ^2}  ,\eqnd{\eCSrhottx} $$
 where $\M$ is a free mass parameter that
 will be later identified with the fermion physical mass.
%and we have defined

From equation \eCSNfersaddle{a}, one then infers
$$\mu_2(\omega)=M_0-i\omega-{i\g \over(2\pi)^3}\int\d^3 p{\left(\M-ip_3\right)\over\left(p_3-\omega\right)\left( p^2+\M^2\right)}\exp[i\g \Theta(p_3)].$$
The integral can be performed by expanding  in powers of $\g$, symmetrizing over all integration momenta and using  the identities \eqns{\eCSbasicIdii{--}\eCSbasicIdiii}, a technique that will be used systematically in many places in this work. \par
Using the identity \eqns{\eCSbasicIdii}, one  verifies that, for $n>0$ ,
$$-{ i  \over(2\pi)^3}\int\d^3 p{\left(\M-ip_3\right)\over\left(p_3-\omega\right)\left( p^2+\M^2\right)}\Theta^n(p_3)= {i\over n+1}\left( \M-i\omega \right)\Theta^{n+1}( \omega) .$$
For $n=0$, an additional term is generated (identity \eqns{\eCSbasicIdiii}) and one finds
$$-{i  \over(2\pi)^3}\int\d^3 p{\left(\M-ip_3\right)\over\left(p_3-\omega\right)\left( p^2+\M^2\right)}=-  \Omega_1(\M)+i \left(\M-i\omega\right)\Theta(\omega),$$
where we   define more generally, for later purpose,
$$\Omega_n(M)= {1\over (2\pi)^3}\int{\d^3 p  \over ( p^2+M^2)^n  } .\eqnd{\eOmegandef}$$
For $n=1$, it is a divergent quantity that has to be regularized. \par
One  obtains
$$\mu_2(\omega)=M_0-\M-\g \Omega_1(\M)+ \left(\M-i\omega\right)\exp[i\g \Theta( \omega)]. $$
We then choose the mass parameter  $\M$   to be the solution of the {\it gap equation}
$$M_0=\M+\g \Omega_1(\M)  \eqnd{\eCSMassrenormalized}$$
and we verify later that $\M$ is then the fermion physical mass.\par
We conclude
$$\mu_2(\omega)=\left(\M-i \omega\right)\exp\left[i \g\Theta(\omega)\right] \eqnd{\eCSfermusol} $$
and, thus,
$$|\mu_2(\omega)|^2=\M^2+\omega^2\,.\eqnn $$
The verification of equation \eCSNfersaddle{b} is then straightforward.
\smallskip
{\it The constants $\Omega_n(M)$.} For $n>1$, $\Omega_n(M)$, defined by Eq.~\eqns{\eOmegandef}  is given by
$$\Omega_n(M)=  {1\over8\pi^{3/2}}{\Gamma(n-3/2)\over (n-1)!}M^{3-2n}\,.\eqnn $$
%%%%%%%%%%%%%%%%%%%%%%%%%%%%%%%%%%%%%%%%%%%%%%%%%%%%%%%%%%%%%%%%%%%%%%%%%%%%
%%%%%%%%%%%%%%%%%%%%%%%%%%%%%%%%%%%%%%%%%%%%%%%%%%%%%%%%%%%%%%%%%%%%%%%%%%%%%
For $n=1$, assuming some regularization, we define  the UV cut-off $\Lambda$ by
$$\Omega_1(0)=\int{\d^3 p \over (2\pi)^3}{1\over p^2}\equiv {\Lambda\over 4\pi}\,.\eqnd\eCSLambdadef $$
In dimensional regularization (which we will not use) this contribution   vanishes.\par
Then,
$$\Omega_1(M)={\Lambda-|M|\over 4\pi}+O(M^2/\Lambda)\,.\eqnd{\eCStadpole} $$
In what follows we will omit everywhere the $\M^2/\Lambda$ correction.
%%%%%%%%%%%%%%%%%%%%%%%%%%%%%%%%%%%%%%%%%
\subsection {The free energy density}

We define the free energy density, which is proportional to the ground state energy, dividing by a factor $N$,  by\sslbl\ssCSfreeenergy
$$W={1\over N V}\ln({\cal Z}/{\cal Z}_0),\eqnd{\eCSfreeenergy}$$
where $\cal Z$ is the partition function, ${\cal Z}_0$ a normalization and $V$ the volume. From Eq.~\eqns{\eCSNactiondensity}, one infers, in the large $N$ limit,
$$\eqalign{ W  &= - 2\pi\int \d \omega\,\tilde\rho_\alpha( \omega)\tilde\lambda_\alpha(\omega)- i\g\int{\d\omega\,\d\omega' \over  \omega-\omega'   }\tilde  \rho_1(\omega)\tilde\rho_2(\omega')\cr&\quad
+{1\over(2\pi)^3}\int\d \omega\,\d^2 p\,\tr\ln\tilde{\bf K}(\omega,p),  \cr}
  $$
where $\tilde\lambda$ and $\tilde\rho$ have to be replaced by the solutions of the saddle point equations
$$\eqalign{\tilde\lambda_1&={1\over 2\pi}(\M+i\omega)\left(\e^{-i\g\Theta(\omega)}-1\right),\cr\tilde\rho_1&=
(\M-i\omega) \e^{ i\g\Theta(\omega)}  {1\over(2\pi)^3}\int{\d^2 k \over k^2+\omega^2+\M^2}\,,\cr}$$
and complex conjugate for $\tilde\lambda_2,\tilde\rho_2$. Moreover,
$$\tr\ln\tilde{\bf K}= \ln(p^2+\omega^2+\M^2).$$
Then, from the saddle point equation,
$$i\g \int\d\omega'{\tilde\rho_2(\omega')\over \omega-\omega'}=
(\M+i\omega)\left(\e^{-i\g\Theta(\omega)}-1\right).$$
Thus,
$$\eqalign{ W =&{1\over(2\pi)^3}\int\d^3 p\,\ln(1+\M^2/p^2)\cr&\quad+i\g  \int {\d \omega\,\d \omega'\over \omega-\omega'}(\M -i\omega )(\M+i\omega') \e^{ i\g[\Theta(\omega)-  \Theta(\omega')]}\cr&\quad\times  {1\over(2\pi)^6}\int{\d^2 p\, \d^2 p'\over \left(p^2+\omega^2+\M^2\right)\left( p'{}^2+\omega'{}^2+\M^2\right)}.\cr}$$
%%%%%%%%%%%%%%%%%%%%%%%%%%%%%%%%%%%%%%%%%%%%%%%%%%%%%%
Expanding in powers of $\g$, replacing the functions $\Theta $ by their integral representation and evaluating each term using identities of the kind explained in section \label{\appCSgenId}, one obtains
$$W={1\over(2\pi)^3}\int\d^3 p\,\ln(1+\M^2/p^2)+\g \M \Omega_1^2(\M)+\frac{1}{3}\g^2 \Omega_1^3(\M),\eqnd{\eCSNfreeenergyf}$$
where the first term has still to be regularized. Differentiating the first
term, one finds
$${\partial \over\partial \M}{1\over(2\pi)^3}\int\d^3 p\,\ln(1+\M^2/p^2)
=2\M \Omega_1 (\M).$$
Using the explicit form \eqns{\eCStadpole} and integrating back, one obtains
$${1\over(2\pi)^3}\int\d^3 p\,\ln(1+\M^2/p^2)={\Lambda\over 4\pi}\M^2-{|\M|^3 \over 6\pi}+\ {\rm const.}\ . $$
We note that $W$ is even in the change $\M \leftrightarrow -\M$, $\g \Leftrightarrow -\g$ or, equivalently $M_0 \leftrightarrow -M_0 $, $\g \Leftrightarrow -\g$.
%%%%%%%%%%%%%%%%%%%%%%%%%%%%%%%%%%%%%%%%%%%%%%%%%%%%%%%%%%%%%%%%%%%%%%%
%%%%%%%%%%%%%%%%%%%%%%%%%%%%%%%%%%%%%%%%%%%%%%%%%%%%%%%%%%%%%%%%%%%%%%
\subsection{The fermion two-point function for $N$ large}

The fermion two-point function has the general form
$$ \left<\psi^i_\alpha( x)\bar\psi^j_\beta(  x')\right>_0=  \delta_{ij} W^{(2)} _{\alpha\beta}( x- x')   $$
with
$$ W^{(2)}( x )
 = {1\over(2\pi)^3} \int \d^3 p\,\e^{i px } \tilde W^{(2)}(p)  .$$
In the large $N$ limit, the fermion two-point function is obtained by inverting the   operator \eqns{\eCSferiiiKop} at the saddle point. Using the expression \eqns{\eCSferNiipt} and replacing the functions $\mu_\alpha$ by the explicit solutions \eqns{\eCSfermusol} of the saddle point equations, one finds
$${\cal D}(p)=p^2+\M^2,$$
and thus
$$  \tilde {\bf K}^{-1}= -{ i p_1\sigma_1+ip_2 \sigma_2+\left( i\omega\sigma_3-\M \right)\exp\left[i\g \sigma_3 \Theta(\omega)\right]\over p^2+\M^2}\,. \eqnd{\eCSferKinverse} $$
%This confirms the Ansatz \eqns{\eCSfermioniipt}.\par
We note that
$$ \tilde {\bf K}^{-1}=-\exp\left[\frac{i}{2}\g \sigma_3 \Theta(\omega)\right]{\left(i\sla{p}-\M\right)\over p^2+\M^2}\exp\left[\frac{i}{2} \g \sigma_3 \Theta(\omega)\right]. $$
It is convenient to define
$$U( \omega)=\exp\left[\ud i\g\sigma_3 \Theta ( \omega)\right].\eqnd{\eCSUmatdef} $$
The dressed fermion fermion propagator then takes the form
$$ \tilde W^{(2)}(p)=-  \tilde {\bf K}^{-1}(p)=U(p_3) {\left(i\sla{p}-\M\right)\over p^2+\M^2}U(p_3).\eqnd\eCSNfermioniipt $$
This expression confirms that the parameter $\M$ is the physical fermion mass.
Moreover, this form suggests that, after mass renormalization and a non-local phase transformation, the fermion theory is equivalent to a free theory.\par
Correspondingly, the two-point vertex function, inverse of $ W^{(2)}$, can be written as
$$\tilde\Gamma^{(2)}(p) =- U^{-1}(p_3)\left(i\sla{p}+\M\right)U^{-1}(p_3). \eqnd{\eCSNvertexiipt}    $$
More explicit expressions, using the explicit form of $\Theta$, are
$$\eqalign{\tilde W^{(2)}(p)&={1\over p^2+\M^2}\left\{ip_1\sigma_1+ip_2\sigma_2-\sqrt{\M^2+p_3^2}\,\exp\left[i\sigma_3(\g-4\pi)\Theta(p_3)\right]\right\} \cr
\tilde\Gamma^{(2)}(p)&=ip_1\sigma_1+ip_2\sigma_2+
\sqrt{\M^2+p_3^2}\,\exp\left[-i\sigma_3(\g-4\pi)\Theta(p_3)\right]
.
}\eqnd{\eGammaTwo}$$
Finally,
 $$\int\d^2 q\,\tilde W^{(2)}(q)= \left(ip_3\sigma_3-\M\right)U^2(p_3) {1\over(2\pi)^3}\int{\d^3 q\over \left(p_3-q_3\right)\left(\M ^2+q^2\right)}  $$
 a result consistent with equation \eqns{\eCSrhottx}.
%%%%%%%%%%%%%%%%%%%%%%%
\subsection{Gauge-invariant observables}

To be able to distinguish gauge artefacts from gauge independent properties, it is necessary to calculate gauge invariant observables.
The $\psi$-field two-point function is not gauge invariant, except in the limit of coinciding points. Two gauge invariant operators    are  (no summation assumed)
$$R_\alpha (x)= {1\over N}\bar{ \psi}_\alpha  (x)\cdot{ \psi}_\alpha (x) \eqnd{\eCSRalphadef}$$
Below, we consider the scalar combination
$$R(x)=R_1(x)+R_2(x)\eqnd{\eCSRsumdef} $$
and the third component of the current
$$J_3(x)=i \bigl(R_1(x)-R_2(x)\bigr).\eqnd\eCSJiiidef $$
\smallskip
{\it Expectation value.} We can already determine the expectation value of the   operator $R_\alpha (x)  $. The equal-time expectation value of $\rho_1$ is given by
$$\left<R_1(x)\right>= \int\d\omega\,\tilde\rho_1(\omega)=
{1\over(2\pi)^3 }\int\d^3 p{ \M-i p_3 \over p^2+\M^2}\exp[i\g\Theta(p_3)]
\eqnd{\eCSRexpectation}.$$
Expanding in powers of $\g$ and replacing $\Theta(p_3)$ by its integral representation \eqns{\eCSTheta}, symmetrizing the integrand with respect to $p $ and all the other integration momenta, one verifies that all terms vanish except the two first ones.
This result, and other  similar ones, rely on the identities \eqns{\eCSomegaSnm}.
Since the result is real,   $\langle J_3 \rangle =0$ and
 $$  \left<R\right> =\left<R_1+R_2\right>=2\left<\rho_1\right>=2\M\Omega_1(\M)+  \g\Omega_1^2(\M)\eqnd{\eCSRexpectFinal}.$$
 We find that the expectation value, which is gauge invariant, is indeed given by an explicitly $O(3)$ symmetric expression, in the sense that it makes no reference to the gauge propagator and is expressed in terms of an obviously covariant integral. In terms of the explicit expression \eqns{\eCStadpole}, for $\M>0$ it reads
 $$\left<R\right>=\g {\Lambda^2\over 16\pi^2}+{\Lambda \M \over 2\pi}\left(1-{\g \over 4\pi}\right)-{\M^2 \over 2\pi}\left(1-{\g \over 8\pi}\right).$$
 \smallskip

{\it $R$ expectation value and ground state energy.}
The  normalized free energy density defined by Eq.~\eqns{\eCSfreeenergy} and calculated in section \label{\ssCSfreeenergy}, is given by (Eq.~\eqns{\eCSNfreeenergyf})
$$W={1\over(2\pi)^3}\int\d^3 p\,\ln(1+\M^2/p^2)+\g \M \Omega_1^2(\M)+\frac{1}{3}\g^2 \Omega_1^3(\M).\eqnd{\eCSWferren} $$
Then,
$$\left<R\right>={\partial W \over \partial M_0}\,,\eqnn $$
which can be rewritten as (Eq.~\eqns{\eCSMassrenormalized})
$${\partial W\over \partial \M}={\partial W\over \partial M_0 }{\partial M_0\over\partial \M}=\left<R\right>\bigl(1-2\g \M \Omega_2(\M)\bigr)
 =\left<R\right> \left(1-{g\over 4\pi}\right) .$$
One then verifies the consistency between the  expressions \eqns{\eCSWferren} and \eqns{\eCSRexpectFinal}.

\medskip

{\it Connected $R$ correlation function at zero momentum.}
The $R$ two-point function, at zero momentum is obtained by differentiating $\left<R\right>$ with respect to $M_0$.
One then finds (the subscript $c$ stands for connected)
$$\eqalignno{ \left<\tilde R ( 0 )\tilde R ( 0)\right>_{\rm c}&= {\partial \left<R\right>\over\partial M_0}={\partial \left<R\right>\over\partial \M}\left/{\partial M_0\over\partial \M}\right.\cr& = 2\Omega_1(\M)-{4 \M^2\Omega_2(\M)
\over 1-2\g \M \Omega_2(\M)}={\Lambda \over 2\pi}-{\M \over \pi}{(1-\g/8\pi)\over (1-\g/4\pi)}\, .\hskip8mm&\eqnd{\eCSRRexpectation} \cr}$$
This quantity requires only an additive renormalization:  $ \langle\tilde R ( 0 )\tilde R ( 0)\rangle_{\rm c}-\Lambda/2\pi$ is finite.
\par
Higher order functions can be obtained by further differentiation.
here differentiating again, one obtains (still $\M>0$)
$$ \left<\tilde R ( 0 )\tilde R ( 0)\tilde R ( 0)\right>_{\rm c}=-{1\over\pi}{(1-\g/8\pi)\over (1-\g/4\pi)^2}  \,,\eqnd{\eCSRRRexpectation}$$
which is finite. All other connected correlation functions then vanish for $\Lambda\to\infty$.
%%%%%%%%%%%%%%%%%%%%%%%%%%%%%%%%%%%
\section{Perturbative calculations at large $N$}

It is interesting to see how some results that we have obtained by field integral techniques emerge from perturbative calculations.
\sslbl\sPerturbative
Therefore, in this section, we calculate a few orders of some of the quantities that we have determined in the preceding sections.
%%%%%%%%%%%%%%%%%%%%%%%%%%%%%%%%

\subsection{The fermion two-point function at two-loop order}

We first expand the fermion two-point function, which is not a gauge-invariant quantity but which has been determined  exactly in the large $N$ limit (equation \eqns{\eCSNfermioniipt}).\sslbl\ssCSpertiipt
\smallskip
{\it The fermion propagator.}
The bare fermion propagator ($\left<\bullet\right>_0$ means Gaussian expectation value) is given by
$$ \left<\psi^i_\alpha( x)\bar\psi^j_\beta(  x')\right>_0=  \delta_{ij} \int{ \d^3 p\over(2\pi)^3}\,\e^{i p(x-x') }[\tilde\Delta_{\rm F}] _{\alpha\beta}( p)   $$
with
$$  \tilde{\bf\Delta}_{\rm F}  ( p )
 ={ i \sla{p} -M_0\over p^2+M_0^2}\,.\eqnn $$
\topinsert
\epsfysize=16.5mm
\epsfxsize=74.9mm
\vbox{\elevenpoint
\centerline{\epsfbox{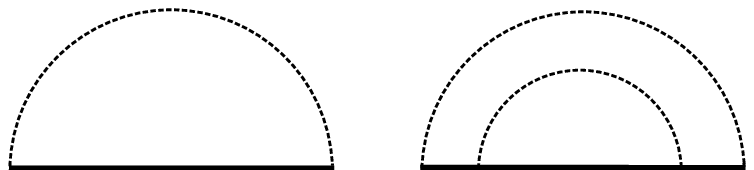}}
\kern-17.2mm
\hbox{\kern36.8mm $k-  p $}
\kern11.2mm
\hbox{\kern40.0mm$p $ }}
\figure{4mm}{One- and two-loop contributions to the two-point vertex function. Dotted lines represent gauge propagators.}
\figlbl\figcsi
\endinsert
\medskip
{\it One-loop calculation: the two-point vertex function.}
Perturbative calculations of the two-point vertex function or mass operator involve the one-loop diagram of figure \label{\figcsi}. It can be written as
$$\Sigma_1(k)= \ud  \int{\d^3 p \over (2\pi)^3}{\left[\sigma_1\left(i\sla{p}-M_0\right) \sigma_2-
\sigma_2\left(i\sla{p}-M_0\right) \sigma_1\right]\over\left(k_3-p_3\right) \left(p^2+M_0^2\right)}\,.\eqnn $$
The evaluation is simple and yields
$$ \Sigma_1(k)=- \Omega_1(M_0) +\left(k_3- i M_0 \sigma_3   \right) \Theta_0 (k_3) ,\eqnd{\eCSpsipsioneloop}$$
where $\Theta_0 $ is the function \eqns{\eCSTheta} in which $\M$ is replaced by $M_0$.  The expression in the case  of $M_0=0$ agrees with the calculation in the temporal gauge in Ref.~\refs{\rWadia}  at one-loop order.
\par
The two-point vertex function is then
$$\eqalign{\tilde\Gamma^{(2)}(k)&=-M_0-i\sla{k}-\g\Sigma_1(k)+O(\g^2)\cr
&=-M_0-i\sla{k}+\g\Omega_1(M_0)-\g(k_3-iM_0\sigma_3) \Theta_0 (k_3)+O(\g^2).\cr}$$
After introduction of the physical fermion mass \eqns{\eCSMassrenormalized}, which amounts to a mass renormalization,
the expression agrees with the expansion of the result \eqns{\eCSNvertexiipt} at order $\g$.
%%%%%%%%%%%%%%%%%%%%%%%%%%%
\medskip
{\it Two-loop order.} After mass renormalization (equation \eqns{\eCSMassrenormalized}), the two-loop contribution $\Sigma_2$ to the   two-point vertex function (two-loop  diagram of  figure \label{\figcsi}) is proportional to
$$\Sigma_2\propto -{1\over(2\pi)^3}\int{\d^3 q \left(iq_3 \sigma_3+M \right)\over (q^2+M ^2)(k_3-q_3)} \Theta  (q_3) =-\ud \left(M +i\sigma_3 k_3 \right)\Theta ^2 (k_3) ,$$
which is again consistent with the expansion of the expression \eqns{\eCSNvertexiipt}.
\topinsert
\epsfysize=21.0mm
\epsfxsize=38.2mm
\vbox{\elevenpoint
\centerline{\epsfbox{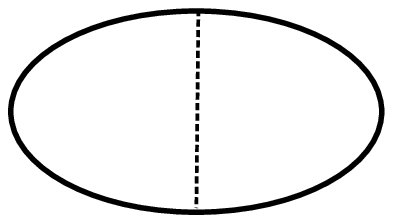}}
}
\figure{4mm}{Two-loop contribution to the vacuum energy. Dotted lines represent gauge propagators.}
\figlbl\figcsvci
\endinsert
%%%%%%%%%%%%%%%%%%%
\subsection The ground state (or vacuum) energy up to three loops

We have defined the normalized, gauge-independent, free energy density by equation \eqns{\eCSfreeenergy}. We  expand it as
$$W={1\over N V}\ln({\cal Z}/{\cal Z}_0)=W_0+\g W_1 +\g^2 W_2 \eqnd\eCSferfreeenergy$$
and keep only the leading terms for  $N$  large.
Note that in view of the remark after Eq.~\eqns{\eCSRexpectation}, the
expectation value $\langle R \rangle$ at leading $N$ contains only terms up to order $g^2$ and thus in view of Eq.~\eqns{\eCSWferren} there is no term of order $O(\g^3)$ in Eq.~\eqns{\eCSferfreeenergy}. This observation was also made in Ref.~\refs{\rWadia} where the free energy has been calculated in the light-cone gauge. \par
We choose ${\cal Z}_0$ such that
$$W_0=\tr\ln(1+ \sla{\partial}/M_0)={1\over(2\pi)^3}\int\d^3 p\,\ln(1+M_0^2/p^2).$$
The two-loop contribution is (Fig.~\label{\figcsvci})
$$W_1=-{1\over 4 }\int{ \d^3 p\,\d^3 q\over(2\pi)^6\left(q_3-p_3\right)} \,\tr\left[\tilde \Delta_{\rm F}(q) \bigl( \sigma_1\tilde \Delta_{\rm F}(p) \sigma_2-\sigma_2\tilde\Delta_{\rm F}(p) \sigma_1 \bigr)\right]. $$
Calculating the trace, one verifies that the gauge propagator cancels and one finds
$$ W_1 =-M_0\Omega_1^2(M_0). \eqnd{\eCSWone} $$
\midinsert
\epsfysize=21.0mm
\epsfxsize=38.2mm
\vbox{\elevenpoint
\centerline{\epsfbox{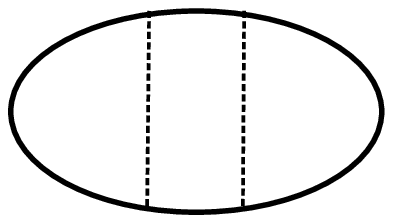}}
}
\figure{4mm}{Three-loop contribution to the vacuum energy. Dotted lines represent gauge propagators.}
\figlbl\figcsvcii
\endinsert
\medskip
{\it Three-loop order.}
The three-loop order calculation (order $\g^2$, Fig.~\label{\figcsvcii}) is  lengthier.
Setting
$$S_2=\Delta_{\rm F}(q)\left(\sigma_1\tilde\Delta_{\rm F}(p) \sigma_2 -\sigma_2\tilde\Delta_{\rm F}(p) \sigma_1\right)\tilde\Delta_{\rm F}(q),$$
one obtains
$$W_2=-{1\over8}\int{\d^3 p\,\d^3 q\,\d^3 r\over(2\pi)^9(r_3-q_3)(q_3-p_3)}\tr\left\{\tilde\Delta_{\rm F}(r)\left[\sigma_1 S_2(p,q) \sigma_2 -\sigma_2 S_2(p,q)  \sigma_1\right]\right\}.$$
After evaluating the trace, the expression decomposes into the sum of two terms,
$$\eqalign{W_{2,1}&=-  \int{\d^3 p\,\d^3 q\,\d^3 r\over(2\pi)^9(r_3-q_3)(q_3-p_3)} {M_0^2+p_3r_3 \over (p^2+M_0^2)(q^2+M_0^2)(r^2+M_0^2)}\cr
W_{2,2}&= -\int{\d^3 p\,\d^3 q\,\d^3 r\over(2\pi)^9 } {2M_0^2 \over (p^2+M_0^2)(q^2+M_0^2)^2(r^2+M_0^2)}\,.\cr}$$
The first contribution can now symmetrized over $r_3,p_3,q_3$. The gauge field poles then cancel and the result is
$$W_2=    -2M_0^2\Omega_1^2(M_0)\Omega_2(M_0)+\frac{1}{3}\Omega_1^3(M_0) .\eqnd{\eCSWdeux}$$
The expansion of the free energy expressed in terms of the fermion physical mass  \eqns{\eCSMassrenormalized} then becomes
$$W=\int{\d^3 p\over(2\pi)^3}  \ln(1+\M^2/p^2)+\M\Omega_1^2(\M)\g+\frac{1}{3}
\Omega_1^3(\M)\g^2
,\eqnd{\eCSWpert}  $$
in agreement with the expression \eqns{\eCSWferren}.
%%%%%%%%%%%%%%%%%%%%
\section{Correlation functions involving $R= \bar\psi\cdot\psi $ at large $N$}

We now determine the $\langle(\bar\psi\cdot\psi)\psi\bar\psi\rangle$ vertex function at leading order for $N$ large. But before a simple remark is useful.\sslbl\ssCSNexpand
%%%%%%%%%%%%%%%%%%%%%%%%%%%%
\subsection {A simple transformation}

To generate insertions of $R_\alpha$ operators, we  introduce non-diagonal, space-depen\-dent, mass terms.
A simple limit corresponds to correlations of $\int\d^2 x\,R_\alpha(x_3,x)$. They can be generated by
$${\cal S}_M=\int\d^3 x\,\sum_\alpha \bar\psi_\alpha(x) M_\alpha(x_3)\cdot \psi_\alpha(x).\eqnd{\eCSfermasst} $$
We then make on $\psi$ a phase transformation of the form
$$\psi(x)\mapsto \e^{\varphi(x_3)}\psi(x),\quad\bar \psi(x)\mapsto \e^{-\varphi(x_3)}\bar\psi(x).$$
In the action \eqns{\eCSpsiaction} (where $M$ is replaced by $M_\alpha(x_3)$), only the $x_3$ derivative is modified:
$${\cal S}_{\rm F}\mapsto {\cal S}_{\rm F} -\int\d^3 x\, \bar\psi (x)\sigma_3\partial_3\varphi(x_3)\cdot\psi(x). $$
Expressing the diagonal matrix $\bf M$ of eigenvalues $M_1,M_2$ in the form
$${\bf M}=\ud(M_1+M_2){\bf 1}+\ud(M_1-M_2)\sigma_3 $$
and choosing $\varphi(x_3)$ such that
$$\partial_3 \varphi(x_3)=\ud\bigl(M_1(x_3)-M_2(x_3)\bigr),$$
we reduce  the mass term \eqns{\eCSfermasst} to
$${\cal S}_M=\ud\int\d^3 x\,\bigl(M_1(x_3)+M_2(x_3)\bigr)\bar\psi_\alpha(x)\cdot \psi_\alpha(x).$$
We conclude that the expectation values of products of $\int\d ^2 x\,R_\alpha(x_3,x)$ operators  are trivially related to the expectation values of $\int\d^2 x\,R (x_3,x)$ and this provides some checks in the calculations.
%%%%%%%%%%%%%%%%%%%%%%%%%
\subsection Field integral formalism: expansion at the saddle point

Connected correlation functions involving the $R$ operator at non-vanishing momenta  can  be calculated for $N$ large by
using the field integral formulation of section \label{\sCSNpath}.
The calculation involves expanding the large $N$ action \eqns{\eCSferSN} at the saddle point to quadratic order and performing a Gaussian integration. The quadratic form depends on the second functional derivatives of the large $N$ action  with respect to $\rho,\lambda$ at the saddle point. The less trivial part is the second functional derivative of the determinant with respect to $\lambda$.
\smallskip
{\it The second functional derivative of\/ $\tr\ln{\bf K}$.} Calculating the second functional derivative of the trace of the logarithm of the operator {\eqns\eCSferiiiKop} with respect to $\{\lambda_\alpha(t_1,t_2,x),  \lambda_\beta(t'_1,t'_2,x')\}$, at the saddle point, we obtain the formal expression (no summation on $\alpha,\beta$ implied)
$$-[{\bf K}^{-1}]_{ \alpha\beta}(t_2,x;t'_1,x')[{\bf K}^{-1}]_{ \beta\alpha}(t'_2,x';t_1,x).$$
\smallskip
{\it Other functional derivatives.}
The second functional derivative of the large $N$ action \eqns{\eCSferSN} with respect to $\{\lambda_\alpha(t_1,t_2,x),  \rho_\beta(t'_1,t'_2,x')\}$  is
$$\delta(t'_1-t_2)\delta(t'_2-t_1)\delta^{(2)}(x-x')\delta_{\alpha\beta} $$
%and, thus, proportional to the identity,
and, finally, with respect to $\{\rho_\alpha(t_1,t_2,x),  \rho_\beta(t'_1,t'_2,x')\}$,
$$\ud\g \sgn(t_1-t_2) \delta(t'_1-t_2)\delta(t'_2-t_1)\delta^{(2)}(x-x')\left(\delta_{\alpha1}\delta_{\beta2}-\delta_{\alpha2}\delta_{\beta1}\right).$$
This determines the $2\times2$ matrix of functional derivatives, which has still to be inverted and this is the non-trivial part of the calculation of $R$ correlation functions.
\smallskip
{\it Fourier representation.}
We   introduce the Fourier representation
$$[{\bf K}^{-1}]_{ \alpha\beta}(t ,x;t' ,x')={1\over(2\pi)^3}\int\d^2p\,\d\omega\,\e^{i\omega(t -t' ) +i p (x-x')}[\tilde{\bf K}^{-1}]_{ \alpha\beta}(\omega,p),$$
where the explicit expression is given by Eq.~\eqns{\eCSferKinverse}.
Thus,
$$\eqalign{& [{\bf K}^{-1}]_{ \alpha\beta}(t_2,x;t'_1,x')[{\bf K}^{-1}]_{ \beta\alpha}(t'_2,x';t_1,x)\cr&\quad={1\over(2\pi)^6}\int\d^2 p\,\d^2 p'\,\d\omega\,\d\omega'\,
\e^{i\omega(t_2-t'_1)+i\omega'(t'_2-t_1)+i(p-p')(x-x')}\cr&\qquad\times [\tilde{\bf K}^{-1}]_{ \alpha\beta}(\omega,p)[\tilde{\bf K}^{-1}]_{ \beta\alpha}(\omega',p').\cr}$$
In a global Fourier representation, this yields
$${1\over(2\pi)^3}\delta^{(2)}(p+p')\delta(\omega_1+\omega'_2)\delta(\omega_2+\omega'_1)\int\d^2 q\, [\tilde{\bf K}^{-1}]_{ \alpha\beta}(\omega_2,q)[\tilde{\bf K}^{-1}]_{ \beta\alpha}(\omega'_2,p-q).\eqnd{\eCSKK} $$
We  set
$${\bf G}_{\alpha\beta}(\omega,\omega',p)={1\over(2\pi)^3}\int\d^2 q\, [\tilde{\bf K}^{-1}]_{ \alpha\beta}(\omega ,q)[\tilde{\bf K}^{-1}]_{ \beta\alpha}(\omega' ,p-q)={\bf G}_{\beta\alpha}(\omega ,\omega',p).$$
In the same way, the $\lambda\rho$ element becomes
$$\delta(\omega_1+\omega'_2)\delta(\omega_2+\omega'_1)\delta^{(2)}(p+p')\delta_{\alpha\beta}\eqnd{\eCSlambdarho} $$
and the $\rho\rho$ element
$$\ud i\g \left(\delta_{\alpha1}\delta_{\beta2}-\delta_{\alpha2}\delta_{\beta1}\right)\delta^{(2)}(p+p')\delta(\omega_1+\omega_2+\omega'_1+\omega'_2){1\over \omega_1+\omega'_2}\,.\eqnd{\eCSrhorho} $$
We note that all expressions are diagonal in the   momentum variables.\par
\smallskip
{\it Remarks.} The calculation of the inverse operator involves solving two coupled integral equations, a problem we discuss in the coming sections. Note that if the inverse operator is expanded in powers of the coupling $\g$, it generates a sum of ladder diagrams with the dressed fermion propagator \eqns{\eCSNfermioniipt}, which sums all propagator corrections.
\smallskip
To calculate correlation functions of $R_\alpha(t,x)= \bar{\Bfg\psi}_\alpha(t,x)\cdot{\Bfg\psi}_\alpha(t,x)/N$, one needs the operator sources for the vectors  $\rho_ \alpha(t,t,x)$  of the form
$$[j_\rho]_\alpha(t_1 ,x)\delta(t_1-t_2).$$
In the Fourier representation, they become
$$[\tilde j_\rho]_\alpha(\omega_1+\omega_2,p).$$
If one wants to  calculate only correlation functions of $R=\rho_1(t,t,x)+ \rho_2(t,t,x)$, one needs  a source in  $t,x$ space that takes the form
$$ j_\rho (t_1 ,x) \delta(t_1-t_2)\left(\delta_{\alpha1}+\delta_{\alpha2}\right).$$
In the Fourier representation, it becomes
$$\tilde j_\rho(\omega_1+\omega_2,p)\left(\delta_{\alpha1}+\delta_{\alpha2}\right).$$
This somewhat simplifies the calculation.
%%%%%%%%%%%%%%%%%%%%%%%%%%%
%%%%%%%%%%%%%%%%%%%%%%%%%%%%%%%%%%%%
\subsection{The $\langle(\bar\psi\psi)\psi\bar\psi\rangle$ vertex function}

Though ultimately we want to calculate the $R$ two-point function, we have first to calculate  the connected three-point correlation function with one $R$ insertion,
$$W^{(1,2)}(x;y,z)=\left<\bar\psi(x)\cdot\psi(x)\psi(y)\cdot\bar\psi(z)\right>_{\rm c},\eqnn $$
because, unlike the $R$ two-point function, it satisfies an integral equation for $N$ large, as the analysis of section \label{\ssCSNexpand} shows.\par
We denote its Fourier transform by $\tilde W^{(1,2)}(k;\ell-k/2,\ell+k/2)$ and it is related to the corresponding vertex function by
$$   \tilde W^{(1,2)}( k;\ell-\ud k,\ell+\ud k) =-\tilde W^{(2)}(\ell-\ud k)\tilde\Gamma^{(1,2)}( k;\ell-\ud k,\ell+\ud k)\tilde W^{(2)}(\ell+\ud k).\eqnn $$
An important property, which   we use systematically later, is the following: one verifies that the vertex function $ \tilde\Gamma^{(1,2)}$ has the following general decomposition:
$$ \tilde\Gamma^{(1,2)}(k;\ell-k/2,\ell+k/2) =- E(k;\ell-k/2,\ell+k/2) -i   \sigma_3 F(k;\ell-k/2,\ell+k/2)\, ,\eqnd\eCSvertexform $$
where $E$ and $F$ are scalar functions.
\smallskip
{\it The $R(x)=\bar\psi(x)\cdot\psi(x)$ insertion at zero momentum.}   A  differentiation of the fermion two-point function \eqns{\eCSNvertexiipt} with respect to the bare mass $M_0$ yields the vertex function \eqns{\eCSvertexform} with an operator insertion at zero momentum. Using the gap equation \eqns{\eCSMfergap}, one finds that the mass insertion is given by
$$\eqalignno{&  {1\over \left(1-2\g M \Omega_2\right)}
{\partial\tilde  \Gamma^{(2)}(\ell)\over \partial M}\cr&\quad=-{1\over \left(1-2\g M \Omega_2\right)}U^{-2}(\ell_3)\left[1+2i\g M \sigma_3\left(M+i \ell_3 \sigma_3 \right) \Xi(0,\ell_3)\right]\cr
&\quad=-{1\over \left(1-2\g M \Omega_2\right)}U^{-2}(\ell_3)\left[1+{\g\over4\pi} \ell_3 { i M\sigma_3 -  \ell_3   \over \ell_3^2+M^2}\right].&\eqnd{\eCSvertexkzero} \cr}$$
The result is consistent with the decomposition \eqns{\eCSvertexform} and one immediately infers the values
\eqna\eCSVertexEFzero
$$\eqalignno{E(0;\ell,\ell)&={1\over  1-2\g M \Omega_2(M) }U^{-2}(\ell_3)\left(1-{\g\over4\pi} {\ell_3^2 \over \ell_3^2+M^2}\right),&\eCSVertexEFzero{a}\cr
F(0;\ell,\ell)&={1\over  1-2\g M \Omega_2 (M)}U^{-2}(\ell_3){\g\over4\pi}{M\over \ell_3^2+M^2}\,.&\eCSVertexEFzero{b} \cr}$$
%%%%%%%%%%%%%%%%%%%%%%%%%%%%%%%%%%%%%%%%%
\topinsert
\epsfysize=25.7mm
\epsfxsize=75.0mm
\vbox{
\centerline{\epsfbox{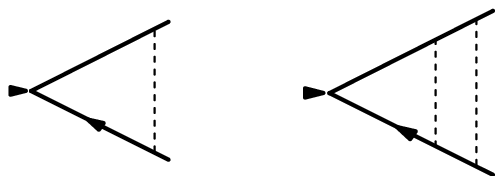}}
\kern-25.6 mm
\moveleft30.0mm\vbox{\elevenpoint\hbox{\kern79.0mm $  \ell+k/2  $}
\kern-0.1mm
\hbox{\kern56.8mm $  p+k/2  $}
\kern2.9mm
\hbox{\kern51.0mm $  k  $\kern15.5mm$  \ell-p $}
\kern3.4mm
\hbox{\kern56.2mm$  p-k/2 $ }
\kern-0.6mm
\hbox{\kern79.0mm $   \ell-k/2  $}}}
\figure{4mm}{One-loop and two-loop contributions to the $\left<(\bar\psi\psi)\psi\bar\psi\right>$ vertex function with dressed propagator. Dotted lines represent gauge fields.}
\figlbl\figcsver
\endinsert
\subsection The perturbative expansion of the $\langle (\bar\psi \psi)\psi\bar \psi\rangle$ vertex function at two loops

To gather some insight about the general structure of correlation functions involving the $\bar\psi\psi$ operator, we begin with a perturbative calculation for $N$ large of the $\langle (\bar\psi \psi)\psi\bar \psi\rangle$ vertex function $\tilde\Gamma^{(1,2)}$. However, we have learned in section \label{\ssCSNexpand} that, for $N$ large, the perturbative expansion reduces to a sum of ladder diagrams (figure \label{\figcsver}) with the dressed propagator \eqns{\eCSNfermioniipt}.\par
In the decomposition \eqns{\eCSvertexform} the functions $E$ and $F$ have a perturbative expansion of the form\sslbl\ssCSrecursion
$$E= \sum_{n=0} E_n \g^n,\quad F=\sum_{n=0}F_n\g^n $$
with
$$E_0=1,\quad F_0=0\,.$$
\smallskip
  We introduce the compact notation
 $$\deqalign{ Dp_1 &=(p+k/2)^2+M^2 ,&  Dm_1  =(p-k/2)^2+M^2,\cr
  Dp_2 &=(q+k/2)^2+M^2 ,&  Dm_2  =(q-k/2)^2+M^2.\cr}$$
\smallskip
{\it Recursion relation.} Since the perturbative expansion reduces to a sum of ladder diagrams, it is possible to generate them by a recursion formula (see section \label{\ssCSNexpand}).
It is given by a $2\times 2$ matrix acting on the vector $(E_n,F_n)$:
$$(E_{n+1},F_{n+1})={\bf T}(E_n,F_n),$$
which corresponds to adding two dressed fermion propagators and a gauge propagator to a diagram.\par
The functions $(E_n,F_n)$  depend  only on $k$ and on $p_3$, if we denote by $p$ the additional integration variables. For $\g=0$ ({\it i.e.}, with the free propagator), we can write the matrix (but a kernel in $(p_3,\ell_3)$) as
$$ {\bf T}(k;\ell_3,p_3)={1\over(2\pi)^3} \int{\d^2 p\over Dp_1Dm_1\left(\ell_3-p_3\right)}{\cal T}( k,p )\eqnd{\eCSTmatzero} $$
with
$${\cal T}(k,p) =\left[\matrix{-2M p_3 & \ud (Dp_1+Dm_1)-2t_2 -2p_3^2  \cr  -\ud(Dp_1+Dm_1)+2t_1  & 2Mp_3\cr}\right] , $$
where we have set
$$ t_1=\frac{1}{4}\left(k^2+4M^2\right),\ t_2=\frac{1}{4}\left(k^2-k_3^2\right) \,.\eqnd\eCSttnotation  $$
Terms proportional to $(k_1p_2-k_2p_1)$, with a vanishing integral  due to rotation symmetry in the $(1,2)$ plane, have been omitted.\par
 It is useful to decompose $\bf T$ into the sum of two terms,
$${\bf T}={\bf T}_1+{\bf T}_2 \eqnd{\eCSTonedef}  $$
with, correspondingly,
$$  {\cal T}_2(k,p ) =\ud  i \sigma_2 \left(  Dp_1 +  Dm_1 \right)\eqnd\eCSTmattwo $$
and
$$  {\cal T}_1(k,p_3)  = 2\left[\matrix{- M p_3 &   - t_2 - p_3^2  \cr    t_1  &  Mp_3\cr}\right]      . \eqnd\eCSTmatone   $$
Note that $ {\cal T}_1(k,p_3) $ depends only on the component $p_3$.\par
With the dressed propagator, ${\bf T}_2$ is not modified. Quite generally, we define
$$ \chi(\ell_3)=\g\left[\Theta(\ell_3+k_3/2)+\Theta(\ell_3-k_3/2) \right] \eqnd\eCSchiphase $$
and, moreover, here we set
$$\chi_1\equiv \chi(p_3).$$
We can then write the dressed matrix \eqns{\eCSTmatone}  as
$$\eqalignno{{\cal T}_1&=-i\sigma_2 \left( t_1+ t_2+p_3^2  \right) +
\left(t_1- t_2-p_3^2 \right)\pmatrix{\sin\chi_1 & \cos\chi_1 \cr
\cos\chi_1 & -\sin\chi_1\cr}\cr&\quad +2Mp_3\pmatrix{-\cos\chi_1 &\sin \chi_1 \cr
\sin\chi_1 & \cos\chi_1 \cr}.&\eqnd{\eCSTdressed}\cr}$$
or
$${\cal T}_1 ={\cal O}(-\chi_1/2)\left[ \left(t_1-t_2 -p_3^2 \right)\sigma_1- i\left(t_1+t_2+p_3^2\right)\sigma_2-2Mp_3\sigma_3\right] {\cal O}( \chi_1/2) , \eqnd{\eCSTdressed} $$
where ${\cal O}$ is the rotation matrix
$${\cal O}(\chi)= \pmatrix{\cos\chi & -\sin\chi \cr \sin\chi & \cos\chi\cr} \equiv \e^{-i\sigma_2 \chi}.\eqnd\eCSOrotation $$
\smallskip
{\it Definitions.} The calculations that follow involve only two new basic functions.  We  first define
$$\eqalignno{{\cal B}_1(k) &= \int {\d^3p\over (2\pi)^3}  {1\over \left(p^2+M ^2\right)\left[(k-p)^2+M ^2\right]}
\cr&=\int {\d^3q\over (2\pi)^3}  \int_0^1 \d s {1\over \left[q^2+s(1-s)k^2+M ^2\right]^2}\cr
&={1\over 4\pi}{1\over |k|}\arctan \left(\left|k \over 2 M \right| \right) .&\eqnd{\eCSBubble}\cr}$$
which is a one-loop scalar function.
Note that  in three dimensions  all one-loop diagrams can be reduced to elementary functions.
\par
We also define
$$\Xi(\omega, k)={1\over(2\pi)^3}\int{\d^3 p\over(\omega-p_3)\left[(p+k/2)^2+M^2\right] \left[(p-k/2)^2+M^2\right]}\,.\eqnd{\eCSXidef}$$
It satisfies the relations
$$\eqalignno{\Xi(k,\omega)&=-\Xi(k,- \omega),\cr
 \Xi(0, \omega)&=-{1\over 2M}  {\partial \Theta( \omega)\over \partial M} = {\omega\over 8\pi M}{1\over \omega^2+M^2},& \eqnd\eCSXiTheta \cr
\Theta( \omega-\ud k_3)-\Theta( \omega+\ud k_3)&=\left.2k_3\left[ \omega \Xi(k, \omega)-   {\cal B}_1(k)\right]\right|_{k_1=k_2=0}\,. \cr}$$
\smallskip
{\it One-loop contribution.} We  need only the free fermion propagator and, thus, the matrix \eqns{\eCSTmatzero} since at one-loop order   the phase  $\chi=O(\g)$ does not contribute.\par
The one-loop integrand is obtained by acting with $\bf T$ on the vector $( 1,0)$.
One finds
$$\eqalign{E_1&=-{1\over(2\pi)^3}\int{\d^3 p \,2Mp_3\over \left(\ell_3-p_3\right)
Dm_1 Dp_1}\cr
F_1&= {1\over(2\pi)^3}\int{\d^3 p \,\left(4t_1-Dp_1-Dm_1\right)\over 2\left(\ell_3-p_3\right)
Dm_1 Dp_1}.\cr}$$
Integrating, one obtains
$$ E_1 =-2M\ell_3 \Xi(k,\ell_3)+2M {\cal B}_1(k)\eqnn $$
and
$$ F_1 =2 t_1\Xi(k,\ell_3)-\ud \left[\Theta(\ell_3+k_3/2)+\Theta(\ell_3-k_3/2)\right],\eqnn $$
where we have used the definitions \eqns{\eCSBubble,\  \eCSXidef} and the notation \eqns{\eCSttnotation}. We note that in $F_1$ the second term comes from the matrix \eqns{\eCSTmattwo}.\par
This result suggests the factorization
$$\eqalignno{\tilde\Gamma^{(1,2)}(k;\ell-k/2,\ell+k/2)&=- U^{-1}(\ell_3-k_3/2) \tilde V^{(1,2)}(k;\ell-k/2,\ell+k/2) \cr&\quad \times U^{-1}(\ell_3+k_3/2),&\eqnd{\eCSVdef}\cr} $$
where at one-loop order  $\tilde V^{(1,2)}$  reduces to
$$\tilde V^{(1,2)}=1+2\g M\left[{\cal B}_1 (k)-\ell_3\Xi(\ell_3, k)\right]+ \ud i\sigma_3\g  (k^2+4M^2) \Xi(\ell_3, k) +O(\g^2).\eqnd{\eCSVertexiloop} $$
The connected function is then given by
$$\eqalignno{&\tilde W^{(1,2)}( k;\ell-\ud k,\ell+\ud k) =-\tilde W^{(2)}(\ell-\ud k)\tilde\Gamma^{(1,2)}( k;\ell-\ud k,\ell+\ud k) \tilde W^{(2)}(\ell+\ud k)\cr
&= -   U (\ell_3-\ud k_3)H^{(1,2)}( k;\ell-\ud k,\ell+\ud k)   U ( \ell_3+\ud k_3)\hskip9mm &\eqnd{\eCSWRpsipsi} \cr}$$
with
$$\eqalign{H^{(1,2)}( k;\ell-\ud k,\ell+\ud k)&=\left[i\sla{\ell}-\ud i\sla{k}+M\right]^{-1}
\,\tilde V^{(1,2)}( k;\ell-\ud k,\ell+\ud k)\cr&\quad\times \left[i\sla{\ell} +\ud i\sla{k}+M\right]^{-1} .\cr}$$
Quite generally, we introduce the decomposition,
$$\tilde V^{(1,2)}(k;\ell-k/2,\ell+k/2) =A(\ell_3,k)+i\sigma_3 B(\ell_3,k) .\eqnn $$
The relation between the vectors $(E,F)$ and $(A,B)$ then is
$$\pmatrix{E(\ell_3,k) \cr F(\ell_3,k)\cr}={\cal O}\bigl(-\chi(\ell_3 )/2\bigr) \pmatrix{A(\ell_3,k) \cr B(\ell_3,k)\cr}\,.\eqnd\eCSEFtoAB $$
We expand $A$ and $B$ in powers of $\g$,
$$A=\sum_{n=0} A_n \g^n,\ B=\sum_{n=0} B_n \g^n, $$
and  obtain
$$A_0=1\,,\ A_1= 2  M\left[{\cal B}_1 (k)-\ell_3\Xi(\ell_3, k)\right],\ B_0=0\,,\ B_1=2 t_1\Xi(\ell_3, k).$$
\smallskip
{\it Two-loop  order.} The two-loop order is the sum of the contribution obtained by acting with the matrix $\bf T$ taken at $\g=0$ on the vector $(E_1,F_1)$ and the contribution of  order $\g$ of $\bf T$ acting on $(1,0)$.\par
Details about the calculation can be found in appendix \label{\appCSvertex}. One obtains
$$\eqalign{E_2&=-\frac{1}{8} \left[\Theta(\ell_3+k_3/2)+\Theta(\ell_3-k_3/2)\right]^2\cr&\quad +t_1 \left[\Theta(\ell_3+k_3/2)+\Theta(\ell_3-k_3/2)\right]\Xi (\ell_3,k)\cr&\quad+2\left[(M^2-t_1)\ell_3^2-t_1t_2\right]\Xi^2 (\ell_3,k)+2\left(t_1+M^2\right){\cal B}_1^2(k),\cr
&\quad -4 M^2\ell_3 {\cal B}_1(k) \Xi (\ell_3,k) \cr
F_2&=   M \left[\Theta(\ell_3+k_3/2)+\Theta(\ell_3-k_3/2)\right]\left[\ell_3 \Xi (\ell_3,k)-{\cal B}_1(k)\right]\cr&\quad +4M t_1\Xi (\ell_3,k) {\cal B}_1(k) .\cr}$$
Then, using the relation \eqns{\eCSEFtoAB}, one verifies that all terms proportional to $\Theta$ functions cancel, justifying to two-loop order the transformation \eqns{\eCSVdef}, and one finds
$$\eqalign{A_2&= 2\left[(M^2-t_1)\ell_3^2-t_1t_2\right]\Xi^2 (\ell_3,k)+2\left(t_1+M^2\right){\cal B}_1^2(k),\cr
&\quad -4 M^2\ell_3 {\cal B}_1(k) \Xi (\ell_3,k) \cr
B_2&=   4M t_1\Xi (\ell_3,k) {\cal B}_1(k) .\cr}$$
\smallskip
{\it Higher orders.} With the definition
$$\tau=\sqrt{t_1t_2+\ell_3^2(t_1-M^2)}=\frac{1}{4}\sqrt{k^4+k^2(4\ell_3^2+4M^2-k_3^2)-4M^2k_3^2 }\,,\eqnd{\eCStaudef} $$
a calculation to order $\g^6$ (appendix \label{\appCSvertex}) then suggests
the general form
\eqna\eCSVertexscal
$$\eqalignno{A(\ell_3,k)&={\cos\left(2\g\tau \Xi(\ell_3,k)\right)-(M\ell_3/\tau)\sin\left(2\g\tau \Xi(\ell_3,k)\right)
 \over \cos\bigl( \g k{\cal B}_1(k)\bigr)-2(M/k)\sin\bigl( \g k{\cal B}_1(k)\bigr) } \,, &\eCSVertexscal{a} \cr B(\ell_3,k)&= {t_1\over  \tau}{ \sin\left(2\g\tau \Xi(\ell_3,k)\right)\over  \cos\bigl(\g k{\cal B}_1(k)\bigr)-2(M/k)\sin\bigl( \g k{\cal B}_1(k)\bigr)}.&\eCSVertexscal{b}\cr}$$
Using Eq.~\eqns{\eCSXiTheta}, one verifies that the coefficients $A$ and $B$  for $k=0$ are consistent with the expressions \eqns{\eCSvertexkzero}.
%%%%%%%%%%%%%%%%%%%%%%%
\subsection{Integral equation}

For $N$ large, the vertex functions   are solutions to   {\it one-dimensional}\/  coupled integral equations, which in terms of the vector ${\cal V}_{EF}\equiv(E,F)$ read\sslbl\ssCSinteqreduction
$$ {\cal V}_{EF}(\ell_3,k)={\cal V}_0 +{\g\over(2\pi)^3}\int{\d^3 p\,{\cal T} (p_3,k){\cal V}_{EF}(p_3,k)\over\left(\ell_3-p_3\right) Dp_1 Dm_1 }  \eqnd{\eCSEFvertexinteq}  $$
with ${\cal V}_0\equiv (E_0,F_0)=(1,0)$.
\smallskip
{\it Preliminary remark.}
Acting with $\g{\bf T}_2$ on both sides of the equation, one obtains
$$\eqalign{&{1\over(2\pi)^3}\int{\d^3 q\, \g {\cal T}_2(q_3,k) {\cal V}_{EF}(q_3,k)\over\left(\ell_3-q_3\right) Dm_2 Dp_2 }=\ud i \chi(\ell_3) \sigma_2{\cal V}_0 \cr &\quad+{\g^2\over(2\pi)^6}{ i \over 2}\sigma_2 \int{\d^3 q\,d^3 p\left(Dm_2+Dp_2\right)  {\cal T} (p_3,k){\cal V}_{EF}(p_3,k)\over\left(\ell_3-q_3\right) \left(q_3-p_3\right)Dm_1 Dp_1 Dm_2 Dp_2}  \cr}   $$
with the notation (Eq.~\eqns{\eCSchiphase}),
$$\chi(\ell_3)=\g\left[\Theta(\ell_3+k_3/2)+\Theta(\ell_3-k_3/2)\right].$$
The usual identity
$${1\over \left(\ell_3-q_3\right) \left(q_3-p_3\right)} = {1\over \left(\ell_3-q_3\right) \left(\ell_3-p_3\right)}+{1\over \left(p_3-q_3\right) \left( p_3-\ell_3\right)} $$
transforms the equation into
$$\eqalignno{ &{\g\over(2\pi)^3}\int{\d^3 q\,  {\cal T}_2(q_3,k) {\cal V}_{EF}(q_3,k)\over\left(\ell_3-q_3\right) Dm_2 Dp_2 }={i\over 2} \chi(\ell_3) \sigma_2
{\cal V}_{EF}(\ell_3,k)\cr&\quad -{\g\over(2\pi)^3}{ i\over 2} \sigma_2 \int{ \d^3 p\, \chi(p_3){\cal T} (p_3,k){\cal V}_{EF}(p_3,k)\over\left(\ell_3-p_3\right)  Dm_1 Dp_1  }
\,.&\eqnd\eCSEFTiiId  \cr}   $$
\smallskip
{\it Rescaled integral equation.}
In the integral equation, we now rescale $\chi\mapsto \varepsilon\chi$, ${\cal T}_2 \mapsto  \varepsilon{\cal T}_2$ and determine the $\varepsilon$ dependence of the solution.
Differentiating with respect to $\varepsilon$, we find
$$\eqalign{ \partial_\varepsilon {\cal V}_{EF}(\ell_3,k)&={\g\over(2\pi)^3}\int{\d^3 p\,\partial_\varepsilon{\cal T} (p_3,k){\cal V}_{EF}(p_3,k)\over\left(\ell_3-p_3\right) Dm_1 Dp_1 }\cr
&+{\g\over(2\pi)^3}\int{\d^3 p\,{\cal T} (p_3,k)\partial_\varepsilon{\cal V}_{EF}(p_3,k)\over\left(\ell_3-p_3\right) Dm_1 Dp_1 }
,\cr } $$
where
$$\eqalign{ \partial_\varepsilon{\cal T} (p_3,k)&=\ud i \chi(p_3)\left(  \sigma_2{\cal T}_1-{\cal T}_1 \sigma_2\right)+{\cal T}_2 \cr
&=\ud i \chi(p_3)\left(  \sigma_2{\cal T} -{\cal T}  \sigma_2\right)+{\cal T}_2  \cr} $$
because ${\cal T}_2$ commutes with $\sigma_2$.\par
We now use the identity \eqns{\eCSEFTiiId} (which is not affected by the rescaling) to eliminate  ${\cal T}_2$.   The equation becomes
$$\eqalign{ \partial_\varepsilon {\cal V}_{EF}(\ell_3,k)
&={\g\over(2\pi)^3}\int{\d^3 p\,{\cal T} (p_3,k)\partial_\varepsilon{\cal V}_{EF}(p_3,k)\over\left(\ell_3-p_3\right) Dm_1 Dp_1 }+{ i\over 2} \chi(\ell_3) \sigma_2
{\cal V}_{EF}(\ell_3,k)
 \cr&\quad +{\g\over(2\pi)^3}{ i\over 2}  \int{\d^3 p\, \chi(p_3)\left[\sigma_2{\cal T} (p_3,k)-{\cal T} (p_3,k) \sigma_2\right]{\cal V}_{EF}(p_3,k)\over\left(\ell_3-p_3\right) Dm_1 Dp_1 }\cr
&\quad -{\g\over(2\pi)^3}{ i\over 2}\sigma_2 \int{ \d^3 p\, \chi(p_3){\cal T} (p_3,k){\cal V}_{EF}(p_3,k)\over\left(\ell_3-p_3\right)  Dm_1 Dp_1  }
  \cr
  &= {\g\over(2\pi)^3}\int{\d^3 p\,{\cal T} (p_3,k)\partial_\varepsilon{\cal V}_{EF}(p_3,k)\over\left(\ell_3-p_3\right) Dm_1 Dp_1 } +{ i\over 2} \chi(\ell_3) \sigma_2 {\cal V}_{EF}(\ell_3,k)\cr
  &\quad -{\g\over(2\pi)^3}{ i\over 2}  \int{\d^3 p\, \chi(p_3) {\cal T} (p_3,k) \sigma_2 {\cal V}_{EF}(p_3,k)\over\left(\ell_3-p_3\right) Dm_1 Dp_1 }\,.\cr
 } $$
Setting
$$X(\ell_3,k)=  \partial_\varepsilon {\cal V}_{EF}(\ell_3,k)-\ud i \chi(\ell_3),
  \sigma_2 {\cal V}_{EF}(\ell_3,k)$$
we obtain the equation
$$X(\ell_3,k)={\g\over(2\pi)^3}\int{\d^3 p\,{\cal T} (p_3,k)X(p_3,k)\over\left(\ell_3-p_3\right) Dm_1 Dp_1 },$$
which has  $X=0$ as the only solution expandable in powers of $\g$.
We infer
$$  \partial_\varepsilon {\cal V}_{EF}(\ell_3,k)=\ud i \chi(\ell_3)
  \sigma_2 {\cal V}_{EF}(\ell_3,k)$$
and thus
$$\left.{\cal V}_{EF}(\ell_3,k)\right|_{\varepsilon=1}=\e^{i\chi(\ell_3)\sigma_2/2}\left.{\cal V}_{EF}(\ell_3,k)\right|_{\varepsilon=0}\,.$$
We  recognize the relation \eqns{\eCSEFtoAB} and thus
$$\left.{\cal V}_{EF}(\ell_3,k)\right|_{\varepsilon=0}=\pmatrix{A \cr B}\,.$$
The vector ${\cal V}=(A,B)$ is thus  solution of Eq.~\eqns{\eCSEFvertexinteq} with $\varepsilon=0$.
%%%%%%%%%%%%%%%%%%%%%%%%%%%%%%%%%
\subsection{Solution of the reduced integral equation}

The vector ${\cal V}=(A,B)$ is solution of the reduced integral equation
$$\eqalignno{&{\cal V}(\ell_3,k)\cr& =(1,0)+{\g\over(2\pi)^3}\int{\d^3 p\,{\cal T}_1(p_3,k) {\cal V}(p_3,k)\over\left(\ell_3-p_3\right) \left[(p+k/2)^2+M^2\right]\left[(p-k/2)^2+M^2\right] }\hskip9mm&\eqnd\eCSVABeq\cr} $$
or in component form,
\eqna\eCSABVertexgen
$$\eqalignno{ A(\ell_3,k) & =1-{2\g\over(2\pi)^3}\int{\d^3 p\, \left[ Mp_3  A (p_3,k)+(p_3^2+t_2)  B (p_3,k) \right] \over\left(\ell_3-p_3\right) \left[(p+k/2)^2+M^2\right]\left[(p-k/2)^2+M^2\right] }\,,\hskip10mm & \eCSABVertexgen{a}\cr
B(\ell_3,k) & =  {2\g\over(2\pi)^3}\int{\d^3 p\, \left[ t_1 A (p_3,k)+Mp_3 B (p_3,k) \right] \over\left(\ell_3-p_3\right) \left[(p+k/2)^2+M^2\right]\left[(p-k/2)^2+M^2\right] }\,. &\eCSABVertexgen{b} \cr}
$$
Inserting the expressions \eCSVertexscal{} in the equations, we note that the numerators simplify since
$$\eqalign{ Mp_3  A (p_3,k)+(p_3^2+t_2)  B (p_3,k)&={M p_3  \cos\left(2\g\tau \Xi(p_3,k)\right)+\tau \sin\left(2\g\tau \Xi(p_3,k)\right)\over
\cos\bigl( \g k{\cal B}_1(k)\bigr)-2(M/k)\sin\bigl( \g k{\cal B}_1(k)\bigr) },\cr
 t_1 A (p_3,k)+Mp_3 B (p_3,k)&= {t_1  \cos\left(2\g\tau \Xi(p_3,k)\right)\over
\cos\bigl( \g k{\cal B}_1(k)\bigr)-2(M/k)\sin\bigl( \g k{\cal B}_1(k)\bigr) }.\cr}
$$
The verification that the functions are the solutions of the integral equation is then simple and relies on a few identities derived in \label{\appCSintXin} concerning integrals of sine and cosine functions. The proofs of the latter identities rely on an expansion order by order in  powers of the coupling $\g$.
\medskip

%%%%%%%%%%%%%%%%%%%%%%%%%%%%%%%%%%%%%%%%%%%%%%%%%%%%%%%%%%%%%%%%%%%%%%%%%%
\section {The $R$ two-point function}

We have defined (Eqs.~\eqns{\eCSRalphadef,\ \eCSRsumdef}) the two gauge-invariant operators\sslbl\ssCSvertextoRR
$$R_\alpha (t,x)= {1\over N}\bar{ \psi}_\alpha  (t,x)\cdot{ \psi}_\alpha (t,x),\quad R(t,x)=R_1(t,x)+R_2(t,x).$$
We have already determined the connected $R$ two-point function $\langle\tilde R(k)\tilde R(-k)\rangle_{\rm c} $ at zero momentum (Eq.~\eqns{\eCSRRexpectation}).
We now calculate it for generic momenta.  We begin with a three-loop perturbative calculation in the large $N$ limit and then derive the exact result from the known vertex function.
\topinsert
\epsfysize=17.4mm
\epsfxsize=72.8mm
\vbox{\elevenpoint\kern2mm
\centerline{\epsfbox{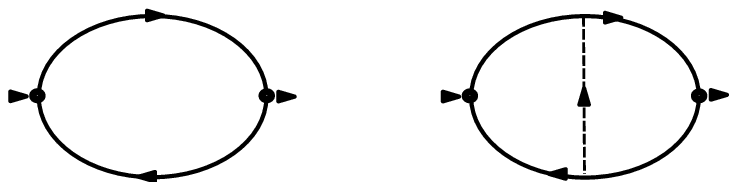}}
\kern-20.8mm
\moveright22.0mm\vbox{\hbox{\kern13.0mm$p+k/2$ }
\kern0.1mm
\hbox{\kern41.3mm$p+  k/2$\kern19.5mm$q+  k/2 $}
\kern2.2mm
\hbox{\kern0.1mm$k$\kern30.6mm$k$\kern9.8mm$k$\kern16.3mm$p-q$\kern7.3mm$k$}
\kern1.6mm
\hbox{\kern40.2mm$p-k/2 $\kern20.5mm$q-k/2  $}
\kern0.7mm
\hbox{\kern13.0mm$p -k/2 $ }}}
\figure{3mm}{The $(\bar\psi\psi)$ two-point function: one- and two-loop diagrams. Dotted lines represent gauge fields.}
\figlbl\figCSeye
\endinsert \subsection{Perturbative calculations}

Since calculations are performed using the dressed fermion propagator \eqns{\eCSNfermioniipt},   only ladder diagrams have to be considered (see section \label{\ssCSNexpand}).
\smallskip
{\it One-loop result.} The one-loop diagram  (first diagram in Fig.~\label{\figCSeye}) with the dressed propagator reads
$$\left<\tilde R(k)\tilde R(-k)\right>_{\rm c}= {1\over(2\pi)^3}\int{\d^3 p\,{\cal N} (p,k)  \over Dp_1Dm_1}\,,$$
where
$$ {\cal N} =-\ud k^2+2p^2-2M^2 - 4Mp_3\sin(\chi_1 )+\ud(4p_3^2 -k_3^2-4M^2 )\bigl(\cos(\chi_1 )-1\bigr)      $$
and $\chi_1\equiv \chi(p_3)$  is the angle \eqns{\eCSchiphase}.\par
We note that
$$2p^2=Dp_1+Dm_1-\ud k^2-2M^2 .$$
Thus,
$$ \left<\tilde R(k)\tilde R(-k)\right>_{\rm c} =2\Omega_1(M)+{1\over(2\pi)^3}\int{\d^3 p\,{\cal N}' (p,k)  \over Dp_1Dm_1} \eqnd\eCSRRone  $$
with
$$ {\cal N}' =-  \left(k^2+4M^2\right) - 4Mp_3\sin(\chi_1 )+\ud(4p_3^2 -k_3^2-4M^2 )\bigl(\cos(\chi_1 )-1\bigr)  .  $$
 \par
At one-loop order,   $\chi_1=0$ and, thus, the contribution is
$$\eqalign{\left<\tilde R(k)\tilde R(-k)\right>_{\rm c} &=2 \Omega_1(M)-{1\over(2\pi)^3}\int{\d^3 p\left(k^2+4M^2\right)  \over \left[\left(p+k/2\right)^2+M^2\right] \left[\left(p-k/2\right)^2+M^2\right]}\cr&\quad=2 \Omega_1(M)-\left(k^2+4M^2\right){\cal B}_1 (k).\cr} $$
 \medskip
{\it Two-loop result.} At two-loop, one needs the second diagram of Fig.~\label{\figCSeye} with the free propagator together with the order $\g$ contribution of expression \eqns{\eCSRRone} (which corresponds to the diagram of Fig.~\label{\figCSprop}). The two-loop ladder diagram yields
$$\eqalign{&{2M\g\over(2\pi)^6}\int{\d^3 p\,\d^3 q\left[(p_3-q_3)(k^2+4M^2)+q_3(Dp_1+Dm_1)-p_3(Dp_2+Dm_2)\right] \over\left(q_3-p_3\right)Dp_1Dm_1Dp_2Dm_2}\cr&\quad= -2M\g\left(k^2+4M^2\right){\cal B}_1^2(k)\cr&\qquad +{2M\g\over(2\pi)^6}\int{\d^3 p\,\d^3 q\left[ q_3(Dp_1+Dm_1)-p_3(Dp_2+Dm_2)\right] \over\left(q_3-p_3\right)Dp_1Dm_1Dp_2Dm_2} \cr
 \cr&\quad= -2M\g\left(k^2+4M^2\right){\cal B}_1^2(k)+{4M  \over(2\pi)^3}\int{\d^3 p\,p_3\chi(p_3) \over Dp_1Dm_1}      \,,\cr}$$
where the $(p,q)$ symmetry has been used.
The last term is exactly cancelled by the order $\g$ of expression  \eqns{\eCSRRone}, which yields
$$-{4M \over(2\pi)^3}\int{\d^3 p\,p_3 \chi(p_3) \over Dp_1Dm_1}.$$
Adding the two contributions, one finds
$$-2M\g\left(k^2+4M^2\right){\cal B}_1^2(k).$$
\midinsert
\epsfysize=17.2mm
\epsfxsize=28.8mm
\vbox{\elevenpoint\kern2mm
\centerline{\epsfbox{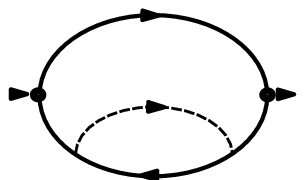}}
\kern-21.8mm
\hbox{\kern57.0mm$q+k/2 $ }
\kern5.8mm
\hbox{\kern44.1mm$k$\kern12mm$p-q$\kern11.5mm$k$}
\kern0.1mm
\hbox{\kern39.3mm$q-k/2 $\kern22.6mm$q-k/2  $}
\kern2.8mm
\hbox{\kern57.8mm$p -k/2 $}}
\figure{3mm}{The $\bar\psi\psi$ two-point function at two loops. Dotted lines represent gauge fields.}
\figlbl\figCSprop
\endinsert
\medskip
{\it Three-loop contribution.}
 Details about the direct three- and four-loop calculations are given in section \label{\appCSRRcalculations}. One finds (Eqs.~\eqns{\eCSRRthreeloops,\ \eCSRRfourloops})
$$-\frac{1}{3}\g^2\left(k^2+12M^2\right)\left(k^2+4M^2\right){\cal B}_1^3(k)-\frac{4}{3}\g^3M\left(k^2+4M^2\right)\left(k^2+6M^2\right){\cal B}_1^4(k).$$
%%%%%%%%%%%%%%%%%%%%%%%%%
\subsection{$R$ two-point function and vertex  three-point function}

The $R$ two-point function is derived from the vertex three point function \eqns{\eCSvertexform} by multiplying it by two additional dressed fermion propagators and taking the trace.\sslbl\ssCSVertextoRR
This leads to the general relation
$$ \left<\tilde R(k)\tilde R(-k)\right>_{\rm c}= 2\Omega_1(M) -{1\over(2\pi)^3}\int{\d^3 q\,{\cal N}(q_3,k) \over Dm_2 Dp_2} \eqnd\eCSRVtoRR $$
with
$$\eqalignno{{\cal N}&=Dm_2+Dp_2+ \left[(4t_1-Dm_2-Dp_2)\cos( \chi_2/2)+4M q_3 \sin( \chi_2/2)\right] A(q_3,k)\cr &\quad +4\left[ Mq_3\cos( \chi_2/2)+\bigl(t_2+  q_3^2-\frac{1}{4}(Dm_2+Dp_2)\bigr) \sin( \chi_2/2)\right] B(q_3,k)  ,\hskip5mm&\eqnd\eCSVtoRRnum\cr}$$
where $\chi_2\equiv \chi(q_3)$ (definition \eqns{\eCSchiphase}), $A$ and $B$ are the coefficients of $\bf 1$ and $i\sigma_3$ in the vertex function $\tilde V^{1,2)}$ (the relation \eqns{\eCSEFtoAB} between $(E,F)$ and $(A,B)$ has been used).\par
We first prove phase factor factorization and cancellation, in such a way that
  the relation \eqns{\eCSRVtoRR} reduces to
$$\eqalignno{&\left<\tilde R(k)\tilde R(-k)\right>_{\rm c}=2\Omega_1(M)\cr&\quad-{1\over(2\pi)^3}\int{\d^3 q \left[\left(k^2+4M^2\right)A(q_3,k)+4Mq_3 B(q_3,k) \right]\over \left[\left(q+k/2\right)^2+M^2\right] \left[\left(q-k/2\right)^2+M^2\right]},&\eqnd{\eCSvertextoRR}\cr} $$
where $A$ and $B$ are the coefficients of $\bf 1$ and $i\sigma_3$ in $\tilde V^{(1,2)}$ given by equations \eCSVertexscal{}.
\smallskip
{\it Proof at first order.} The first contribution of the additional terms in
Eq.~\eqns{\eCSRVtoRR} with respect to \eqns{\eCSvertextoRR} is
$$\eqalign{&\int{\d^3 q  \over(2\pi)^3}  {\left[ A(q_3,k)-1\right]\left( Dp_2+Dm_2\right)\over Dm_2 Dp_2} \cr&\quad -2\int{\d^3 q \over(2\pi)^3} { \chi_2 \left[  M q_3  A(q_3,k) +\left(t_2+ q_3^2\right)  B(q_3,k)\right]\over Dm_2 Dp_2}.\cr}$$
We transform the first term by using the integral equation \eCSABVertexgen{}:
$$\eqalign{&\int{\d^3 q  \over(2\pi)^3}  {\left[ A(q_3,k)-1\right]\left( Dp_2+Dm_2\right)\over Dm_2 Dp_2} \cr
&= -{2\g \over(2\pi)^6}\int{\d^3p\,\d^3 q\left( Dp_2+Dm_2\right)\left[Mp_3 A(p_3,k)+(p_3^2+t_2)B(p_3,k)\right]\over Dp_1 Dm_1 Dp_2 Dm_2 \left(q_3-p_3\right)}.\cr}$$
Integrating over $q$, we find
$$ {1\over(2\pi)^3}\int{\d^3 p\,\chi(p_3)\left[2Mp_3 A(p_3,k)+2(p_3^2+t_2)B(p_3,k)\right]\over Dp_1 Dm_1}  $$
and, thus, the first order difference vanishes. \par
\smallskip
{\it A general identity.}
For the general proof, we need
$${\cal I}_n(p_3)=\g\int{\d^3 q\over(2\pi)^3}{\left(Dp_2+Dm_2\right)\chi ^n(q_3) \over \left(p_3-q_3\right)Dp_2 Dm_2 }\,.$$
We introduce the integral representation of $\chi(q_3)$,
$$\chi(q_3)=\g\int{\d^3 r^i\over(2\pi)^3}{\left(Dp_i+Dm_i\right)  \over \left(q_3-r^i_3\right)Dp_i Dm_i } $$
with $i>2$ and
$$Dp_i=(r^i+k/2)^2+M^2,\ Dm_i=(r^i-k/2)^2+M^2.$$
The integral becomes
$${\cal I}_n=\g^{n+1}\int{\d^3 q\over(2\pi)^3}
{\left(Dp_2+Dm_2\right)  \over \left(p_3-q_3\right)Dp_2 Dm_2 }\prod_{i=3}^{n+2}{\d^3 r^i\over(2\pi)^3}{\left(Dp_i+Dm_i\right)  \over \left(q_3-r^i_3\right)Dp_i Dm_i } \,.$$
Except for the gauge propagator, the integrand in symmetric in $(q,r^i)$.
Symmetrizing the whole integrand and using the identity \eqns{\eCSbasicIdii},   we find
$${\cal I}_n(p_3)={1\over\left(n+1\right) }\chi^{ n+1}(p_3).\eqnn $$
\smallskip
{\it General proof.}
We now consider the whole contribution coming from the numerator \eqns{\eCSVtoRRnum} and proportional to $(Dp_2+Dm_2)$,
$$\left(Dp_2+Dm_2\right)\left[1- \cos( \chi_2/2)  A(q_3,k)  - \sin( \chi_2/2) B(q_3,k)\right].$$
We again   use the integral equation \eCSABVertexgen{} both for $A$ and $B$ and obtain
$$\left(Dp_2+Dm_2\right)\left[1-\cos(\chi_2/2)+  {2\g\over(2\pi)^3}\int{\d^3 p\,{\cal N}'(p_3,k)  \over\left(q_3-p_3\right) Dp_1 Dm_1 }\right]$$
with
$$\eqalign{{\cal N}'(p_3,k) &=\left[ Mp_3  A (p_3,k)+(p_3^2+t_2)  B (p_3,k) \right]\cos(\chi_2/2)\cr&\quad + \left[ t_1 A (p_3,k)+Mp_3 B (p_3,k) \right]\sin(\chi_2)/2).\cr}$$
We substitute this expression into Eq.~\eqns{\eCSRVtoRR}, integrate over $q$ and find the numerator
$$\eqalign{&-4\left[ Mp_3  A (p_3,k)+(p_3^2+t_2)  B (p_3,k) \right]\sin(\chi_1/2)\cr
&\quad -4 \left[ t_1 A (p_3,k)+Mp_3 B (p_3,k) \right]\left[\cos(\chi_1 /2)-1\right] \cr}$$
with a denominator $Dp_1 Dm_1$.
We have still to add the remaining part of \eqns{\eCSVtoRRnum} (replacing $q$ by $p$ as integration variable),
$$\eqalign{&\left[ 4t_1  \cos( \chi_1/2)+4M p_3 \sin( \chi_1/2)\right] A(p_3,k)\cr &\quad +\left[4Mp_3\cos( \chi_1/2)+(4 t_2+4 p_3^2 ) \sin( \chi_1/2)\right] B(p_3,k).\cr}$$
The $\chi$-dependence cancels and one recovers the anticipated  expression \eqns{\eCSvertextoRR},
$$4 \left[ t_1 A (p_3,k)+Mp_3 B (p_3,k) \right].$$
%%%%%%%%%%%%%%%%%%%%%%%%%%%
\subsection{The exact $R$ two-point function at large $N$}

We now derive the connected two-point function  \sslbl\ssExactRR
 $\langle\tilde R(k)\tilde R(-k)\rangle_{\rm c}$
from the expression \eCSVertexscal{} of the vertex function.
In section \label{\ssCSVertextoRR}, we have proved the reduced relation \eqns{\eCSvertextoRR},
$$\eqalignno{&\left<\tilde R(k)\tilde R(-k)\right>_{\rm c}=2\Omega_1(M)\cr&\quad-{1\over(2\pi)^3}\int{\d^3 q \left[\left(k^2+4M^2\right)A(q_3,k)+4Mq_3 B(q_3,k) \right]\over \left[\left(q+k/2\right)^2+M^2\right] \left[\left(q-k/2\right)^2+M^2\right]}, \cr} $$
where $A$ and $B$ are the coefficients of $\bf 1$ and $i\sigma_3$ in $\tilde V^{(1,2)}$ given by equations \eCSVertexscal{}.\par
Inserting into the equation
the expressions \eCSVertexscal{}, one finds a factor
$$-{  k^2+4M^2 \over  \cos\bigl( \g k{\cal B}_1(k)\bigr)-2(M/k)\sin\bigl( \g k{\cal B}_1(k)\bigr) } $$
multiplied by the integral
 $$\int{\d^3 q  \cos\left(2\g\tau\Xi(q_3,k)\right)  \over \left[\left(q+k/2\right)^2+M^2\right] \left[\left(q-k/2\right)^2+M^2\right]}.$$
Using the result \eqns{\eCSintRRa}, one obtains the two-point function
$$\left<\tilde R(k)\tilde R(-k)\right>_{\rm c} =2 \Omega_1(M)-{k^2+4M^2\over k\g}{ \tan\bigl(\g k{\cal B}_1(k)\bigr)\over 1 -2 M \tan\bigl(\g k{\cal B}_1(k)\bigr)/k}\,,\eqnd\eCSRRkexact$$
where from \eqns{\eCSBubble},
$$\g k {\cal B}_1(k)={\g \over 4\pi}\arctan(k/2M)\,.$$
This expression   agrees with the known values at $k=0$ (Eq.~\eqns{\eCSRRexpectation}) and $M=0$ and the perturbative expansion to the calculated orders. It requires only a constant additive renormalization.

 %%%%%%%%%%%%%%%%%%%%%%%%%%%%%%%%%%%%%
\section{Mass gap and critical coupling}

{\it Mass gap equation.} We now examine the gap equation \eqns{\eCSMassrenormalized},\sslbl{\ssSCmassgap}
$$M_0=\M+\g \Omega_1(\M),\eqnd{\eCSMfergap}$$
where $\M$ is the   fermion physical mass and $\Omega_1(\M)$ is a divergent quantity  defined by equation  \eqns{\eOmegandef}.\par
Introducing the explicit result \eqns{\eCStadpole},
we infer from Eq.~\eqns{\eCSMfergap}  that   $M_0$ can be written as
$$M_0=M_c+m \eqnn $$
with
$$M_c=\g{\Lambda \over 4\pi} \,,\eqnn $$
where $m$, which is finite for $\Lambda\to\infty$ since $\M $ is finite,  provides a physical mass scale. The gap equation \eqns{\eCSMfergap} then reads
$$m=\M\left(1-{\g \over 4\pi}\sgn(\M) \right) , \eqnd{\eCSMR} $$
which has solutions in the domain connected with $\g=0$ only for $|\g|<4\pi$.
\par
We  note that if $\M$ is a solution, $-\M$ is a solution of the gap equation obtained by changing $M_0\to-M_0$ together with $\g\to-\g$, a general symmetry of the problem. Therefore, from now on {\it we  restrict the discussion to} $\M\ge 0$.
\par
For $g\ne 4\pi$, the fermion mass, solution of the gap equation, is
$$\M ={m \over 1-g/4\pi}\,.\eqnn $$
If $m\ne 0$ the fermions are massive and scale symmetry is explicitly broken.
For $m>0$, the equation has a solution only for $g<4\pi$ while  for $m<0$ it has a solution  only for $g>4\pi$. For $m\ne 0$, since $\M$ diverges at $g=4\pi$, crossing the values $g=\pm 4\pi$ seems impossible and one may suspect that the field theory makes sense only for $|g|<4\pi$.
\par
For the special value $m=0$ or $M_0=M_c$, Eq.~\eqns{\eCSMR} reduces to
$$ \M  =  {\g\over 4\pi}|\M|. \eqnd{\eCSgapEC} $$
For $\g\ne \g_c=\pm 4\pi$, the equation implies $\M=0$, that is, fermions are massless and the theory is thus conformal invariant.
\par
By contrast, for $m=0$ and $\g=\pm 4\pi$, the equation is always satisfied and $\M$ remains   undetermined.
We then note that if indeed $\M\ne 0$ in a theory with no renormalized mass scale, $m=0$  indicates a spontaneous breaking of scale invariance. As in similar cases (like in \refs{\rBM}~and earlier works \refs{\rBMB})  it  requires the appearance of a dilaton pole at $g=\pm 4\pi$.
This issue is discussed further below where we show that no dilaton pole appears, yet another indication that, indeed, the physical range is limited to $-4\pi \le g \le 4\pi$.
\par
Returning to the definition \eqns{\eCScouplingdef} of the coupling constant $\g$, one verifies that the critical values $\g_c$ are acceptable since they imply the relation between integers
$$\kappa=\pm N\,.$$
\medskip
{\it The massless or large momentum limit.} In the limit $M\to0$ or, equivalently, $k\to\infty$,   the expression \eqns{\eCSRRkexact} reduces to
$$\left<\tilde R(k)\tilde R(-k)\right>_{\rm c}\mathop{=}_{M=0}{\Lambda \over 2\pi}-{\tan(\g/8) \over\g} k\,,\eqnd\eCSRRconformal $$
which, up to a constant, is simply proportional to the two-fermion phase space. For all values of $k$, the correlation function diverges for $\g=\pm 4\pi$ and for $4\pi<|\g|<8\pi$ it violates positivity. Therefore, in the domain containing the origin, the physical range  is limited to $|\g|\le 4\pi$. The special case $|\g|=4\pi$ has to be examined separately.
\par
Eq.~\eqns{\eCSRRconformal} is in agreement with  Ref.~\refs{\GurAri} where  calculations were done in the light-cone gauge (See Eqs.~(32) and (33) in Ref.~\refs\GurAri). A detailed analysis of this result in different regularization schemes in the light-cone gauge can be found in the appendix of Ref.~\refs{\BardeenFermions}.
\medskip
{\it Singularity of the two-point function and critical couplings.}
It is now interesting to look at the zeros of the denominator in expression \eqns{\eCSRRkexact}. The denominator vanishes for
$$\tan\left({\g\over 4\pi}\arctan(k/2M)\right)={k \over 2M} $$
and, thus,
$${\g\over 4\pi}\arctan(k/2M)=\arctan(k/2M) \pmod{\pi}.$$
The denominator vanishes for all values of $k$ for $\g=4\pi$,  that is, for one critical value  of $\g$ that is singled out by the fermion gap equation.  More precisely,
$$\left<\tilde R(k)\tilde R(-k)\right>_{\rm c}\mathop{\sim}_{\g\to 4\pi}-{1\over (4\pi-\g)}{k \over \arctan(k/2M)}\,.\eqnd\eCSRRgivpi $$
This expression shows that even after a renormalization of the $R$-field to remove the singularity at $\g=4\pi$, the $R$ two-point function has no massless pole at $k=0$.
\par
Other solutions are $k$-dependent. Another family of solutions
is given by
$${\g\over4\pi}-1=-{\pi \over \arctan(k/2M) },$$
which, for $k\to\infty$ converge toward $\g=-4\pi $ from below. Since in the Euclidean formulation poles with $k$ real are unphysical, this confirms again that the admissible range of values of $\g$ is restricted to $-4\pi\le \g\le 4\pi$. In this range, the physical fermion mass is always an increasing function of the bare mass (Eq.~\eqns{\eCSMR}), which sounds reasonable.
\par
We thus note that the case $m=M_0-M_c=0$ and $g=\pm 4\pi$, which allowed for a possible non vanishing fermion mass  $M   $, is inconsistent. Such a spontaneously broken scale invariance phase would require the appearance of a massless pole in Eq.~\eqns{\eCSRRkexact}. However, this  does not happen.  After a renormalization of the $R$-field by a factor $\sqrt{4\pi-g}$ to remove the singularity at $\g=4\pi$, one verifies from Eq.~\eqns{\eCSRRgivpi} that the correlation function has no pole at $k^2=0$. Moreover, expression \eqns{\eCSRRRexpectation} immediately shows that this is not a suitable renormalization to remove singularities at $\g=4\pi$. A more suitable renormalization would seem to be a multiplication by $(4\pi-\g)$ but then the two-point function,  as well as the three-point function at zero-momentum,  vanish in the $\g=4\pi$ limit.
%%%%%%%%%%%%%%%%%%%%%%%%%%%%%%%%%%%%%%%%%%%%%%%%%%%%%%%%%%
\section{The current two-point function}

The current associated to fermion number conservation is
$$J_\mu(x)=i\bar\psi(x)\sigma_\mu\cdot\psi(x).\eqnd\eCSfermioncurrent $$
The current is gauge-invariant and conserved.\par \sslbl\sJJcorrel
The third component is related to quantities already defined since
$$J_3(x)=i\bigl(R_1(x)-R_2(x)\bigr).\eqnn $$
%%%%%%%%%%%%%%%%%%%%%%%%%%%
\subsection{The $\langle J_3\psi\bar\psi\rangle$ vertex}

Following the same strategy as for the $R$ two-point function, we first determine the $\langle J_3\psi\bar\psi\rangle$ vertex, which again has a simple decomposition
of the form \eqns{\eCSvertexform}.
It satisfies the integral equation \eqns{\eCSEFvertexinteq} but with the different initial conditions
$$E_0=0\,,\quad F_0=1\,.$$
The proof of section \label{\ssCSinteqreduction} does not depend on the explicit form of the inhomogeneous term and, therefore, the vertex function
  factorizes as in expression \eqns{\eCSVdef},
$$\left< J_3\psi\bar\psi \right>=U^{-1} V^{(1,2)}_3U^{-1},\eqnd\eCSvertexVJdef $$
where $  V^{(1,2)}_3$ can be written as
$$\tilde V^{(1,2)}_3=A+i\sigma_3 B\ {\rm with}\ A=\sum_{n=1}A_n \g^n\,,\ B=1+\sum_{n=1}B_n \g^n  \eqnd{\eCSViiidef} $$
and, in Fourier representation, the vector ${\cal V}_3\equiv (A,B)$ satisfies the reduced integral equation
\eqns{\eCSVABeq} but with the vector $(1,0)$ replaced by the vector $(0,1)$.
%%%%%%%%%%%%%%%%%%%%%
\subsection{Perturbative calculations}

To gain some insight about the structure of the current vertex function, we calculate a few terms of the perturbative expansion.
\smallskip
{\it One- and two-loop order.} At one-loop order, one finds
$$\eqalign{A(\ell_3,k)&=2\g\left[- \left(\ell_3^2+t_2\right) \Xi(\ell_3,k)+ \ell_3 {\cal B}_1(k)\right]+O(\g^2) \cr
B(\ell_3,k)&=1+ 2M\g\left[\ell_3\Xi(\ell_3,k)-{\cal B}_1(k)\right]+O(\g^2). \cr}$$
The two-loop order contributions are
$$\eqalign{A_2(\ell_3,k)&=4Mt_2{\cal B}_1(k)\Xi(\ell_3,k), \cr
B_2(\ell_3,k)&=2\left[\left((M^2-t_1)\ell_3^2-t_1 t_2 \right)\Xi^2(\ell_3,k) +2\ell_3(t_1-M^2){\cal B}_1(k)\Xi(\ell_3,k)\right.\cr&\quad\left.+(M^2-t_1){\cal B}_1^2(k)\right] .\cr}$$
\smallskip
{\it All orders: the massless limit.}
In the limit $M=0$, the form of the vertex can be guessed to all orders.
In terms of the quantity \eqns{\eCStaudef},
$$\tau=\sqrt{t_1t_2+\ell_3^2(t_1-M^2)}, $$
for $M=0$, these expressions can  be written as
$$\eqalignno{A&=-{\ell_3^2+t_2\over  \tau} \cos\left(\g k{\cal B}_1 \right)\sin\left(2\g \tau \Xi \right)-(\ell_3/\sqrt{t_1})\sin\left(\g k{\cal B}_1 \right) \cos\left(2\g \tau \Xi \right)
   \cr
B&= \cos\left(\g k{\cal B}_1 \right)\cos\left(2\g\tau \Xi \right) -(\ell_3\sqrt{t_1}/\tau)\sin\left(\g k{\cal B}_1 \right) \sin\left(2\g \tau \Xi \right)
  . \cr   }$$
 \smallskip
{\it All orders.} Calculating more terms, and using the experience gained with the scalar function, one guesses that in terms of the quantity \eqns{\eCStaudef},
the exact expressions are
\eqna\eCSVertexvect
$$\eqalignno{A(\ell_3,k)&= {2\ell_3 \over k}\sin(kg{\cal B}_1)\cos(2\g\tau \Xi )- {t_2+\ell_3^2\over \tau}\cos(kg{\cal B}_1)\sin(2\g\tau \Xi )\cr&\quad +{2M t_2\over k\tau} \sin(kg{\cal B}_1)\sin(2\g\tau \Xi ),&\eCSVertexvect{a} \cr
B(\ell_3,k)&= \cos\left(\g k{\cal B}_1 \right)\cos\left(2\g\tau \Xi \right) +{\ell_3 k \over2\tau}\sin\left(\g k{\cal B}_1 \right) \sin\left(2\g \tau \Xi \right)
 \cr&\quad-{2M \over k}\sin\left(\g k{\cal B}_1 \right) \cos\left(2\g \tau \Xi \right)+{M\ell_3\over\tau}\cos\left(\g k{\cal B}_1 \right) \sin\left(2\g \tau \Xi \right) .\hskip5mm &\eCSVertexvect{b}
\cr}$$
In this form, one finds linear combinations of the same trigonometric functions as in expressions \eCSVertexscal{}.
%%%%%%%%%%%%%%%%%%%%%%%%%%%%%%
\subsection{Integral equation}

The function $A$ and $B$ are solutions of integral equations of the form \eCSABVertexgen{}, but with different boundary conditions, which read
\eqna\eCSJJVertexgen
$$\eqalignno{ A(\ell_3,k) & = -{2\g\over(2\pi)^3}\int{\d^3 p\, \left[ Mp_3  A (p_3,k)+(p_3^2+t_2)  B (p_3,k) \right] \over\left(\ell_3-p_3\right) \left[(p+k/2)^2+M^2\right]\left[(p-k/2)^2+M^2\right] }\,,\hskip9mm & \eCSJJVertexgen{a}\cr
B(\ell_3,k) & = 1+ {2\g\over(2\pi)^3}\int{\d^3 p\, \left[ t_1 A (p_3,k)+Mp_3 B (p_3,k) \right] \over\left(\ell_3-p_3\right) \left[(p+k/2)^2+M^2\right]\left[(p-k/2)^2+M^2\right] }\,. &\eCSJJVertexgen{b} \cr}
$$
Inserting the expressions \eCSVertexvect{}, one obtains the numerators
$$\eqalign{& Mp_3  A (p_3,k)+(p_3^2+t_2)  B (p_3,k)=  {2\tau p_3\over k}\sin\left(\g k{\cal B}_1 \right)\sin\left(2\g\tau \Xi \right)\cr&\quad+\left[(p_3^2+t_2)\cos\left(\g k{\cal B}_1 \right)  -{2M t_2\over k} \sin\left(\g k{\cal B}_1 \right)\right]\cos\left(2\g\tau \Xi \right),\cr
&t_1 A (p_3,k)+Mp_3 B (p_3,k)\cr&\quad =\left[ M p_3 \cos\left(\g k{\cal B}_1 \right) + \ud k p_3 \sin\left(\g k{\cal B}_1 \right)\right]\cos\left(2\g\tau \Xi \right)\cr &\quad+\left[
-\tau \cos\left(\g k{\cal B}_1 \right) +{2M\tau\over k} \sin\left(\g k{\cal B}_1 \right)\right]\sin \left(2\g\tau \Xi \right) .\cr}$$
Using the same identities as in the scalar case (for details see \label{\appCSintXin}), one can check that the functions \eCSVertexvect{} are indeed the solutions of  the integral equations with the proper boundary conditions.
%%%%%%%%%%%%%%%%%%%%%%%%%%%%%%%%%%%%%%%%%%%%%%%%%%%%%%
\subsection{From the $J_3$ vertex to the $J_3$ two-point function}

We follow directly the method of section \label{\ssCSVertextoRR}.
The general relation between the $J_3$ vertex and the $J_3$ two-point function is\sslbl\ssCSVertextoJJ
$$ \left<\tilde J_3(k)\tilde J_3(-k)\right>_{\rm c}= 2\Omega_1(M) -{1\over(2\pi)^3}\int{\d^3 q\,{\cal N}(q_3,k) \over Dm_2 Dp_2}  $$
with
$$\eqalign{{\cal N}&=Dm_2+Dp_2+ \left[ 4M q_3\cos( \chi_2/2)+\left(Dp_2+Dm_2 -4 t_1\right)\sin( \chi_2/2)\right] A(q_3,k)\cr &\quad +\left[4 t_2+4q_3^2-Dp_2-Dm_2)\cos( \chi_2/2) -4Mq_3 \sin( \chi_2/2)\right] B(q_3,k)  ,\cr}$$
where $\chi_2\equiv \chi(q_3)$ (definition \eqns{\eCSchiphase}), $A$ and $B$ are the coefficients of $\bf 1$ and $i\sigma_3$ in the vertex function $\tilde V^{1,2}_3$ (defined in \eqns{\eCSViiidef}) and the relation \eqns{\eCSEFtoAB} between $(E,F)$ and $(A,B)$ has been used).\par
We first prove that the relation  reduces to
$$\eqalignno{\left<\tilde J_3(k)\tilde J_3(-k)\right>_{\rm c}&= 2\Omega_1(M)\cr&\quad-{4\over(2\pi)^3}\int{\d^3 q \left[ M q_3A(q_3,k)+\left(t_2+  q_3^2 \right)B(q_3,k) \right]\over \left[\left(q+k/2\right)^2+M^2\right] \left[\left(q-k/2\right)^2+M^2\right]},\hskip7mm&\eqnd{\eCSvertextoJJ}\cr} $$
We use a strategy similar to the one used in section \label{\ssCSVertextoRR}.\par
We have to transform the combination
$$\left(Dp_2+Dm_2  \right)\left[1+\sin( \chi_2/2)  A(q_3,k)- \cos( \chi_2/2)     B(q_3,k)\right].$$
We   use the integral equation \eCSJJVertexgen{} both for $A$ and $B$ and obtain
$$\left(Dp_2+Dm_2\right)\left[1-\cos(\chi_2/2)+  {2\g\over(2\pi)^3}\int{\d^3 p\,{\cal N}'(p_3,k)  \over\left(q_3-p_3\right) Dp_1 Dm_1 }\right]$$
with
$$\eqalign{{\cal N}'(p_3,k) &=-\left[ 2Mp_3  A (p_3,k)+(p_3^2+t_2)  B (p_3,k) \right]\sin(\chi_2/2)\cr&\quad - \left[ t_1 A (p_3,k)+Mp_3 B (p_3,k) \right]\cos(\chi_2)/2).\cr}$$
We then integrate over $q$ and find the numerator
$$\eqalign{&4\left[  Mp_3  A (p_3,k)+(p_3^2+t_2)  B (p_3,k) \right]\left[\cos(\chi_2/2)-1\right]\cr&- 4\left[ t_1 A (p_3,k)+Mp_3 B (p_3,k) \right]\sin(\chi_2)/2) .\cr} $$
All $\chi$ dependent terms then cancel and the remaining term is
$$-4\left[  Mp_3  A (p_3,k)+(p_3^2+t_2)  B (p_3,k) \right],$$
in agreement with expression \eqns{\eCSvertextoJJ}.
%%%%%%%%%%%%%%%%%%%%%%%%%%%%%%%%%%%%%%%%%%%%
\subsection{The $J_3$ two-point function}

Due to current conservation, the current two-point function has the general form\sslbl\ssCSJJtwopoint
$$\left<J_\mu(k)J_\nu(-k)\right>=\left(\delta_{\mu\nu}-{1\over k^2}k_\mu k_\nu\right)J(k^2).\eqnd\eCSJJthree $$
The current two-point function can be determined by calculating only the $J_3$ two-point function since
$$\left<J_3(k)J_3(-k)\right>={1\over k^2}\left(k^2-k_3^2\right)J(k^2).\eqnd\eCSJJgeneral $$
The $J_3 $  two-point function is also given by
$$\left<J_3(x)J_3(y)\right>=-2\left<R_1(x)R_1(y)\right>+2\left<R_1(x)R_2(y)\right>, $$
where we have used the property that all functions are real and
$$\left<R_1(x)R_1(y)\right>=\left<R_2(x)R_2(y)\right>\ {\rm and}\ \left<R_1(x)R_2(y)\right>=\left<R_2(x)R_1(y)\right>.$$
Since $\langle RR \rangle $ is already known,   only one function remains to be determined.\par
%%%%%%%%%%%%%%%%%%%%%%%%%%%%%%%%%%%%%%%%%%%%%%%%%%%%%
Equation \eqns{\eCSvertextoRR} is  modified. We have proved in section \label{\ssCSVertextoJJ} that after subtraction of the terms that cancel the phase factors,  the relation becomes (Eq.~\eqns{\eCSvertextoJJ})
$$\eqalignno{\left<\tilde J_3(k)\tilde J_3(-k)\right>_{\rm c}&= 2\Omega_1(M)\cr&\quad-{4\over(2\pi)^3}\int{\d^3 q \left[ M q_3A(q_3,k)+\left(t_2+  q_3^2 \right)B(q_3,k) \right]\over \left[\left(q+k/2\right)^2+M^2\right] \left[\left(q-k/2\right)^2+M^2\right]}.\hskip7mm \cr} $$
\smallskip
{\it Leading (one-loop) order.} It is given by
$$ \eqalign{ \left<\tilde J_3(k)\tilde J_3(-k)\right>_{\rm c} &= 2\Omega_1(M)-\left(k^2-k_3^2\right){\cal B}_1(k)\cr&\quad -  {1\over(2\pi)^3}\int{\d^3 p\, 4p_3^2\over\left[(p+k/2)^2+M^2\right]\left[(p-k/2)^2+M^2\right]}.\cr}$$
The calculation of the remaining integral can be found in appendix \label{\ssRRcalculation}. One finds (Eq.~\eqns{\eCSFocalculf})
$$\eqalignno{&{1\over(2\pi)^3}\int{\d^3 p\,  p_3^2\over\left[(p+k/2)^2+M^2\right]\left[(p-k/2)^2+M^2\right]}=\frac{1}{4}\Omega_1(M)+{\Lambda\over48\pi} \cr &\quad -\frac{1}{8}\left(k^2+4M^2\right){\cal B}_1(k)+{k_3^2\over k^2}
\left[\frac{1}{8}\left(k^2+4M^2\right){\cal B}_1(k)+\frac{1}{4}\Omega_1(M)-{\Lambda\over16\pi} \right].\hskip7mm &\eqnd{\eCSintqiiib}   \cr}$$
Therefore,
$$ \eqalign{ \left<\tilde J_3(k)\tilde J_3(-k)\right>_{\rm c}
&=-{1\over2}\left(1 -{k_3^2 \over k^2}\right) \left((k^2-4M^2){\cal B}_1(k)+{M\over 2\pi}\right),   \cr}$$
where the divergent additive constant is fixed by current conservation.
\smallskip
{\it Higher-loop calculations.} After some algebra, one finds
$$\eqalign{&\left(k^2-k_3^2\right)\left[2\g M{\cal B}_1^2(k)+\frac{1}{3}\g^2\left(k^2-4M^2\right){\cal B}_1^3(k)-\frac{2}{3}\g^3 Mk^2{\cal B}_1^4(k)\right.\cr&\quad-\frac{1}{15}\g^4k^2\left(k^2-4M^2\right){\cal B}_1^5(k)+\frac{4}{45}\g^5 M k^4{\cal B}_1^6(k)\cr&\quad\left. +\frac{2}{315}\g^6k^4\left(k^2-4M^2\right){\cal B}_1^7(k)+ O(\g^7)\right].\cr}$$
The known terms sum in the form
$$\eqalignno{& \left<\tilde J_3(k)\tilde J_3(-k)\right>=\left(1-{k_3^2\over k^2}\right)\cr&\quad\times \left[M\left({1\over \g}-{1\over 4\pi}\right)-{M\over \g} \cos(2\g k{\cal B}_1)- {1\over4\g k}  (k^2-4M^2) \sin(2\g k{\cal B}_1) \right] .\hskip9mm&\eqnd\eCSJJexact \cr} $$
The result is consistent with expressions \eCSVertexvect{}. Indeed,
inserting the expressions \eCSVertexvect{} into equation \eqns{\eCSvertextoJJ},
one finds in the numerator the combination (we have factorized $-4$)
$$\eqalign{&{-2M t_2 \over k}\sin(kg{\cal B}_1)\cos(2\g\tau \Xi ) +{2 \tau  q_3  \over  k }  \sin(kg{\cal B}_1)\sin(2\g\tau \Xi )\cr&\quad +(t_2+q_3^2) \cos\left(\g k{\cal B}_1 \right)\cos\left(2\g\tau \Xi \right)
  \cr} $$
One then expands for $q_3\to\infty$ and keeps the two leading terms, proportional to $q_3^2$ and $q_3^0$. The term proportional to $q_3^2$ can be integrated using the results \eqns{\eCSomegaSnnii} and \eqns{\eCSintqiiib}. The integration of the term proportional to $q_3^0$ relies on $S_{n,n}$ in equation \eqns{\eCSomegaSnm}.
One then recovers the expression \eqns{\eCSJJexact}.
\smallskip
{\it The large momentum or zero mass limit.} In the limit $M\to0$, the two-point function reduces to
$$\left<\tilde J_3(k)\tilde J_3(-k)\right>=-\left(1-{k_3^2\over k^2}\right){k\over 4\g}\sin(\g/4).\eqnd\eCSJJConformal$$
We note that positivity is only satisfied if $\sin(\g/4)/\g$ is positive.
This again implies $|\g|\le 4\pi$.
\par
More generally, one verifies that for $g= 4\pi$, the two-point function vanishes linearly, another confirmation that the field theory does not make sense for $|\g|>4\pi$.
\par
Using Eqs.~\eqns{\eCSJJthree,\ \eCSJJgeneral} we have in the conformal limit of $M\to 0$  also
$$\left<\tilde J_-(k)\tilde J_+(-k)\right>=-\left(1-{k_-k_+\over k^2}\right){k\over 4\g}\sin(\g/4).\eqnd\eCSJJPM  $$
  The current correlation functions in the $M=0$ conformal phase in Eq.~\eqns{\eCSJJConformal}  and \eqns{\eCSJJPM} that are calculated here in the $A_3=0$ gauge  are in agreement with the results of Ref.~\refs{\GurAri} (see there Eqs.~(33),(36) and (37)), which were calculated in the light-cone gauge.

%%%%%%%%%%%%%%%%%%%%%%%%%%%%%%%%%%%%%%%%%%%%%%
\subsection The $R$ three-point function

We have already evaluated the three-point function at zero momentum (expression \eqns{\eCSRRRexpectation}). We can now evaluate it with only one zero momentum by differentiating expression \eqns{\eCSRRkexact} with respect to the bare mass or, in terms of the physical fermion mass
$$\eqalignno{ &\left<\tilde R ( k )\tilde R (  - k)\tilde R(0)\right>_{\rm c}={1\over1- 2\g M \Omega_2(M)}{\partial\over \partial M} \left<\tilde R ( k )\tilde R (-k)\right> \cr&\quad=
-{4M \Omega_2\over1-2\g M   \Omega_2}-{8M\over  \g\left( 1-2\g M   \Omega_2\right)}{ \tan\bigl(\g k{\cal B}_1(k)\bigr)/k\over 1 -2 M \tan\bigl(\g k{\cal B}_1(k)\bigr)/k}
\cr& \qquad - {k^2+4M^2 \over   \g \left(1-2\g M   \Omega_2\right)} {
(2/k^2)\tan^2\bigl(\g k{\cal B}_1(k)\bigr)-4\g M  {\cal B}_2(k)\left[1+\tan^2\bigl(\g k{\cal B}_1(k)\bigr)\right] \over \left[1 -2 M \tan\bigl(\g k{\cal B}_1(k)\bigr)/k  \right]^2  }  ,\cr&&\eqnd{\eCSRRRkk} \cr} $$
 where
$${\cal B}_2(k)= = \int {\d^3q\over (2\pi)^3}  {1\over \left(q^2+M ^2\right)^2\left[(k-q)^2+M ^2\right]}.\eqnn $$
%%%%%%%%%%%%%%%%%%%%%%%%%%%%%%
\smallskip
{\it The general connected $RRR$ three-point at leading order.} We find
$$\eqalign{ \left<\tilde R ( k )\tilde R ( p)\tilde R(-p-k)\right>_{\rm c}
&=\int{\d^3 q\over(2\pi)^3}{4M\left(M^2-3q^2+2p\cdot q-2k\cdot q+p\cdot k\right)\over \left(q^2+M^2\right)\left[(q+k)^2+M^2\right]\left[(q-p)^2+M^2\right]}\cr
&=\int{\d^3 q\over(2\pi)^3} {2M\bigl(8M^2+p^2+k^2+(p+k)^2 \bigr)\over \left(q^2+M^2\right)\left[(q+k)^2+M^2\right]\left[(q-p)^2+M^2\right]}       \cr&\quad -4 M \left[{\cal B}_1(p+k)+{\cal B}_1(p)+{\cal B}_1(k)\right] .  \cr}$$
We define the scalar three-point vertex as
$$T(k,p)=\int{\d^3 q\over(2\pi)^3}{1\over \left(q^2+M^2\right)\left[(q+k)^2+M^2\right]\left[(q-p)^2+M^2\right]}.\eqnd{\eiiiptdef} $$
In particular,
$$T(k,0)={\cal B}_2(k).$$
Then,
$$\eqalignno{\left<\tilde R ( k )\tilde R ( p)\tilde R(-p-k)\right>_{\rm c}&=
2M\left[8M^2+p^2+k^2+(p+k)^2 \right] T(k,p)\cr&\quad-4 M \left[{\cal B}_1(p+k)+{\cal B}_1(p)+{\cal B}_1(k)\right].&\eqnd{\eCSRRRkpqo}\cr}$$
The generalization to all orders of the denominator in \eqns{\eCSRRRkk} is straightforward. Setting
$$ \Delta_R(k)=1 -2 M \tan\bigl(\g k{\cal B}_1(k)\bigr)/k\,, $$
it reads
$$\Delta_R (k)\Delta_R(p)\Delta_R(p+k)\,.$$
However, the generalization to all orders of the function
$\left<\tilde R ( k )\tilde R ( p)\tilde R(-p-k)\right>_{\rm c}$ is not straightforward and requires more complicate calculations.
%%%%%%%%%%%%%%%%%%%%%%%%%%%%%%%%%%%%%%%%%%%%
\section {Adding a deformation to the Chern--Simons fermion action}

We now introduce a scalar field $\sigma(x)$ and add two new terms to the action \eqns{\eCSUNApsi} of the form \sslbl\sSigmaTerm
$${\cal S}_\sigma=\int\d^3 x\left[-\sigma(x)\bar{\Bfg\psi}(x)\cdot{\Bfg\psi}(x)+{N\over 3 g_\sigma}\sigma^3(x)-N {\cal R}\sigma(x)\right],\eqnd\eCSsigmaaction $$
where the new parameters $g_\sigma$ and ${\cal R}$ are fixed when $N\to\infty$.\par
The extra $\sigma(x)$  terms in Eq.~\eqns{\eCSsigmaaction} are analogous to the triple-trace deformation $\lambda_6\phi(x)^6$ added to the Chern--Simons boson theory \refs{\AharonyIII,\ \rBM}. The action in Eq.~\eqns{\eCSsigmaaction}  differs from the action in ~\refs{\AharonyIII} ~by the term  ${\cal R}\sigma(x)$ added here in order to have a proper perturbative meaning to the $\sigma$  term in the action.\par
From dimensional analysis, one concludes that $\sigma$ has mass dimension $1$ and that no other term of dimension $3$ or less and odd in $\sigma$ can be added. Moreover, $g_\sigma$ is dimensionless and ${\cal R}$ has dimension two.\par
For ${\cal R}\ne0$, in the classical limit $\sigma(x)$ has a non-vanishing expectation value $\sigma$, obtained by varying $\sigma(x)$, which is given by
$$ \sigma^2=   g_\sigma  {\cal R}\,.\eqnn  $$
Then, setting
$$\sigma(x)= \sigma+\varsigma(x)  $$
we note that the expectation value $\sigma$ adds to the bare mass $M_0$ and
$${\cal S}_\sigma=\int\d^3 x\left[-\bigl(\sigma+\varsigma(x)\bigr)  \bar{\Bfg\psi}(x)\cdot{\Bfg\psi}(x)+{N\over 3 g_\sigma}\varsigma^3(x)+{N \sigma\over   g_\sigma} \varsigma^2(x)\right].$$
Since the propagator of $\varsigma(x)$ is a constant,
the dimension of $\varsigma(x)$, from the viewpoint now of power counting, is $\frac{3}{2}$ in such a way that $\varsigma^3(x)$ has dimension $\frac{9}{2}$ and, thus, is irrelevant (non-renormalizable).
We  show below that, by contrast,  in the large $N$ limit the situation is different.
%%%%%%%%
\subsection{Large $N$ limit: gap equation and free energy}

We note that the reflection symmetry $\g\to-\g$, $M_0\to-M_0$ is now extended if
$g_\sigma\to-g_\sigma$ and ${\cal R}\to-{\cal R}$ and then the saddle point value $\sigma\to-\sigma$ and also $\M\to -\M$. {\it From now on  we thus again choose $\M\ge 0$}.  \par
First, we perform the transformations of section \label{\sslargeN} in
the addition \eqns{\eCSsigmaaction}. In particular,
$$\int\d^3 x \,\sigma(x)\bar{\Bfg\psi}(x)\cdot{\Bfg\psi}(x)\mapsto
\int\d^2 x\,\d t \,\sigma(t,x)\left[\rho_1(t,t,x)+\rho_2(t,t,x)\right].$$
By varying $\sigma$, we now obtain the additional saddle point equation
$$\eqalignno{ \sigma^2&= g_\sigma  {\cal R}+   g_\sigma \left[\rho_1(0)+\rho_2(0) \right]\cr
&=g_\sigma  {\cal R}+  g_\sigma \int\d\omega\left[\tilde\rho_1(\omega)+\tilde\rho_2(\omega)\right]=g_\sigma \left({\cal R}+\left<R\right>\right).&\eqnd\eCSsigmasaddle \cr}$$
In the other saddle point equations, the bare mass $M_0$ is simply replaced by
$M_0+\sigma$. The gap equation \eqns{\eCSMfergap} is thus replaced by
$$M_0+\sigma=\M+\g \Omega_1(\M)\eqnd\eCSgapsigma $$
and $\sigma$ is   solution of (using Eq.~\eqns{\eCSRexpectFinal})
$$\eqalignno{ \sigma^2 &= g_\sigma\left({\cal R} +\left<R\right>\right)=   g_\sigma \left[{\cal R}+2 \M \Omega_1(\M)+\g\Omega_1^2(\M)\right],&\eqnd\eCSsigmasaddleb \cr
&=  g_\sigma \left[{\cal R}+\M {\Lambda- \M  \over 2\pi}+ \g\left( {\Lambda- \M  \over 4\pi}\right)^2 \right].\cr}$$
Eliminating $\sigma$ between the two equations, we transform the gap equation into
$$\eqalignno{& g_\sigma \left[{\cal R}+2 \M \Omega_1(\M)+\g\Omega_1^2(\M)\right]\cr&\quad=M_0^2-2M_0\left[\M+\g \Omega_1(\M)\right]+\left[\M+\g \Omega_1(\M)\right]^2 &\eqnd\eCSEgapmodified \cr}$$
or, more explicitly,
$$\eqalignno{&\left[1-{(\g-g_\sigma)\over 2\pi}\left(1-{\g \over 8\pi}\right)\right]\M^2+\left(1-{\g\over 4\pi}\right)\left[{(\g-g_\sigma)\over 2\pi}\Lambda-2M_0\right] \M \cr &\quad - g_\sigma  {\cal R}+M_0^2 -       {\g\over 2\pi}M_0 \Lambda +\g(\g-g_\sigma){\Lambda^2 \over 16\pi^2}=0\, ,  &\eqnd\eCSEgapmodifiedb \cr}$$
an  equation quadratic in $\M$.\par
\smallskip
{\it The effective $\sigma$ potential.} After taking into account the $\lambda$ and $\rho$ saddle point equations, the action density for constant $\sigma$ fields normalized as in Eq.~\eqns{\eCSfreeenergy}  becomes
 %$$V_{\rm eff.}(\sigma)= -
%W(M_0+\sigma)
%W(M(M_0,\sigma)) -{\cal R}\sigma+{\sigma^3 \over 3g_\sigma}\,, \eqnd\eCSVsigmaeff $$
 $$V_{\rm eff.}(\sigma)= -W(M(M_0,\sigma)) -{\cal R}\sigma+{\sigma^3 \over 3g_\sigma}\,, \eqnd\eCSVsigmaeff $$
where $W$ is given by   Eq.~\eqns{\eCSNfreeenergyf} and  $M(M_0,\sigma)$  is solution of Eq.~\eqns{\eCSgapsigma}.
%\par
%Note that the effective potential for the $\sigma$ field is given by
%$$V_{\rm  eff.}(\sigma)=W(M(\sigma),\sigma)\eqnd\eCSVeffDefined$$
%where $M(\sigma)$ is the solution of \eqns{\eCSgapsigma}.
Since
 $${\del\over \del\sigma}
%W(M_0+\sigma)
W(M(M_0,\sigma))
={\del\over \del M_0}
%W(M_0+\sigma)
W(M(M_0,\sigma))=\left<R\right> , \eqnn  $$
the expression for $\sigma$ in Eq.~\eqns{\eCSsigmasaddleb} is then obtained from
$${\del\over \del\sigma}V_{\rm eff.}(\sigma)=-\left<R\right>-{\cal R}+{\sigma^2 \over g_\sigma}=0\,.  \eqnd\eCSdelVsigma  $$
Furthermore, by calculating the successive derivatives of $V_{\rm eff.}$ at the saddle point, one obtains the vertex (or 1PI) functions at zero momentum.
%The bare mass parameter $M_0$ is kept fixed and using
%$$ {\del\over \del\sigma}= {1\over  1- g/ 4\pi  }{\del\over \del M}$$
%we have
%$${\del\over \del\sigma }V_{\rm eff.}(\sigma)= {\cal R}- {\sigma^2\over g_\sigma}+2M\Omega_1(M)+g\Omega_1^2(M) = 0\,, \eqnd\eCSEffSigma$$
%which gives \eqns{\eCSsigmasaddleb}.
\smallskip
{\it Divergences and counter-terms.}
We now assume that $\g$  and $g_\sigma$ need not to be renormalized, which we have   checked to some extent. Then the cancellation of divergences in the coefficient of $\M$ implies that $M_0$ has the form
$$M_0=M_c+m\,,\eqnn $$
where $m$ is finite and
$$M_c=(\g-g_\sigma){\Lambda\over 4\pi}\,.\eqnd\eCSMcritical $$
Similarly, the cancellation of divergences in the constant term yields the condition
$${\cal R}={\cal R}_c-{m\Lambda\over 2\pi}+{\eta \over4\pi} m^2, \eqnn
$$
where $\eta$ is a constant parameter and
$${\cal R}_c= ( g_\sigma-\g){\Lambda^2\over 16\pi^2}\,.\eqnd\eCScalRcritical $$
The gap equation then reads
$$ \left[1-{(\g-g_\sigma)\over 2\pi}\left(1-{\g \over 8\pi}\right)\right]\M^2-2  \left(1-{\g\over 4\pi}\right)m  \M   + m^2\left(1-\eta {g_\sigma\over 4\pi}\right)=0\, .   \eqnd\eCSEgapmodifiedc
$$
Note also that the relation \eqns{\eCSgapsigma}  becomes
$$\sigma={\Lambda \over 4\pi}g_\sigma -m +\M\left(1-{\g \over 4\pi}\right) \eqnd\eCSsigmam $$
and, therefore,
$${ 2\sigma \over  g_\sigma}={\Lambda \over 2\pi}  -{2m \over g_\sigma} +{2\M \over g_\sigma}\left(1-{\g \over 4\pi}\right). \eqnd\eCSsigmaoverg $$
\medskip
{\it Critical limit.} If
$m=0$ or
$\eta=4\pi/g_\sigma$
the gap equation has the solution $\M=0$, which corresponds to massless fermions.
For $m\ne 0$   and $\eta=4\pi/g_\sigma$, the gap equation has then also
a non-zero solution
$$\M=m {2 \left(1-{\g/ 4\pi}\right)\over   1-{(\g-g_\sigma)\left(1-{\g / 8\pi}\right)/2\pi} }.$$
Since the solution can also be positive, checking which is the
leading
solution requires calculating the corresponding free energy.\par
Demanding that the coefficient of $\M$ also vanishes yields either $\g=\pm 4\pi$, which here also is a singular point, or $m=0$. Then, both solutions of the gap equation coalesce and $\M=0$.
\par
Finally, the gap equation is satisfied for any value of $\M\ge 0$ if the coefficient of $\M^2$ also vanishes, that is, for
$$\left(\g-g_\sigma \over 2\pi\right)\left(1-{\g \over 8\pi}\right)=1 \ \Leftrightarrow\  g_\sigma=-{(4\pi-\g)^2\over 8\pi-g}\,.\eqnd\eCScubicggs  $$
We examine this special situation in section \label{\ssCSdilaton}.
\par
For $\M<0$, the analysis is the same;  one has just to change $\g\to-\g$, $g_\sigma\to -g_\sigma$.\par
%%%%%%%%%%%%%%%%%%%%%%%%%%%%%%%%%%%%%%%%%%%
\subsection {Connected scalar two-point functions at zero momentum}

Since $R=\bar\psi\psi$ and $\sigma$ are both scalar $U(N)$-invariant fields, they are coupled. To calculate the  two-point functions at zero momentum, we have to differentiate the general saddle point equations twice with respect to $M_0$ at ${\cal R}$ fixed, with respect to $M_0$ and ${\cal R}$ and, finally, twice with respect to ${\cal R}$.
\smallskip
{\it The $R$ two-point function.}
The two-point function is given by
$$ \left<\tilde R(0)\tilde R(0)\right>_{\rm c} ={\partial\left<R\right> \over \partial M_0}= {2\sigma \over g_\sigma}{\partial \sigma \over\partial M_0} \,,\eqnd\eCSRRcubzero $$
where Eq.~\eqns{\eCSsigmasaddleb} has been used.\par
Differentiating Eq.~\eqns{\eCSgapsigma} with respect to $M_0$ and combining it with Eq.~\eqns{\eCSRRcubzero}  to  eliminate $\partial \sigma/\partial M_0$, we obtain
$$\left(1- {\g \over 4\pi}   \right){\partial \M \over \partial M_0}=1+{g_\sigma \over 2\sigma}{\partial \left<R\right> \over\partial M_0}=1+ {g_\sigma \over 2\sigma}{\partial \left<R\right> \over \partial \M}{\partial \M \over \partial M_0}   \,.\eqnd\eCSsigmadM $$
We infer
$${\partial \M \over\partial M_0}={2\sigma\over g_\sigma}{1\over (1-\g/4\pi)} {\cal D}^{-1}(0)\eqnd\eCSdMrdM $$
with
$${\cal D}(k=0)  = {2 \sigma\over g_\sigma} -{1\over 1-\g/4\pi}  {\partial \left<R\right> \over \partial \M}\,.  $$
We note that
$${1\over 1-\g/4\pi}  {\partial \left<R\right> \over \partial \M}=\left<\tilde R(0) \tilde R(0)\right>_{\rm c,0}\,, $$
where $\left<\tilde R(0) \tilde R(0)\right>_{\rm c,0}0$ is given by the expression \eqns{\eCSRRexpectation}. Therefore,
$${\cal D}(0)  = {2 \sigma\over g_\sigma}- \left<\tilde R(0) \tilde R(0)\right>_{\rm c,0}\,. \eqnd\eCScalDdef $$
It follows
$$\eqalignno{\left<\tilde R(0)\tilde R(0)\right>_{\rm c} &={\partial \left<R\right> \over\partial M_0}={\partial \left<R\right> \over \partial \M}{\partial \M \over \partial M_0} \cr
 &= { 2\sigma \over g_\sigma}{1\over (1-\g/4\pi)} {\partial \left<R\right> \over \partial \M}   {\cal D}^{-1}(0)\cr
&={ 2\sigma \over g_\sigma}\left<\tilde R(0) \tilde R(0)\right>_{\rm c,0}{\cal D}^{-1}(0) =   -{2 \sigma \over g_\sigma}+ \left(2 \sigma \over g_\sigma\right)^2{\cal D}^{-1}(0) .\hskip9mm
  &\eqnd\eCScubicRRo \cr
  }$$
More explicitly,
$$\eqalignno{{\cal D}(0) &= {\Lambda \over 2\pi}   -2{m \over g_\sigma} +2{\M \over g_\sigma}\left(1-{\g \over 4\pi}\right) - {1\over 2\pi}\left[ \Lambda-2 \M {\left(1-{\g / 8\pi}\right)\over(1-\g/4\pi)} \right]\cr
&=-{2m \over g_\sigma}+2 \M\left[{1\over g_\sigma }\left(1-{\g \over 4\pi}\right)+{1\over 2\pi}{\left(1-{\g / 8\pi}\right)\over(1-\g/4\pi)} \right].&\eqnd\eCSsigmasigma   \cr}$$
Then, for $m=0$, the equation ${\cal D}=0$ directly implies the relation
  \eqns{\eCScubicggs}. This result is consistent with the existence of a massless pole of the $R$ two-point function when the relation  \eqns{\eCScubicggs} is satisfied but the example of the point $\g=4\pi$ shows that we still have to verify that the coefficient of $k^2$ in the inverse of the $\tilde R(k)$ two-point function does not vanish.
\smallskip
{\it The  $\langle R\varsigma\rangle$  two-point function.}
$$\left<\tilde\varsigma(0)\tilde R(0)\right>_{\rm c}={\partial \sigma \over \partial M_0}\, . $$
Then, from \eqns{\eCSsigmadM,\ \eCSdMrdM,\ \eCScalDdef},
$$ {\partial\sigma \over \partial M_0} =-1+\left(1- {\g \over 4\pi}      \right){\partial \M \over \partial M_0}
 =  g_\sigma {\cal D}^{-1}(0)  {\partial \left<R\right> \over \partial \M}
  . $$
Thus,
$$\left<\tilde\varsigma(0)\tilde R(0)\right>_{\rm c} = {g_\sigma \over 2 \sigma}\left<\tilde R(0)\tilde R(0)\right>_{\rm c}\, . $$
This relation will be seen to be a direct consequence of the relation \eqns{\eCSsigmaRrelation}.
\smallskip
%\Red{
{\it The $\varsigma$   vertex functions at zero momentum.}
The $\varsigma$ two-point vertex function at zero-momentum can be directly calculated by differentiating expression \eqns{\eCSdelVsigma} with respect to $\sigma$.
%}
\par
%\Red{
First from Eq.~\eqns{\eCSgapsigma}, one infers
$${\del M\over\del \sigma}={1\over 1-\g/4\pi}\,.$$
Then, from
$${\del^2 V_{\rm eff}(\sigma) \over (\del\sigma)^2}  ={2\sigma \over g_\sigma}-{\del\left< R\right>\over \del \sigma}  ={2\sigma \over g_\sigma}-{\Lambda \over2\pi}+{M\over\pi}{1-\g/8\pi \over 1-\g/4\pi}$$
and Eq.~\eqns{\eCSsigmaoverg}, one derives
$$\left<\tilde\varsigma(0)\tilde \varsigma(0)\right>_{\rm 1PI} ={\del^2 V_{\rm eff}(\sigma) \over (\del\sigma)^2}  ={2\sigma \over g_\sigma}-{\del\left< R\right>\over \del \sigma}
 ={\cal D}(0)\,.  $$
 %}
%\Red{
Similarly, the third derivative yields the three-point function:
%$$={\partial \sigma \over \partial {\cal R}}={g_ \sigma \over 2\sigma}\left(1+{\partial\left<R\right> \over\partial \M}  {\partial \M \over \partial {\cal R}}\right),$$
%where Eq.~\eqns{\eCSsigmasaddleb} has been used.
%Then,  from Eq.~\eqns{\eCSgapsigma} we infer
%$${\partial \sigma \over\partial {\cal R}}=  \left(1-{\g\over 4\pi}\right){\partial \M \over\partial {\cal R}}\,,$$
%which we use to eliminate ${\partial \M /\partial {\cal R}}$ and calculate $
%{\partial \sigma / \partial {\cal R}}$. We find
%$$\left<\tilde\varsigma(0)\tilde \varsigma(0)\right>_{\rm 1PI}={\partial \sigma \over \partial {\cal R}}={\cal D}^{-1}\,.$$
%The property that the three functions have the same denominator will be explained by a simple equation of motion.\par
%Finally, we can also calculate the $\sigma$ three-point function:
$$ \left<\tilde\varsigma(0)\tilde \varsigma(0)\tilde \varsigma(0)\right>_{\rm 1PI}={\del^3 V_{\rm eff}(\sigma) \over (\del\sigma)^3}= {2\over g_\sigma}+{1 \over\pi} {1-\g/8\pi \over(1-\g/4\pi)^2}\,. \eqnd\eCSsigmathree             $$
Finally, all other $n$-point functions with $n>3$ vanish at zero momentum.
%}
%$$\eqalignno{&\left<\tilde\varsigma(0)\tilde \varsigma(0)\tilde \varsigma(0)\right>  =-{\cal D}^{-2}{\partial {\cal D}\over \partial{\cal R}}=-2{\cal D}^{-2}\left[{1\over g_\sigma }\left(1-{\g \over 4\pi}\right)+{1\over 2\pi}{\left(1-{\g / 8\pi}\right)\over(1-\g/4\pi)} \right]{\partial \M\over \partial{\cal R}}\cr
%&\quad =-2{\cal D}^{-3}\left[{1\over g_\sigma } +{1\over 2\pi}{\left(1-{\g / 8\pi}\right)\over(1-\g/4\pi)^2} \right]=-2{\cal D}^{-3}{{\cal D}+2m/g_\sigma \over 2M(1-\g/4\pi)}\, . & \eqnd\eCSthreePvarsigma \cr }
 %$$

%%%%%%%%%%%%%%%%%%%%%%%%%%%%%%%%%%%%%%%%%%%%%%%%%%%%%%%%
\subsection{The generic $R$ two-point function}

Expanding around the saddle point at second order in  the fields,
$$\sigma(x)= \sigma+\varsigma(x) ,\quad \rho_\alpha(t,t',x)=\rho_\alpha(t-t')+\varrho_\alpha(t,t',x),$$
we find that
the modified action generates the two additional quadratic terms
$$N\int\d^2 x\,\d t\,\left[-\varsigma(x)\bigl( \varrho_1(t,t,x)+\varrho_2(t,t,x)\bigr) +{ \sigma \over g_\sigma}\varsigma^2(x)\right].$$
%\Red{
To calculate the $R$ correlation function, we can integrate over $\varsigma$, which amounts to replacing $\varsigma$ by solution of the field equation obtained by differentiating with respect $\varsigma(x)$.
One finds the contribution
$$-{N g_\sigma \over 4\sigma}\int\d^2 x\,\d t\,\left[\varrho_1(t,t,x)+\varrho_2(t,t,x)\right]^2 ,$$
which adds to the quadratic term in $\varrho$,
$$\ud \g N \int\d^2 x\,\d t \,\d t'\sgn(t'-t)\rho_1(t',t ,x)\rho_2(t ,t',x).$$
An expansion in powers of $g_\sigma$  then leads to
$$\left<\tilde R(k)\tilde R(-k)\right>_{\rm c}={\left<\tilde R(k)\tilde R(-k)\right>_{\rm c,0}
\over 1- (g_\sigma/2\sigma)\left<\tilde R(k)\tilde R(-k)\right>_{\rm c,0}}\,,\eqnd\eCScubicRRkexact $$
where $\langle\tilde R(k)\tilde R(-k)\rangle_{\rm c,0}$ is the connected two-point function \eqns{\eCSRRkexact}.
%}
\par
It is convenient to rewrite the expression as
$$\left<\tilde R(k)\tilde R(-k)\right>_{\rm c} =-{2 \sigma \over g_\sigma} +{4 \sigma^2 /g_\sigma^2
\over 2\sigma/ g_\sigma-  \left<\tilde R(k)\tilde R(-k)\right>_{\rm c,0}}\,.\eqnd\eCScubicRRkexactb $$
The poles of the two-point function are thus solution  of
$${\cal D}(k)\equiv {2\sigma\over g_\sigma}-2 \Omega_1(\M)+{k^2+4\M^2\over k\g}{ \tan\bigl(\g k{\cal B}_1(k)\bigr)\over 1 -2 \M \tan\bigl(\g k{\cal B}_1(k)\bigr)/k}  =0\,. \eqnd\eCScubicpole $$
%or
%$$\eqalign{&2\left( \Omega_1(\M)-{ \sigma\over g_\sigma}\right)\left[1 -2 \M \tan\bigl(\g k{\cal B}_1(k)\bigr)/k\right]\cr&\quad ={k^2+4\M^2\over k\g}\tan\bigl(\g k{\cal B}_1(k)\bigr).\cr}$$
%$$ 2\left( \Omega_1(\M)-{ \sigma\over g_\sigma}\right)\left[1 -2 \M \tan\bigl(\g k{\cal B}_1(k)\bigr)/k\right] ={k^2+4\M^2\over k\g}\tan\bigl(\g k{\cal B}_1(k)\bigr). $$
Using Eq.~\eqns{\eCSsigmam}, we note that
$$\eqalign{\Omega_1(\M)-{ \sigma\over g_\sigma}&={1\over 4\pi}(\Lambda-\M)-
 {\Lambda \over 4\pi} -{1\over g_\sigma}\left[ -m +\M\left(1-{\g \over 4\pi}\right)\right]\cr
 &={m\over g_\sigma}-{\M \over g_\sigma}\left(1-{\g-g_\sigma \over 4\pi}\right) .\cr} $$
The denominator of the expression \eqns{\eCScubicRRkexactb} is thus a finite quantity, as it should. Moreover, we see that the $R$ two-point function now requires an additive but also a   multiplicative renormalization $\sigma/g_\sigma$ as Eq.~\eqns{\eCScubicRRkexactb} explicitly shows.
\smallskip
%\Red{
{\it The $\varsigma$ two-point function.} To calculate the $\varsigma$ two-point function, it is convenient to first integrate over $\varrho$ at $R$ fixed. Since we have determined the $R$ two-point function, we know the result of the integral. We introduce the notation
$$R_{\rm s}(x)=R(x)-\langle R \rangle .$$
 The effective quadratic action then reads
$$\eqalign{ {1\over N}{\cal S}_{\rm eff.}&=\ud \int\d^3 x\,\d^3 y\, R_{\rm s}(x)R_{\rm s}(y)\left<R(x)R(y)\right>_{\rm c,0}^{-1}\cr&\quad+ \int\d^3 x \left[-\varsigma(x) R_{\rm s}(x)+{\sigma\over g_\sigma}\varsigma^2(x)\right],}$$
where in $\left<R(x)R(y)\right>_{\rm c,0}^{-1}$ the inverse has to be understood in the sense of kernels.
%}
\par
%\Red{
Note that if we integrate over $\varsigma$, we recover the inverse  connected $R$ two-point function in the form
$$\left<R(k)R(-k)\right>_{\rm c}^{-1}=\left<R(k)R(-k)\right>_{\rm c,0}^{-1}-{g_\sigma \over 2\sigma}\,,$$
which is equivalent to Eq.~\eqns{\eCScubicRRkexactb}. However, we can now instead integrate over $R_{\rm s}$ and find
$$ {1\over N}{\cal S}_{\rm eff.}=-\frac{1}{4}\int\d^3 x\,\d^3 y\,\varsigma(x)\varsigma(y)\left<R(x)R(y)\right>_{\rm c,0} + {\sigma\over g_\sigma} \int\d^3 x\,\varsigma^2(x).$$
Thus,
$$\left<\tilde \varsigma(k)\tilde \varsigma(-k)\right>=  {1
\over 2\sigma/ g_\sigma-  \left<\tilde R(k)\tilde R(-k)\right>_{\rm c,0}}\equiv {\cal D}^{-1}(k) \,.\eqnd\eCSverSigCorrel $$
with (Eq.~\eqns{\eCScubicpole})
$${\cal D}(k)=-{2m\over g_\sigma}+{2\M \over g_\sigma}\left(1-{\g-g_\sigma \over 4\pi}\right) +{k^2+4\M^2\over k\g}{ \tan\bigl(\g k{\cal B}_1(k)\bigr)\over 1 -2 \M \tan\bigl(\g k{\cal B}_1(k)\bigr)/k}  \,. \eqnd\eCScubicpoleb $$
We see that this correlation function is finite in the infinite cut-off limit.
%}
We show in section \label{\ssCSDzero} that the function $\cal D$ has a zero for $k$ pure imaginary and $|k|<2M$ if the two conditions are satisfied:
$$ -{M \over 4\pi-\g}\le -{ m\over g_\sigma}+{ \M \over g_\sigma}\left(1-{\g-g_\sigma \over 4\pi}\right)<0\,,$$
the first condition, equivalent to ${\cal D}(0)\ge 0$, being necessary for the theory to be physical. This zero corresponds to a massive scalar bound state.
%%%%%%%%%%%%%%%%%%%%%
\smallskip
{\it Small $k$ expansion.}
Concentrating on the situation where ${\cal D}(0)$ is   positive and small and $\M\ne0$, we can expand the denominator for $k$ small.
 Using
$$\eqalignno{&  {2 M\over k}\tan \bigl((gkB_1(k))\bigr)\cr &\quad = {g\over 4\pi } \left[ 1  +{1\over 3}
  \left( k\over 2M \right)^2   \left(\left( g\over 4\pi \right)^2-1\right)  \sum_0^\infty \left({k\over {2 M}}\right)^{2n} P_{n}\bigl(({g/ 4\pi}\bigr)^2)\right]
\,,\hskip7mm&  \eqnd\eCSTanBoneExpand  \cr}$$
where
$$\eqalignno{& P_0=1\,, \quad
P_1\bigl(( g/ 4\pi )^2\bigr)= \frac{1}{5}\bigl(({g/4\pi})^2-3\bigr)
 \cr  & P_2\bigl(( g/ 4\pi )^2\bigr)=\frac {1}{ 105}
\left[17( g/ 4\pi )^4-53( g/ 4\pi )^2 +45\right], \ldots
\hbox{{\it etc.}}\ . &  \eqnd\eCSPn  \cr}$$
%%%%%%%%%%%%
We find
$$\eqalignno{{\cal D}(k)&={\M(4\pi-\g)-4\pi m \over 2\pi g_\sigma}+{\M (8\pi-\g) \over 2\pi(4\pi-\g)}+{\M \over 6\pi}  {(8\pi-\g)  \over  (4\pi-\g)}\left({k\over 2\M}\right)^2
\cr &\quad +{\M \over 90\pi}  {(4\pi+\g)  \over  (4\pi-\g)} \left[ 2\left({g\over 4\pi}\right)^2
+5\left({g\over 4\pi}\right)-6\right]\left({k\over 2\M}\right)^4+O(k^6).\hskip3mm  \eqnd\eCSDexpand }$$
We note that the positivity of the coefficient of $k^2$ implies  $g<4\pi$.
Then, ${\cal D}$ has a zero corresponding to a scalar particle with mass given by
 $$M_S^2\sim 12 M^2+{12
M\left[\M(4\pi-\g)-4\pi m\right](4\pi-\g)\over (8\pi-\g)g_\sigma} \,,\eqnd\eCscalarMass $$
where ${M/ m}= f(g,g_\sigma,\eta)$ is the solution of the gap equation  \eqns{\eCSEgapmodifiedc}.
%%%%%%%%%%%%%%%%%%%%%%%%%%%%%%%%%%%%%%%
\smallskip
%\Red{
{\it Remark.} From the correlation functions, we note that in the large cut-off limit
$$\varsigma(x)\mathop{\sim}_{\Lambda\to\infty}{g_\sigma \over 2\sigma} R_s(x)=[R_s(x)]_{\rm ren.}, \eqnd\eCSsigmaRrelation $$
and thus $\varsigma$ is the renormalized $R_s$ field.
%}
\subsection{The critical limit}

In the limit $\M\to 0_+$, the expression
reduces to
$${\cal D}(k)=-{2 m\over g_\sigma}+k{\tan(\g/8) \over \g}\,,\eqnd{\eCSMzerolimit}$$
where either  $m=0$ or
$\eta= 4\pi/g_\sigma$.
In the latter case, $m/g_\sigma$ must be negative.
In both cases, for $g>0$ the positivity condition $\tan(\g/8)>0$ implies $\g<4\pi$.
\par
 In the case $m=0$, the connected $\sigma$ two-point function becomes identical to the propagator of a free massless scalar particle:
$$ \left<\tilde \varsigma(k)\tilde \varsigma(-k)\right>={\g \over \tan(\g/8)}\,{1\over k}\,.\eqnn $$
%%%%%%%%%%%%
\smallskip
{\it A special limit.}
A special limit is $\g=g_\sigma$, $m=0$ where all physical masses $M$ and $M_S$ vanish but also
all bare dimensional parameters,
$$M_0=0\,,\quad{\cal R}=0\,,$$
%%%%%%%%%%%%%%%%%%%%%%%%%%%%%%%%%%%%%%%%%%%%%%%%
\subsection{Situation with vanishing coefficient of $\M^2$ in the gap equation}

We are especially interested in the case where the relation \eqns{\eCScubicggs} is satisfied. Then,\sslbl{\ssCSdilaton}
$${\cal D}(k)={2m(8\pi-\g)\over (4\pi-\g)^2}+{(8\pi-\g)\over24\pi  \M (4\pi-\g)}k^2+O(k^4).\eqnn $$
The physical parameters correspond to $m>0$ and, since $g_\sigma$ then is negative, also implies $m/g_\sigma<0$.
\par
When   the relation \eqns{\eCScubicggs} is satisfied, quite generally  Eq.~\eqns{\eCSsigmam} leads to
 $$\M={ m\over 2}\left({4\pi\over 4\pi-\g}+ \eta{(4\pi-\g)\over (8\pi-\g)}\right). \eqnn  $$
Inserting the value of $\M$ in ${\cal D}$, we obtain
$$ {\cal D}(k)={2m(8\pi-\g)\over (4\pi-\g)^2}+{(8\pi-\g)^2\over 12\pi m \left[4\pi(8\pi-g)+\eta(4\pi-g)^2\right]}k^2+O(k^4).\eqnn  $$
This expression is physical only if
$$   \left[4\pi(8\pi-g)+\eta(4\pi-g)^2\right]\ge 0\,.$$
Then, the mass $M_S$ of the scalar
%\Red{
$\bar\psi\psi$
%}
bound state is given by
$$M_S^2= 24\pi m^2 \left({4\pi\over (4\pi-\g)^2}+{\eta\over 8\pi-\g}\right),$$
where this expression assumes that
$${4\pi\over (4\pi-\g)^2}+{\eta\over 8\pi-\g}\propto| \eta-1/g_\sigma|\ll 1\,.  $$
If in addition $m=0$, $\M$ is undetermined and the two-point function has the form
%\Red{
$$\left<\tilde \varsigma  (k)~\tilde \varsigma (-k)\right> \mathop{\sim} _{k\to 0} {24\pi  (4\pi-\g)\over    \left(8\pi-\g\right) }{\M \over k^2}\,.\eqnd\eCSdilatonPole   %}
$$
% factor \pi
 A pole is found for $k=0$, which corresponds to a massless particle provided the residue is positive, which again implies, starting from $g=0$, that $|\g|<4\pi$. Since the mass of the fermions $M$ is arbitrary, this pole may have the interpretation of a dilaton pole.
We then expect that all vertex functions vanish at zero momentum. Indeed, one verifies that this condition is satisfied
by using the relation \eqns{\eCScubicggs} in the three-point function \eqns{\eCSsigmathree}, since, as noted, other vertex functions vanish automatically.
\smallskip
%%%%%%%%%%%%%%%%%%%%%%%%%%%%%%%%%%%%%%%%

%\Red{
{\it The dilaton effective action}. At large distance or small momentum limit, for $M\ne 0$ the fermions decouple and one can derive an effective action for the massless dilaton field $\varsigma(x)$ (which is a $U(N)$ singlet) in the form of a derivative expansion.
%}
\par
%\Red{
It is convenient to  normalize the dilaton field,
$$ \varsigma(x)=f_D^{-1}  D(x)\,, \eqnd\eCSdilaton$$
where $f_D$ is defined from Eq.~\eqns{\eCSdilatonPole},
$$\left<\tilde \varsigma  (k)~\tilde \varsigma (-k)\right> \mathop{\sim} _{k\to 0}  {f_D^2 \over k^2}\,, \quad  f_D=\sqrt{24 \pi}\sqrt{ { 1- g/ 4\pi  \over 2- g/ 4\pi} }\sqrt{M}\,.\eqnd\eCSfD   $$
As noted above, all vertex functions vanish at zero momentum and thus  the effective action contains no derivative-free terms.
The   action of the renormalized dilaton field $D(x)$
has then the form
$$ {\cal S}(D ) =\half \int\d^3 x\,\del_\mu D(x) \del_\mu D(x) F(D(x))
+{\cal O}(\hbox{higher  order  terms  in}\ \del_\mu)\,.
 \eqnd\eCSLagrangianDulaton  $$
%The function $F(D )$ can be  calculated by acting repeatedly with $\del/\del {\cal R}$ on the  $k^2$  term of  $D(k)$ in Eq.~\eqns{\eCScubicpoleb}.
If in the action one keeps only the terms with two derivatives, by a simple change of field coordinates  the action can be reduced to a free massless field action, like in the example of the $O(2)$ non-linear $\sigma$ model.
%}
\par
In order to obtain the full interacting dilaton effective action, one has then to calculate the  $n$-point vertex (or 1PI) function of $\varsigma(x)$. More on the dilaton effective potential in these three dimensional Chern--Simons boson and fermion theories can be found in \refs{\rBM} and \refs{\BardeenFermions}.
%%%%%%%%%%%%%%%%%%%%%%%%%%%%%%%%%%%%%%%%%%%
\subsection{Spontaneously broken scale invariance in boson and fermion  theories}

The relation between 3D $U(N)$ fermion and boson theories
each coupled to Chern--Simons gauge fields, at large $N$, has raised
recently much\sslbl{\ssCSbosonization} interest
\refs{\rWadia,\ \AharonyII,\ \GurAri,\ \AharonyIII}. In particular, a precise mapping
between the boson and fermion theories was exhibited.  Previous
published results were obtained within the light-cone gauge  and
we compared them to our results derived within the temporal gauge
 in preceding sections. The comparison was done in the
case of the massless conformal phase. It is interesting to
compare our results also in the case of the  massive phase with spontaneously broken scale invariance. The comparison will check
the consistency of our results with the light-cone results and at
the same time will confirm the boson-fermion duality mapping
in the case of the massive phase.\par

The condition we found in Eq.~\eqns{\eCScubicggs} for the existence of a
massive phase in the fermion case was,
$$\left(\g-g_\sigma \over 2\pi\right)\left(1-{\g \over 8\pi}\right)=1\,. \eqnd\eCSmassiveFermionic  $$
In the boson theory, the existence of a massive ground state
requires \refs{\AharonyIII,\ \rBM}
$$\lambda_b^2+{\lambda_6\over 8\pi^2}=4
\,, \eqnd\eCSmassiveBosonic $$
where $\lambda_b$ is the Chern--Simons gauge coupling in the
boson theory and $\lambda_6$ is the marginal coupling of the
boson $\lambda_6  ({\Bfg \phi}^\dagger\cdot{\Bfg\phi})^3 / 6 N^2$ interaction.
\par
It is now interesting to find out  whether the boson  and fermion  conditions in Eqs.~\eqns{\eCSmassiveFermionic} and \eqns{\eCSmassiveBosonic}  are dual copies when the dual values of the couplings are taken into account.
Indeed, using in Eq.~\eqns{\eCSmassiveBosonic} the  mapping between the fermion  and the boson theories of
Ref. \refs{\GurAri,\ \AharonyIII}, which was derived in the
light-cone gauge,
$$\lambda_b=1-{g\over 4\pi}    \ , \quad
\lambda_6=8\pi^2\left(1-{g\over4\pi}\right)^2\left(3-4{g\over g_\sigma}\right),
\eqnd\eCSFBmapping
$$
one finds that the boson  condition in
Eq.~\eqns{\eCSmassiveBosonic} and the fermion  condition in
Eq.~\eqns{\eCSmassiveFermionic} derived in this article in a temporal gauge
are identical.
\vfill\eject
%%%%%%%%%%%%%%%%%%%%%%%%%%%%%%%%%%%%%%%%%%%%%%%%%
\section{Summary and conclusions}

In this article,\sslbl\sconclusions we have presented a detailed analysis of the large $N$
limit of three-dimensional $U(N)$ symmetric field theory with fermion in the fundamental representation coupled to a Chern--Simons gauge field. As mentioned in the introduction, this model rose interest in recent years \refs{\Klebanov {--}\AharonyIII,\ \rBM,\ \BardeenFermions } since its singlet sector  has been conjectured  through AdS/CFT  correspondence to be related to Vasiliev's \refs{\Fradkin} higher spin gravity theory on ${\rm AdS}_4$. Previous analysis of this system were done in the light-cone gauge whereas we worked in the axial $A_3=0$ gauge, which is somewhat more difficult to analyse but  lacks the peculiarity of the light-cone gauge.
We  determine the large $N$ properties by using a field integral formalism and the steepest descent method \refs{\rMMZJ}. The saddle point equations, which take here the form of integral equations for non-local fields, determine the ground state phase structure. \par
 The vertex functions and correlators were calculated at leading order in $1/N$. We derived exact solutions to the integral equations. From the vertex functions, we have inferred two-point correlation functions for scalar and current operators. The calculation were done first in the massive phase away from criticality in the phase in which scale invariance is explicitly broken. Later, the critical massless conformal phase was considered. The results were checked by calculating several orders in perturbation theory.
We have compared our results with  results obtained in previous works for gauge invariant physical quantities in the conformal massless limit.\par
In particular, we have thoroughly investigated the conditions for the occurrence  of a massive phase in which scale invariance is spontaneously broken  and its detailed properties. Indeed, the issue of a possible spontaneously broken massive phase was recently debated in several papers \refs{\AharonyIII,\rBM,\BardeenFermions}.
As described in  section \label{\ssExactRR}, we do not find such a massive spontaneously broken phase in our results:  the gap equation \eqns{\eCSgapEC} allows the physical fermion mass $M \ne 0 $  provided $g=\pm 4\pi$, however, Eq.~\eqns{\eCSRRkexact} clearly shows that the range of physical values for the coupling is $-4\pi < g < 4\pi $  leaving the solution  $M \ne 0 $ outside the physical range. Moreover, spontaneously broken scale symmetry would have required a massless dilaton to appear in the theory when
$g=\pm 4\pi$. This does not happen since the $\langle \tilde R(k)\tilde R(-k) \rangle_{\rm c}$ correlator in Eq.~\eqns{\eCSRRkexact} diverges at $g=\pm 4\pi$ for any value of the momentum $k$ and no massless dilaton pole appears there. Thus,
 in the critical theory,
the conformal invariant massless ground state is the only possible ground state in the Chern--Simons fermion matter system.
 A massive ground state appears only away from criticality $M_0-M_c=m\neq 0$ were we have an explicit breaking of scale invariance.

\par
The absence of a massive ground state agrees with the results of Ref.~\refs{\AharonyIII} calculated in the light-cone gauge. However, in the case of the bosons interacting with Chern--Simons gauge fields a massive  spontaneously broken scale invariant phase exists only when a marginal deformation is added of the form $(\lambda_6/6N^2 )( {\Bfg\phi}^\dagger \cdot  {\Bfg\phi})^3$. The massive phase and the dilaton massless pole appears \refs{\AharonyIII,\ \rBM} in this case when  $\lambda^2+\lambda_6/8\pi^2 =4$. The range of the Chern--Simons coupling $\lambda $ is in this case $-1<\lambda<1$ and the added self interaction marginal operator allows a massive ground state. In the absence of the Chern--Simons gauge field, this result agrees with Ref.~\refs{\rBMB}.
\par
As noted in Ref.~\refs{\AharonyIII}, in order for a massive, spontaneously broken scale invariance phase to appear in the fermion case, here too a marginal deformation of the type $\sigma{\bar\psi}\psi +(N/3!)\lambda^f_6\sigma^3$ has to be added.
\par
In section \label{\sSigmaTerm}, we added to the action  the above mentioned self-interacting auxiliary field $\sigma(x)$ coupled to the fermions, which effectively induces a marginal deformation to the Chern--Simons fermion action. Indeed, in our $A_3=0$ gauge calculations  a spontaneously broken massive phase appears and a massless dilaton pole appears in the $\langle\tilde R(k)\tilde R(-k)\rangle_{\rm c}$ correlator. We found that the massive phase appears in the spectrum provided
the $\sigma$ self interacting coupling $g_\sigma$ is related to the Chern-Simons gauge coupling $g$ by $ g_\sigma/4\pi = - (1-g/4\pi)^2 / (2-g/4\pi)$
 in analogy to the relation between $\lambda_6 $ and $\lambda$  that ensures a massive ground state in the Chern--Simons boson case \refs{\AharonyIII,\ \rBM}. 
 Moreover, in section \label{\ssCSbosonization} we have shown that the  conditions for spontaneous breaking of scale invariance in the boson  and fermion theories are dual copies of each other.

\vskip 1cm
\vfill\eject

\centerline{\it Acknowledgements}
\medskip
\noindent
We would like to
thank  O.~Aharony, W.A.~Bardeen  and Ph.~Di Francesco for
useful remarks, discussions and correspondence on some of the issues
covered in this paper. M.~Moshe thanks Saclay theoretical institute and CERN theory group  for their
warm hospitality while different parts of this work were done.
%%%%%%%%%%%%%%%%%%%%%%%%%%%%%%%%%%%%%%%%%%%%%%%
\vfill\eject
%%%%%%%%%%%%%%%%%%%%%%%%%%%%%
\appendix{Some calculations and additions}

\section{Some explicit expressions}

We  give some technical details about explicit expressions that appear in this work.
\sslbl\appCSzero

\subsection{The  function $\Theta(\omega)$ and regularization}

To exhibit more clearly the problems arising from the regularization, we calculate   the function \eqns{\eCSTheta} first and Fourier transform only after.\sslbl{\ssThetacalcul}
Since in momentum space the function is given by a convolution, it becomes a product  ($M>0$)
of the form
$${\e^{-M|x|}\over|x|}\sgn(x_3).
$$
The integration over $p_1,p_2$ yields $x_1=x_2=0$. Thus,
$$\Theta(\omega)={1\over8i\pi}\int\d x_3\,{\e^{i\omega x_3}\e^{-M|x_3|}\over|x_3|}\sgn(x_3).$$
We see that the singularity at $x_3=0$ requires a regularization. If we choose a {\it reflection symmetric} regularization, $|x_3|\ge \varepsilon>0$, we can divide the integral into $x_3>0$ and $x_3<0$, change $x_3$ into $-x_3$ for $x_3<0$ and add again the two contributions. The result is
$$\Theta(\omega)={1\over8i\pi}\int_{x_3 =\varepsilon}^{+\infty}\d x_3\,{\left(\e^{i\omega x_3}- \e^{-i\omega x_3}\right)\e^{-M x_3 }\over x_3 } .$$
The limit $\varepsilon=0$ can then be taken and the integral calculated. One recovers
expression \eqns{\eCSThetaexplicit}. By contrast, a non-symmetric short distance cut-off yields a logarithmically divergent limit.
\par
However, we note that in a Feynman parametrization such a symmetrization is automatically achieved when the order of  integrations is interchanged and   the logarithmic divergence then does not show up:
 $$\eqalignno{\Theta(\omega) &= \int {d^3p\over (2\pi)^3} {1\over (\omega-p_3)(p^2+M ^2)}\quad
{\rm  where }\ {1\over (\omega-p_3)}\quad {\rm  means\  PP}{1\over (\omega-p_3)}
\cr&=\mathop{\lim}_{\varepsilon\to0}\int {d^3p\over (2\pi)^3}  \int_0^1 \d x  {\omega-p_3\over [x(\omega-p_3)^2+x\varepsilon^2+(1-x)(p^2+M ^2)]^2}\cr
&= {1\over 4\pi}\arctan \left( \omega\over M  \right).
&\eqnd\Bubbl \cr}$$
%%%%%%%%%%%%%%%%%%%%%%%%%%%%%%%%%%%%%
\subsection{The function ${\cal D}$ and the bound state problem}

We consider the function \eqns{\eCScubicpoleb}\sslbl\ssCSDzero
$${\cal D}=-{2m\over g_\sigma}+{2\M \over g_\sigma}\left(1-{\g-g_\sigma \over 4\pi}\right) +{k^2+4\M^2\over k\g}{ \tan\bigl(\g k{\cal B}_1(k)\bigr)\over 1 -2 \M \tan\bigl(\g k{\cal B}_1(k)\bigr)/k}  $$
and assume that the parameters are such that ${\cal D}(0)$ is positive.
We write it as
$${\cal D}(k)={\cal D}(0)+ \left[{\cal D}(k)-{\cal D}(0)\right] $$
with
$$\left[{\cal D}(k)-{\cal D}(0)\right]={k^2+4\M^2\over k\g}{ \tan\bigl(\g k{\cal B}_1(k)\bigr)\over 1 -2 \M \tan\bigl(\g k{\cal B}_1(k)\bigr)/k}  -{2M\over 4\pi -\g}\,\eqnd{\eCSD}.$$
We are interested in the real time region below the $\psi\bar\psi$ threshold. We thus set
$$k=2i M z\,,\quad 0\le z\le 1\,,\quad \g=4\pi \lambda\,,\quad |\lambda|<1\,.  $$
and
 $$\eqalign{\phi(z)&={2\pi\over M}\left[{\cal D}(k)-{\cal D}(0)\right]\cr
&=-{1\over  1- \lambda } +{1\over \lambda}
{{\left(1-z^2\right) \tanh \left(\frac{\lambda}{2}  \log
   \left(\frac{z+1}{1-z}\right)\right)}\over {z-\tanh \left(\frac{\lambda}{2}  \log
   \left(\frac{z+1}{1-z}\right)\right)}}\cr
&=-{1\over  1- \lambda } - {1\over  \lambda  }
\,
{(1+z)^\lambda-(1-z)^\lambda \over (1+z)^{\lambda-1}- (1-z)^{\lambda-1}  }  \,.
\cr} \eqnd{\eCSphiz}$$
The function $\phi$ is monotonous decreasing between $0$ and $1$, which varies between $0$ and $-1/(1-\lambda)$. Thus, if the condition
$${\cal D}(0)<{2M\over 4\pi-\g}\ \Leftrightarrow\  -{ m\over g_\sigma}+{ \M \over g_\sigma}\left(1-{\g-g_\sigma \over 4\pi}\right) <0 \ \eqnd{\eCSDinequal}
$$
is satisfied, one finds a massive scalar bound state.
%%%%%%%%%%%%%%%%%%%%%%%%%%%%%%%%%%%%%%%%%%%%%%%%%%%%%%
\section{Relevant algebraic identities}

We now describe the algebraic identities that have been systematically used in this work to solve the various integral equations that were encountered.
\sslbl\appCSone
\subsection{Some elementary though useful identities}

In this work, we use systematically identities   based on the partial fraction decomposition of rational functions of the form\sslbl\appCSgenId
$$\prod_{i=1}^n {P(z)\over  z- \omega_i}\,,$$
where $P$ is a polynomial. If $P$ is of degree smaller than $n$, then
$$\prod_{i=1}^n {P(z)\over  z- \omega_i}=\sum_{i=1}^n{P( \omega_i)\over  z- \omega_i}\prod_{j\ne i =1}^n{1\over  \omega_i- \omega_j} \,.\eqnd{\eCSbasicIdi} $$
In this form, the identity is equivalent to Lagrange interpolation formula.\par
For higher degrees, additional terms are generated and for the relevant cases one finds
$$\eqalignno{ \prod_{i=1}^n {z^n\over  z- \omega_i}&=1+ \sum_{i=1}^n { \omega_i^n\over  z- \omega_i}\prod_{j\ne i =1}^n{1\over  \omega_i- \omega_j}\,,&\eqnd{\eCSbasicIdii} \cr\prod_{i=1}^n {z^{n+1}\over  z- \omega_i}&=
z+\sum_{i=1}^n \omega_i+\sum_{i=1}^n { \omega_i^{n+1}\over  z- \omega_i}\prod_{j\ne i =1}^n{1\over  \omega_i- \omega_j}\,,&\eqnd{\eCSbasicIdiii} \cr \prod_{i=1}^n {z^{n+2}\over  z- \omega_i}&=
z^2+z\sum_{i=1}^n  \omega_i+\sum_{i\ge j=1}^n \omega_i\omega_j+\sum_{i=1}^n { \omega_i^{n+2}\over  z- \omega_i}\prod_{j\ne i =1}^n{1\over  \omega_i- \omega_j}\,.\hskip7mm&\eqnd{\eCSbasicIdiv} \cr
 }
$$
A reformulation of the identities \eqns{\eCSbasicIdi{--}\eCSbasicIdiv} where $z\equiv \omega_{n+1}$, is also directly useful. We define
$$S_{n,m}={\goth S}_{\{\omega_1,\omega_2,\ldots,\omega_{n+1}\}}\left[ \omega_{n+1}^m \prod_{i=1}^n{1\over  \omega_{n+1}-\omega_i} \right], $$
where $\goth S$ symmetrizes over $\{\omega_1,\omega_2,\ldots,\omega_{n+1}\}$.
One then finds
$$S_{n,m}=0\ {\rm for}\ m<n\,,\ S_{n,n}={1\over n+1}\,,\ S_{n,n+1}={1\over n+1}\sum_{i=1}^{n+1}\omega_i\,. \eqnd\eCSomegaSnm$$
Finally, in section \label{\ssCSJJtwopoint} we need
$$S_{n,n+2}={1\over n+1}\left(\sum_{i=1}^{n+1}\omega_i^2+\sum_{i<j=1}^{n+1}\omega_i\omega_j \right).\eqnd\eCSomegaSnnii$$
%%%%%%%%%%%%%%%%%%%%%%%%%%%%%%%%%%%%%%%%%%%%%%%%%%%%%%%%%%%%%%%
\subsection{Some relevant integrals}

We evaluate here a few integrals needed in this work. They involve
explicitly the quantity \eqns{\eCStaudef},\sslbl\appCSintXin
$$\tau=\ud\sqrt{(k^2+4M^2)t_2+p_3^2k^2}=\ud p_3 k+ {t_2\over4 p_3 k}\left(k^2+4M^2\right)+O(1/p_3^3)  . $$
\medskip
{\it From vertex functions to two-point functions.}
A  first set of integrals occurs in the derivation of  two-point functions from the corresponding vertex functions.\par
By expanding in powers of the coupling $\g$, the simplest integrals can be inferred  from integrals of the form
$$  {1\over(2\pi)^3} \int{\d^3 p \, \left( \tau\Xi(p_3,k)\right)^{2n}  \over \left[\left(p+k/2\right)^2+M^2\right] \left[\left(p-k/2\right)^2+M^2\right]}.   $$
We then replace the function $\Xi$ by its integral representation,
$$\Xi(p_3,k)={1\over(2\pi)^3}\int{\d^3 q\,   \over\left(p_3-q_3\right) \left[(q+k/2)^2+M^2\right]\left[(q-k/2)^2+M^2\right]} .$$
We   symmetrize the integrand over the $(2n+1)$ integration variables $p,q^i$.
The numerator has the same degree as the denominator. We use the identities \eqns{\eCSomegaSnm}. Only the term of highest degree in the numerator contributes. Using the value of $S_{n,n}$, we obtain
$${1\over 2n+1}\left(k{\cal B}_1/2\right)^{2n+1}   $$
and, thus, the general equation
$$    {1\over(2\pi)^3} \int{\d^3 p \, F\left( \tau\Xi(p_3,k)\right)  \over \left[\left(p+k/2\right)^2+M^2\right] \left[\left(p-k/2\right)^2+M^2\right]}=\int_0^{{\cal B}_1}F(kz/2)\d z\,, \eqnd{\eCSintRRa} $$
for any function $F(z)$  expandable in powers of $z$.\par
A second type of integrals has the form
$$    {1\over(2\pi)^3} \int{\d^3 p \, p_3^2 \left(\tau\Xi(p_3,k)\right)^{2n}  \over \left[\left(p+k/2\right)^2+M^2\right] \left[\left(p-k/2\right)^2+M^2\right]}.   $$
We need the two first terms of the expansion for $p_3$ large:
$$p_3^2 \tau^{2n}=(k/2)^{2n}p_3^{2n+2}+\frac{1}{4}n t_2\left(k^2+4M^2\right)(k/2)^{2n-2}p_3^{2n} +O(p_3^{2n-2}).$$
Symmetrizing over all integration momenta and using equations \eqns{\eCSomegaSnnii,\eCSomegaSnm} and noting that the remaining factors in the integrand are even functions, we find
$$\eqalign{&\left(k{\cal B}_1/2\right)^{2n} {1\over(2\pi)^3} \int{\d^3 p \,p_3^2    \over \left[\left(p+k/2\right)^2+M^2\right] \left[\left(p-k/2\right)^2+M^2\right]}\cr&\quad+{n\over 4(2n+1)} t_2\left(k^2+4M^2\right)(k/2)^{2n-2}{\cal B}_1^{2n+1}\,,\cr} $$
 and, thus, the general equation
$$\eqalignno{&{1\over(2\pi)^3} \int{\d^3 p \,p_3^2 F\left( \tau\Xi(p_3,k)\right)  \over \left[\left(p+k/2\right)^2+M^2\right] \left[\left(p-k/2\right)^2+M^2\right]} =  t_2{k^2+4M^2\over 4k } \int_0^{{\cal B}_1}zF'(kz/2)\d z  \cr &\qquad  +F\left(k{\cal B}_1/2\right){1\over(2\pi)^3} \int{\d^3 p \,p_3^2    \over \left[\left(p+k/2\right)^2+M^2\right] \left[\left(p-k/2\right)^2+M^2\right]}
  .\hskip5mm & \eqnd{\eCSintRRb}\cr} $$
The remaining integral is calculated in section \label{\ssRRcalculation}.\par
A third relevant family of integrals is
$$    {1\over(2\pi)^3} \int{\d^3 p \, p_3\tau^{2n} \Xi^{2n-1}(p_3,k) \over \left[\left(p+k/2\right)^2+M^2\right] \left[\left(p-k/2\right)^2+M^2\right]}.   $$
Then, for $p_3\to\infty$,
$$p_3\tau^{2n}= (k/2)^{2n}p_3^{2n+1}+\frac{1}{4}n t_2\left(k^2+4M^2\right)(k/2)^{2n-2}p_3^{2n-1} +O(p_3^{2n-3}).$$
Following the same lines, one obtains
$$\eqalignno{&{1\over(2\pi)^3} \int{\d^3 p \,p_3 \tau F\left( \tau\Xi(p_3,k)\right)  \over \left[\left(p+k/2\right)^2+M^2\right] \left[\left(p-k/2\right)^2+M^2\right]} = \frac{1}{8} t_2 \left(k^2+4M^2\right)  {\cal B}^2_1 F (kz/2)  \cr &\qquad  +F\left(k{\cal B}_1/2\right){1\over(2\pi)^3} \int{\d^3 p \,p_3^2    \over \left[\left(p+k/2\right)^2+M^2\right] \left[\left(p-k/2\right)^2+M^2\right]}
  .\hskip5mm & \eqnd{\eCSintRRc}\cr} $$

\medskip
{\it Integral equations and corresponding integrals.}
A second set contains   integrals relevant for the integral equations.
They can be split into two families.
\smallskip
{\it Even functions.}
A first family has the form
$${\cal I}_n  ={1\over(2\pi)^3}\int{\d^3 p\, \tau^{2n}\Xi^{2n}(p_3,k) \over\left(\ell_3-p_3\right) \left[(p+k/2)^2+M^2\right]\left[(p-k/2)^2+M^2\right]} \,.$$
We then replace the function $\Xi$ by its integral representation,
$$\Xi(p_3,k)={1\over(2\pi)^3}\int{\d^3 q\,   \over\left(p_3-q_3\right) \left[(q+k/2)^2+M^2\right]\left[(q-k/2)^2+M^2\right]} .$$
We   symmetrize the integrand over the $(2n+1)$ integration variables $p,q^i$ and use the identities \eqns{\eCSbasicIdi}. We infer
$${\cal I}_n ={1\over 2n+1}\tau^{2n}(\ell_3)\Xi^{2n+1}(\ell_3,k).$$
In particular, if $F(z)$ is an even function, expandable at $z=0$,
and
$${\cal I}(F)  ={1\over(2\pi)^3}\int{\d^3 p\, F\left(\tau\Xi (p_3,k)\right) \over\left(\ell_3-p_3\right) \left[(p+k/2)^2+M^2\right]\left[(p-k/2)^2+M^2\right]} \,,$$
then
$${\cal I}(F)= \int_0^{ \Xi}\,F(\tau z )\d z \,.\eqnd\eCSXinna $$
A related integral is
$${\cal J}_n  ={1\over(2\pi)^3}\int{\d^3 p\,p_3 \tau^{2n}\Xi^{2n}(p_3,k) \over\left(\ell_3-p_3\right) \left[(p+k/2)^2+M^2\right]\left[(p-k/2)^2+M^2\right]} \,.$$
We write it a sum, using $p_3=\ell_3-(\ell_3-p_3)$. The first term yields $\ell_3{\cal I}_n$.
We symmetrize the second term and  use the identities \eqns{\eCSomegaSnnii}. Only the term of highest degree in $p_3$ contributes and one finds
$$ -{2\over (2n+1)k} \left(k{\cal B}_1/2\right)^{2n+1}.$$
For an even function $F$, one then obtains
$${\cal J}(F)= \ell_3 \int_0^{ \Xi}\,F(\tau z )\d z - \int_0^{  {\cal B}_1 }\,F(kz /2)\d z \, .\eqnd\eCSXinnb $$
A last integral of the family is
$${\cal K}_n  ={1\over(2\pi)^3}\int{\d^3 p\, p_3^2\tau^{2n}\Xi^{2n}(p_3,k) \over\left(\ell_3-p_3\right) \left[(p+k/2)^2+M^2\right]\left[(p-k/2)^2+M^2\right]} \,.$$
We substitute this time $p_3^2=\ell_3[\ell_3-(\ell_3-p_3)]-p_3(\ell_3-p_3)$. The first term yields $\ell_3 {\cal J}_n$. For the second term, after cancellation of the gauge propagator, we use the identities \eqns{\eCSomegaSnnii}. The degree of the denominator is $2n$ and of the numerator $(2n+1)$. For $p_3\to\infty$,
$$p_3 2\tau^{2n}=(k/2)^{2n}p_3^{2n+1}+O(p_3^{2n-1}).$$
Thus, only the term of highest degree contributes. But since it is odd and the remaining integrand is even, it does not contribute. We conclude that if $F$ is an even function,
$${\cal K}(F)= \ell_3^2 \int_0^{ \Xi}\,F(\tau z)\d z- \ell_3 \int_0^{  {\cal B}_1 }\,F(kz/2)\d z\, .\eqnd\eCSXinnc $$
\medskip
{\it Odd functions.}
We now examine integrals of odd functions.
%%%%%%%%%%%%%%%%%%%%%%%%%%%%%%%%%%%%%
We first need
$${\cal L}_n= {1\over(2\pi)^3}\int{\d^3 p\, \tau^{2n}\Xi^{2n-1}(p_3,k) \over\left(\ell_3-p_3\right) \left[(p+k/2)^2+M^2\right]\left[(p-k/2)^2+M^2\right]} \,.$$
Once the function $\Xi$ is replaced by its integral representation, as a function of $p_3$, numerator and denominator have the same degree $2n$. We symmetrize and now use the identities \eqns{\eCSbasicIdi, \eCSbasicIdii}.
The result is the sum of two contributions. The first one is
$${1\over 2n}\tau^{2n}(\ell_3)\Xi^{2n }(\ell_3,k).$$
The second one comes only from the term of highest degree in $p_3$ of $\tau^{2n}$:
$$-{1\over 2n} \left(k{\cal B}_1/2\right)^{2n}.$$
Then, if $F$ is an odd function and
$${\cal L}(F)={1\over(2\pi)^3}\int{\d^3 p\, \tau F\left(\tau\Xi (p_3,k)\right) \over\left(\ell_3-p_3\right) \left[(p+k/2)^2+M^2\right]\left[(p-k/2)^2+M^2\right]} \,,$$
the result is
$${\cal L}(F)=\int_{k{\cal B}_1/2}^{\tau\Xi}\,F(z )\d z=\tau \int_0^{ \Xi}\,F(\tau z)\d z-\ud k \int_0^{  {\cal B}_1 }\,F(kz/2)\d z\,   .\eqnd\eCSXinnd $$
A related integral is
$${\cal M}_n= {1\over(2\pi)^3}\int{\d^3 p\,p_3 \tau^{2n}\Xi^{2n-1}(p_3,k) \over\left(\ell_3-p_3\right) \left[(p+k/2)^2+M^2\right]\left[(p-k/2)^2+M^2\right]} \,.$$
We substitute $p_3=\ell_3-(\ell_3-p_3)$. The first term yields $\ell_3{\cal L}_n$.
The second term leads to a ratio of a polynomial of degree $2n$ over a polynomial of degree $(2n-1)$. Using the identities \eqns{\eCSomegaSnnii}, we find a sum of odd terms while the integrand is even.
$${\cal M}(F)=\ell_3{\cal L}(F). \eqnd\eCSXinne $$
 %%%%%%%%%%%%%%%%%%%%%%%%%%%%%%%%%%%%%%%%%%
\section{The $\langle(\bar\psi\psi)\psi\bar\psi\rangle$ vertex: two-loop calculations}

We  calculate the $\langle(\bar\psi\psi)\psi\bar\psi\rangle$ vertex (or 1PI) function up to two loops for $N$ large.\sslbl\appCSvertex
\smallskip
{\it Notation.}
To simplify all expressions, we introduce the compact notation\sslbl\appvertex
$$\deqalign{Dp_1&=(p+k/2)^2+M^2 ,& Dm_1 &=(p-k/2)^2+M^2, \cr
Dp_2&=(q+k/2)^2+M^2 ,& Dm_2 &=(q-k/2)^2+M^2 .\cr
}$$
%%%%%%%%%%%%%%%%%%%%%%%%%%%%%%%%%%%%%%%%%%%%%%%%%%%%%%%%%%%%%%%%%%%%%%%%%%%%%%%%
\subsection{Perturbative calculations}

{\it One-loop results.}
We define
$$W^{(1,2)}(x;y,z)=\left<\bar\psi(x)\cdot\psi(x)\psi(y)\bar\psi(z)\right>.$$
We now calculate the corresponding vertex function in the Fourier representation, setting
$$\tilde\Gamma^{(1,2)}( k;\ell-k/2,\ell+k/2)=-E(\ell_3,k)-i\sigma_3 F(\ell_3,k) $$
with
$$E=\sum_{n=0}E_n \g^n,\quad F=\sum_{n=0}F_n \g^n $$
and the boundary conditions $E_0=1$, $F_0=0$.\par
We use the dressed fermion propagator \eqns{\eCSNfermioniipt} in such a way that only ladder diagrams have to be summed.\par
The order $\g$ reduces to
$$\eqalign{E_1&=-{1\over(2\pi)^3}\int{\d^3 p \,2Mp_3\over \left(\ell_3-p_3\right)
Dm_1 Dp_1}\cr
F_1&= {1\over(2\pi)^3}\int{\d^3 p \,\left(4t_1-Dp_1-Dm_1\right)\over 2\left(\ell_3-p_3\right)
Dm_1 Dp_1}.\cr}$$
Integrating, one obtains
$$\eqalign{E_1&=-2M\ell_3 \Xi(k,\ell_3)+2M {\cal B}_1(k)\cr
F_1&=2 t_1\Xi(k,\ell_3)-\ud \left[\Theta(\ell_3+k_3/2)+\Theta(\ell_3-k_3/2)\right].\cr}$$
We have also defined (Eq.~\eqns{\eCSVdef})
$$\tilde\Gamma^{(1,2)}( k;\ell-\ud k ,\ell+\ud k )=- U^{-1}(\ell_3-\ud k_3 )
\,\tilde V^{(1,2)}( k;\ell-\ud k,\ell+\ud k) U^{-1}( \ell_3+\ud k_3)   $$
with
$$\tilde V^{(1,2)}( k;\ell-\ud k,\ell+\ud k)=A(\ell_3,k)+i\sigma_3 B(\ell_3,k).$$
We expand
$$A=\sum_{n=0}A_n \g^n,\quad B=\sum_{n=0}B_n \g^n .$$
Then $A_0=1$, $B_0=0$ and
$$\eqalign{A_1&=-2M\ell_3 \Xi(k,\ell_3)+2M {\cal B}_1(k)\cr
B_1&=2 t_1\Xi(k,\ell_3),  \cr}$$
where  all quantities are expressed in terms of the fermion physical mass $M$.\par
%%%%%%%%%%%%%%%%%%%%%%%%%%%%%%%%%%%%%%%%%%%%%%%%%%%%%%%%%%%%%%%%%%%%%%
The expansion of $ \tilde V^{(1,2)}$ to order $\g$ inserted into expression \eqns{\eCSVdef} then generates contributions of order $\g^2$ in the expansion of $\tilde \Gamma^{(1,2)}$. They are given by
\eqna\eCSRVertexcorrection
$$\eqalignno{ \delta E_2 &=  t_1\left[\Theta (\ell_3-\ud k_3) +\Theta (\ell_3+\ud k_3)\right]   \Xi(k,\ell_3)&\eCSRVertexcorrection{a} \cr &
 -\frac{1}{8}\left[\Theta (\ell_3-\ud k_3)+ \Theta (\ell_3+\ud k_3)\right]^2 ,\cr
 \delta F_2&=- M\left[ {\cal B}_1(k)-\ell_3 \Xi(k,\ell_3)\right]\left[ \Theta(\ell_3-\ud k_3)+\Theta (\ell_3+\ud k_3)\right]
  .&\eCSRVertexcorrection{b}\cr}$$
The vector $(A_2,B_2)$ is then obtained by subtracting them from $(E_2,F_2)$.
%%%%%%%%%%%%%%%%%%%%%%%%%%%%%%%%%%%%%%%%%%%%%%%%%%%%%%%%%%%%%%%%%%%%%%%%%
\medskip
{\it Two-loop calculation}.
The two-loop contribution corresponds to a sum of diagrams,  a vertex correction and two renormalizations of the fermion propagator, which are generated by expanding the one-loop term calculated with the dressed propagator.\par
 Denoting by $q$ the additional integration vector, we obtain the vertex correction by first acting with the product ${\cal T}(\ell_3,q_3){\cal T}(q_3,p_3)$, where ${\cal T}$  is the matrix \eqns{\eCSTmatzero}, on the vector $(1,0)$. \par
The propagator renormalizations are inferred from the order $\g$ in the expansion of $\bf T$ to order $\g$. In the latter contribution, we replace the function $\Theta$ by its integral representation. With the choice of momenta of figure \label{\figcsver}, we substitute
$$\Theta(p_3+k_3/2)+\Theta(p_3-k_3/2)={1\over(2\pi)^3}\int{\d^3 q\left(Dm_2+Dp_2\right)\over \left(p_3-q_3\right)Dm_2 Dp_2} \,.$$
The common denominator of all contributions then is
$$ D =Dm_1Dp_1Dm_2Dp_2    \left(\ell_3-q_3\right)\left(q_3-p_3\right).$$
After symmetrization over $(p,q)$, the factor $(q_3-p_3)$ cancels and the denominators of the integrands reduce to
$$D'=Dm_1Dp_1Dm_2Dp_2    \left(\ell_3-p_3\right) \left(\ell_3-q_3\right)$$
and the numerators become
$$\eqalign{{\cal N}(E_2)&=-\frac{1}{8}\left(Dp_1+Dm_1\right)\left(Dp_2+Dm_2\right)+\ud t_1\left(Dp_1+Dm_1+Dp_2+Dm_2\right)\cr&\quad +2\left[M^2 p_3q_3 -t_1t_2 -t_1[\ell_3(p_3+q_3)-p_3q_3]\right]\cr
{\cal N}(F_2)&=\ud M p_3\left(Dp_2+Dm_2\right)+\ud M q_3\left(Dp_1+Dm_1\right) +2M t_1\left(2\ell_3-p_3-q_3\right). \cr}$$
The integration then is simple and yields
$$\eqalign{E_2&=-\frac{1}{8} \left[\Theta(\ell_3+k_3/2)+\Theta(\ell_3-k_3/2)\right]^2\cr&\quad +t_1 \left[\Theta(\ell_3+k_3/2)+\Theta(\ell_3-k_3/2)\right]\Xi (\ell_3,k)\cr&\quad+2\left[(M^2-t_1)\ell_3^2-t_1t_2\right]\Xi^2 (\ell_3,k)+2\left(t_1+M^2\right){\cal B}_1^2(k),\cr
&\quad -4 M^2\ell_3 {\cal B}_1(k) \Xi (\ell_3,k) \cr
F_2&=   M \left[\Theta(\ell_3+k_3/2)+\Theta(\ell_3-k_3/2)\right]\left[\ell_3 \Xi (\ell_3,k)-{\cal B}_1(k)\right]\cr&\quad +4M t_1\Xi (\ell_3,k) {\cal B}_1(k) .\cr}$$
Finally, the correction terms \eCSRVertexcorrection{} cancel all terms proportional to $\Theta$ functions, justifying to two-loop order the transformation \eqns{\eCSVdef}, and one finds
$$\eqalign{A_2&= 2\left[(M^2-t_1)\ell_3^2-t_1t_2\right]\Xi^2 (\ell_3,k)+2\left(t_1+M^2\right){\cal B}_1^2(k),\cr
&\quad -4 M^2\ell_3 {\cal B}_1(k) \Xi (\ell_3,k) \cr
B_2&=   4M t_1\Xi (\ell_3,k) {\cal B}_1(k) .\cr}$$

%%%%%%%%%%%%%%%%%%%%%%%%%%%%%%%%%%%%%%%%%%%%%%%%
$$\eqalignno{A_2&=    -2\left\{  \left[ t_1t_2+ \ell_3^2( t_1-M^2)\right]\Xi^2(\ell_3,k)  + 2\ell_3 M^2{\cal B}_1(k) \Xi (\ell_3,k)\right.\cr &\quad \left.- (M^2+t_1) {\cal B}^2_1(k)\right\},\cr B_2&= 4M t_1  {\cal B}_1(k)\Xi(\ell_3,k) .&\eqnn  \cr
A_3&= \frac{4}{3}M \left\{\ell_3\Xi^3 \left[t_1t_2+\ell_3^2(t_1-M^2)\right]      -3{\cal B}_1\Xi^2\left[t_1t_2+\ell_3^2(t_1-M^2)\right]  \right.\cr
&\quad\left.-3 {\cal B}_1^2\Xi\ell_3(M^2+t_1)+5t_1{\cal B}_1^3   \right\},\cr
B_3&=\frac{4}{3}t_1\Xi \left\{- \left[t_1t_2+\ell_3^2(t_1-M^2)\right]\Xi^2      +3  {\cal B}_1^2(t_1+M^2)+4\Xi{\cal B}_1^3      \right\}, \cr
A_4&=\frac{2}{3}\left\{ \left[t_1t_2+\ell_3^2(t_1-M^2)\right]^2\Xi^4 +4M^2\ell_3
  \left[t_1t_2+\ell_3^2(t_1-M^2)\right]\Xi^3{\cal B}_1         \right.\cr
&\quad -6\Xi^2{\cal B}_1^2(t+M^2)  \left[t_1t_2+\ell_3^2(t_1-M^2)\right]    -4\ell_3 M^2 \Xi{\cal B}_1^3(M^2+5t_1)\cr&\quad\left. +{\cal B}_1^4(5 t_1^2+18 t_1M^2+M^4)                \right\} ,\cr
B_4&=\frac{8}{3}t_1 M{\cal B}_1\Xi\left\{ \left[t_1t_2+\ell_3^2(t_1-M^2)\right] \Xi^2  +(M^2+5 t_1){\cal B}_1^2     \right\}
 \cr} $$

\section{The $J_3$ vertex function at two loops}

We calculate now the $J_3=i\bar\psi\sigma_3\psi$ vertex function.
The calculation is very similar to the scalar vertex function but the boundary conditions are now $E_0=0$, $F_0=1$.\par
At one loop  the calculation is simple and one finds
$$\eqalign{E_1&=2 \left[- \left(\ell_3^2+t_2\right) \Xi(\ell_3,k)+ \ell_3 {\cal B}_1(k)\right]+\ud\g \left[ \Theta(\ell_3-k_3/2)+\Theta( \ell_3+k_3/2)\right]  \cr
F_1&=  2M \left[\ell_3\Xi(\ell_3,k)-{\cal B}_1(k)\right] . \cr}$$
This suggests introducing  the transformation \eqns{\eCSvertexVJdef},
$$\left<J_3\psi\bar\psi\right>_{1\,{\rm PI}}=-U^{-1}V_{J_3}^{(2)}U^{-1}  $$
with
$$\tilde V_{J_3}^{(2)}(k;\ell-\ud k,\ell+\ud k)=A(\ell_3,k)+i\sigma_3 B(\ell_3,k).$$
Then, $A_0=0$, $B_0=1$ and
$$\eqalign{A_1&=2 \left[- \left(\ell_3^2+t_2\right) \Xi(\ell_3,k)+ \ell_3 {\cal B}_1(k)\right]   \cr
B_1&=  2M \left[\ell_3\Xi(\ell_3,k)-{\cal B}_1(k)\right] . \cr}$$
The expansion of expression \eqns{\eCSvertexVJdef} then generates the additional contributions
$$\eqalign{\delta E_2 &= M \left[\ell_3\Xi(\ell_3,k)-{\cal B}_1(k)\right] \left[ \Theta(\ell_3-k_3/2)+\Theta( \ell_3+k_3/2)\right]  \cr
\delta F_2&=  \left[ \left(\ell_3^2+t_2\right) \Xi(\ell_3,k)- \ell_3 {\cal B}_1(k)\right] \left[ \Theta(\ell_3-k_3/2)+\Theta( \ell_3+k_3/2)\right] \cr&\quad -\frac{1}{8}\left[ \Theta(\ell_3-k_3/2)+\Theta( \ell_3+k_3/2)\right]^2   . \cr}$$
At two-loops, after some algebra, one obtains
$$\eqalign{E_2(\ell_3,k)&=M \left[\ell_3\Xi(\ell_3,k)-{\cal B}_1(k)\right]\left[ \Theta(\ell_3-k_3/2)+\Theta( \ell_3+k_3/2)\right]\cr&\quad +4M   t_2 \ell_3\Xi(\ell_3,k) {\cal B}_1(k) \cr
F_2(\ell_3,k)&=-\frac{1}{8} \left[ \Theta(\ell_3-k_3/2)+\Theta( \ell_3+k_3/2)\right]^2\cr&\quad + \left[ \Theta(\ell_3-k_3/2)+\Theta( \ell_3+k_3/2)\right]\left[(\ell_3^2+t_2) \ell_3 \Xi(\ell_3,k)- {\cal B}_1(k)\right] \cr&\quad +2\left((M^2-t_1)\ell_3^2-  t_1t_2\right) \Xi^2(\ell_3,k)-4\left(M^2-t_1\right)\ell_3\Xi(\ell_3,k) {\cal B}_1(k)\cr
&\quad+2\left(M^2-t_1\right) {\cal B}_1^2(k).\cr
}$$
Subtracting $\delta E_2$ and $\delta F_2$, we obtain
\eqna\eCSJiiiAB
$$\eqalignno{A_2(\ell_3,k)&= 4M   t_2 \ell_3\Xi(\ell_3,k) {\cal B}_1(k)&\eCSJiiiAB{a} \cr
B_2(\ell_3,k)&=  2\left((M^2-t_1)\ell_3^2-  t_1t_2\right) \Xi^2(\ell_3,k)-4\left(M^2-t_1\right)\ell_3\Xi(\ell_3,k) {\cal B}_1(k)\cr
&\quad+2\left(M^2-t_1\right) {\cal B}_1^2(k).&\eCSJiiiAB{b}\cr
}$$

 %%%%%%%%%%%%%%%%%%%%%%%%%%%%%%%%%%%%%%%
\section The $R$ two-point function

First, we give now some details about the perturbative calculations of the $R$ two-point function.\sslbl\appCSRRcalculations
\subsection{The $R$ two-point function at three loops}

The first diagram is represented in figure \label{\figcsrriii} while the other diagrams all correspond to propagator corrections applied to one-loop and two-loop diagrams. To display the different contributions, we complete the compact notation introduced in section \label{\appvertex}. We define\sslbl\appCSRRthreeloops
$$
Dp_3 =(r+k/2)^2+M^2 ,\quad  Dm_3 =(r-k/2)^2+M^2 .$$
Then, the denominator in the diagram is the product
$$Dp_1Dm_1Dp_2Dm_2Dp_3Dm_3(q_3-p_3)(r_3-q_3),$$
which, up to the factor $(q_3-p_3)(r_3-q_3)$, is totally symmetric in $(p,q,r)$.\par
%%%%%%%%%%%%%%%%%%%%%%%%%%%%%%%%%%%%%%%%%%%%%%%%%%%%%%%%%%%%%%%%%%%%%%%
After some algebra,  the terms that give vanishing contributions due to rotation symmetry in the $(1,2)$ plane being omitted, the numerator can be written as
$$\eqalign{&-\frac{1}{4}\left(Dp_1+Dm_1\right)\left(Dp_2+Dm_2\right)\left(Dp_3+Dm_3\right)\cr&
+\left(t_2+q_3^2\right)\left(Dp_1+Dm_1\right)\left(Dp_3+Dm_3\right)\cr&
+t_1\left(Dp_2+Dm_2\right)\left(Dp_1+Dm_1+Dp_3+Dm_3\right)\cr
&-4\left(t_2+q_3^2\right)t_1 \left(Dp_1+Dm_1+Dp_3+Dm_3\right)
\cr& -4t_1^2\left(Dp_2+Dm_2\right)-4M^2p_3r_3\left(Dp_2+Dm_2\right)\cr
&+4M^2q_3r_3\left(Dp_1+Dm_1\right)+4M^2p_3q_3\left(Dp_3+Dm_3\right)\cr
& +16\left(t_2+q_3^2\right)t_1^2+16M^2t_1 \left(p_3r_3-p_3q_3-q_3r_3\right).\cr}$$
\topinsert
\epsfysize=16.8mm
\epsfxsize=38.9mm
\vbox{\elevenpoint
\centerline{\epsfbox{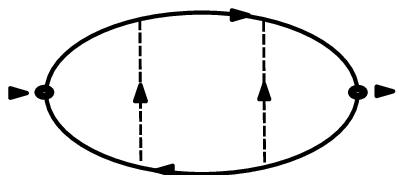}}
\kern-21.7mm
\hbox{\kern56.0mm$q+k/2$}
\kern-0.7mm
\hbox{\kern40.2mm$p+  k/2$\kern21.6mm$r+  k/2 $}
\kern3.1mm
\hbox{\kern39.4mm$k$\kern40.0mm$k$}
\kern3.2mm
\hbox{\kern38.9mm$p-k/2 $\kern22.8mm$r-k/2  $}}
\kern-0.9mm
\hbox{\kern56.0mm$q-k/2$}
\figure{4mm}{The $\bar\psi\psi$ two-point function at two loops. Dotted lines represent gauge fields.}
\figlbl\figcsrriii
\endinsert
The identity \eqns{\eCSomegaSnm} implies that, in the numerator, all quantities symmetric in $(p,q,r)$  give  a vanishing contribution. This leads to a notable reduction of the number of terms. Moreover, this implies that one can substitute
$$  \left(Dp_2+Dm_2\right)\left(Dp_1+Dm_1+Dp_3+Dm_3\right) \mapsto-  \left(Dp_1+Dm_1\right)\left(Dp_3+Dm_3\right) $$
as well as
$$\eqalign{&4M^2q_3r_3\left(Dp_1+Dm_1\right)+4M^2p_3q_3\left(Dp_3+Dm_3\right)\mapsto -4M^2p_3   r_3\left(Dp_2+Dm_2\right)\cr
& -k^2 t_1 \left(Dp_1+Dm_1+Dp_3+Dm_3\right)\mapsto k^2 t_1 \left(Dp_2+Dm_2\right).\cr} $$
One then obtains
$$\eqalignno{ &
 \left(t_2-t_1+q_3^2\right)\left(Dp_1+Dm_1\right)\left(Dp_3+Dm_3\right)\cr
&-4\left(  q_3^2-\frac{1}{4}k_3^2\right)t_1 \left(Dp_1+Dm_1+Dp_3+Dm_3\right)
\cr&  -4M^2\left(t_1+2p_3r_3\right)\left(Dp_2+Dm_2\right) \cr
& +4 t_1 k^2q_3^2+16M^2t_1\left(q_3-p_3\right)\left(q_3-r_3\right).&\eqnd{\eCSRRiiiloopb} \cr}$$
Introducing the integral representation,
$$\Theta(\omega)={1\over(2\pi)^3}\int{\d^3 q\over\left(\omega-q_3\right)\left(q^2+M^2\right)},$$
one verifies that the first line  then yields
$$-{1\over  Dp_2Dm_2} \left(q_3^2-\frac{1}{4}k_3^2 -M^2 \right){\chi^2(q_3)\over \g^2}.$$
The second line can be expressed in the form
 $$- {\left( q_3^2-\frac{1}{4}k_3^2 \right)\left(k^2+4M^2\right)\chi(q_3)\over \g Dp_2Dm_2}\left[{1\over\left(p_3-q_3\right) Dp_1Dm_1}+{1\over\left(r_3-q_3\right) Dp_3Dm_3}\right].$$
Exchanging in the first term above  $(r\longleftrightarrow q)$ and in the second $(p\longleftrightarrow q)$, one  obtains
$$ {\left(k^2+4M^2\right)\over\g Dp_1Dm_1Dp_3Dm_3\left(p_3-r_3\right)}\left[ \left( p_3^2-\frac{1}{4}k_3^2\right) \chi(p_3) -\left(  r_3^2-\frac{1}{4}k_3^2 \right)\chi(r_3)\right],\eqnd{\eCSexpressionii} $$
a  form useful later.\par
Using   the identity \eqns{\eCSbasicIdi}, one verifies that the third line can be rewritten as
$$  -{M^2  \left(k^2+4M^2+8p_3r_3\right)\over\g Dp_1Dm_1Dp_3Dm_3 \left(p_3-r_3\right)}\bigl(\chi(p_3)-\chi(r_3)\bigr).$$
Finally,   the two last  terms  yield
$$-\frac{1}{3}\left(k^2+12 M^2\right)\left(k^2+4 M^2\right){\cal B}_1^3(k),$$
where we have   used the identities \eqns{\eCSbasicIdi, \eCSbasicIdii}.\par
Collecting all   contributions, except the last line of expression \eqns{\eCSRRiiiloopb}, one obtains
\eqna\eCSRRiiiloopc
$$\eqalignno{&-{1\over Dp_2Dm_2} \left(q_3^2-\frac{1}{4}k_3^2 -M^2 \right){\chi^2(q_3)\over\g^2} &\eCSRRiiiloopc{a}\cr
&-{\left(k^2+4M^2\right)\over Dp_1Dm_1Dp_3Dm_3\left(r_3-p_3\right)\g}\left[ \left( p_3^2-\frac{1}{4}k_3^2\right) \chi(p_3) -\left(  r_3^2-\frac{1}{4}k_3^2 \right)\chi(r_3)\right]\hskip11mm&\eCSRRiiiloopc{b}\cr
&+  {M^2  \left(k^2+4M^2+8p_3r_3\right)\over Dp_1Dm_1Dp_3Dm_3 \left(r_3-p_3\right)\g}\bigl(\chi(p_3)-\chi(r_3)\bigr).  &\eCSRRiiiloopc{c}\cr}$$
\smallskip
{\it Propagator renormalizations in the one-loop diagram.}
Two insertions of the one-loop propagator corrections in the one-loop diagram (first diagram of Fig.~\label{\figCSeye}), which can be also been inferred by using the dressed propagator and expanding to order $\g^2$,  yield
$${1 \over Dp_2Dm_2}\left(M^2+\frac{1}{4}k_3^2-q_3^2\right){\chi^2(q_3)\over \g^2}\,.  \eqnd{\eCSRRiiiloopd}   $$
 \smallskip
{\it Propagator renormalizations in the two-loop diagram.}
The sum of all possible insertions of the one-loop propagator correction  into the two-loop diagram contribution to $\langle RR\rangle$ (second diagram in figure \label{\figCSeye}) can be written as the ratio of
\eqna\eCSRRiiib
$$\eqalignno{&M^2\left[\chi(r_3)-\chi(p_3)\right]\left(  k^2+4M^2  +8 p_3r_3\right)&\eCSRRiiib{a} \cr
& +\left(k^2+4M^2\right)\left[ \chi(p_3) \left(p_3^2-\frac{1}{4}k_3^2\right)- \chi(r_3  )\left(r_3^2-\frac{1}{4}k_3^2\right)\right]\hskip5mm  &\eCSRRiiib{b} \cr
& +\chi(p_3) \left(\frac{1}{4}k_3^2+ M^2- p_3^2\right)\left(Dp_3+Dm_3\right)   -\chi(r_3) \left(\frac{1}{4}k_3^2+ M^2- r_3^2\right)\left(Dp_1+Dm_1\right) \cr&&\eCSRRiiib{c} \cr} $$
and
$$\g\left(r_3-p_3\right)Dp_1Dm_1Dp_3Dm_3 \,.$$
The expression \eCSRRiiib{a} exactly cancels the  expression
 \eCSRRiiiloopc{c}.\par
The expression \eCSRRiiib{b}  then cancels  expression
 \eCSRRiiiloopc{b}.\par
Moreover, the expression \eCSRRiiib{c} can be rewritten as
$$-{\chi^2(p_3)\over\g^2}  \left(\frac{1}{4}k_3^2+ M^2- p_3^2\right)
-{\chi^2(r_3)\over\g^2}  \left(\frac{1}{4}k_3^2+ M^2- r_3^2\right).$$
After an appropriate permutation  of variables, it cancels the sum of expressions \eCSRRiiiloopc{a} and \eqns{\eCSRRiiiloopd}. The final result is
$$-\frac{1}{3}\g^2\left(k^2+12 M^2\right)\left(k^2+4 M^2\right){\cal B}_1^3(k),\eqnd\eCSRRthreeloops $$

%%%%%%%%%%%%%%%%%%%%%%%%%%%%%%%%%%%%%%%%%%%%%%%%%%%%
\subsection The $R$ two-point function at four loops

Using now the formalism developed in section \label{\ssCSrecursion}, we can calculate the $R$ two-point function at four loops. We set\sslbl\appCSRRfour
$$t_1=\ud\left(k_3^2-k^2\right), \ t_2=\ud\left(k^2+4M^2\right),$$
and factorize for each of the four set of integration variables the integral
$${1\over(2\pi)^3}\int{\d^2 v\over\left[(v-k/2)^2+M^2\right]\left[(v-k/2)^2+M^2\right]}\,.$$
The product of the four integrals is the symmetric function of the remaining four integration variables, which we denote by $I(\omega) $.\par
Denoting $\omega_1,\omega_2,\omega_3,\omega_4$ the third components of the four vectors, we obtain an integral of the form
$$\left< \tilde R(k) \tilde R(-k)\right>_4=2\g^3 M t_2 \int{\d\omega_1\,\d\omega_2\,\d\omega_3\,\d\omega_4\,N(\omega)\over
\left(\omega_4-\omega_3\right)\left(\omega_3-\omega_2\right)\left(\omega_2-\omega_1\right)}I(\omega) $$
with
$$\eqalign{N(\omega)&=-2t_1t_2\left(\omega_4-\omega_3+\omega_2-\omega_1\right)+4t_2\left[(\omega_4-\omega_3)\omega_2^2+\omega_3^2(\omega_2-\omega_1)\right]
\cr
&\quad +8M^2\left(\omega_1\omega_2\omega_3-\omega_4\omega_2\omega_1+\omega_4\omega_3\omega_1-\omega_4\omega_3\omega_2 \right). \cr} $$
The integral of the first term vanishes and this restores $O(3)$ symmetry. The second term yields
$$-\frac{16}{3}Mt_2^2{\cal B}_1^4(k) =-\frac{4}{3}M\left(k^2+4M^2\right)^2{\cal B}_1^4(k).$$
The last term yields
$$-\frac{8}{3}M^3\left(k^2+4M^2\right){\cal B}_1^4(k).$$
Thus,
$$\left< \tilde R(k) \tilde R(-k)\right>_4=-\frac{4}{3}\g^3M\left(k^2+4M^2\right)\left(k^2+6M^2\right){\cal B}_1^4(k).\eqnd\eCSRRfourloops $$
Under the assumption that the phase factor cancel, as observed up to three loops and later proved, the successive contributions have been determined up to eight loops.
%%%%%%%%%%%%%%%%%%%%%%%%%%%%%%%%%
 %%%%%%%%%%%%%%%%%%%%%%%%%%
\section{Two-point $J_3$ calculations}

In this calculation, to keep the notation simple, we assume $M>0$.\sslbl{\appCSJthree}
%%%%%%%%%%%%%%%%%%%
\subsection{One-loop calculation}

The complete the one-loop calculation of $\langle J_3 J_3\rangle$, we need\sslbl{\ssRRcalculation}
$$X_0={1\over(2\pi)^3}\int{\d^3 p\, p_3^2\over\left[(p+k/2)^2+M^2\right]\left[(p-k/2)^2+M^2\right]}.$$
Then,
$${\partial X_0\over\partial M}=-4M {1\over(2\pi)^3}\int{\d^3 p\, p_3^2\over\left[(p+k/2)^2+M^2\right]^2\left[(p-k/2)^2+M^2\right]}.$$
$$ {1\over(2\pi)^3}\int{\d^3 p\, p_\mu p_\nu\over\left[(p+k/2)^2+M^2\right]^2\left[(p-k/2)^2+M^2\right]}=A\left(\delta_{\mu\nu}-{k_\mu k_\nu \over k^2}\right)+B{k_\mu k_\nu \over k^2}\,.$$
Taking the trace, we obtain
$$\eqalign{2A+B&={1\over(2\pi)^3}\int{\d^3 p\, p^2\over\left[(p+k/2)^2+M^2\right]^2\left[(p-k/2)^2+M^2\right]}\cr} $$
Since
$$p^2=\ud  \left[(p+k/2)^2+M^2\right]+\ud \left[(p-k/2)^2+M^2\right]-k^2/4-M^2, $$
then,
$$2A+B=\ud {\cal B}_1(k)+\ud \Omega_2(M)-\frac{1}{4}\left(k^2 +4M^2\right)X_1  $$
with
$$X_1={1\over(2\pi)^3}\int{\d^3 p\,  \over\left[(p+k/2)^2+M^2\right]^2\left[(p-k/2)^2+M^2\right]}\,.$$
Moreover,
$${\partial {\cal B}_1\over\partial M}=-4M {1\over(2\pi)^3}\int{\d^3 p\,  \over\left[(p+k/2)^2+M^2\right]^2\left[(p-k/2)^2+M^2\right]}.$$
Thus,
$$X_1=-{1\over 4M}{\partial {\cal B}_1\over\partial M}\,.$$
A second relation is obtained by acting with $k_\nu$:
$$Bk^2={1\over(2\pi)^3}\int{\d^3 p\, (k  p)^2\over\left[(p+k/2)^2+M^2\right]^2\left[(p-k/2)^2+M^2\right]}\,.$$
Then,
$$kp=\ud \left[(p+k/2)^2+M^2\right]-\ud \left[(p-k/2)^2+M^2\right]$$
and, thus,
$$Bk^2=-{1\over 2}{1\over(2\pi)^3}\int{\d^3 p\, kp\over\left[(p+k/2)^2+M^2\right]^2}=\frac{1}{4}k^2 \Omega_2(M).$$
We conclude
$$B=\frac{1}{4}  \Omega_2(M),\quad A=\frac{1}{4}{\cal B}_1(k)+\frac{1}{8}\Omega_2(M) +{1\over 32M}\left(k^2+4M^2\right){\partial {\cal B}_1\over\partial M}.$$
Finally,
$$\eqalign{{\partial X_0\over\partial M}&=-4M\left[A(1-k_3^2/k^2)+Bk_3^2/k^2\right]\cr
&=-\frac{1}{8}\left(k^2+4M^2\right){\partial {\cal B}_1\over\partial M}-M{\cal B}_1(k)-\ud M\Omega_2(M)+\cr&\quad {k_3^2\over k^2}\left[ \frac{1}{8}\left(k^2+4M^2\right){\partial {\cal B}_1\over\partial M}+M{\cal B}_1(k)-\ud M\Omega_2(M)\right]. \cr }$$
Then, we notice
$${\partial  \over\partial M}\left[\frac{1}{8}\left(k^2+4M^2\right){\cal B}_1(k)\right]=
\frac{1}{8}\left(k^2+4M^2\right){\partial {\cal B}_1\over\partial M}+M{\cal B}_1(k)$$
and
$$\ud M\Omega_2(M)=-\frac{1}{4}{\partial\Omega_1  \over\partial M}\,.$$
We conclude that
$$\eqalignno{X_0&=-\frac{1}{8}\left(k^2+4M^2\right){\cal B}_1(k)+\frac{1}{4}\Omega_1(M)+{k_3^2\over k^2}
\left[\frac{1}{8}\left(k^2+4M^2\right){\cal B}_1(k)+\frac{1}{4}\Omega_1(M)\right]\cr&\quad +\ {\rm const.}\,,&\eqnd{\eCSFocalcul} \cr}$$
where the constant is a function of $k$ independent of $M$ and the whole expression should have no pole at $k=0$. The explicit residue of the pole is
$$k_3^2\left({M\over 16\pi}+{1\over 16\pi}(\Lambda-M)\right).$$
Therefore, the integration constant should cancel the residue and the coefficient of    $k_3^2/k^2$ becomes
$$\frac{1}{8}\left(k^2+4M^2\right){\cal B}_1(k)+\frac{1}{4}\Omega_1(M)-{ \Lambda\over16\pi}\,.$$
To fix the complete integration constant, we can then take the $k=0$ limit. Then,
$$X_0=\frac{1}{3}\left[\Omega_1(M)-M^2\Omega_2(M)\right]={\Lambda\over12\pi}-{M\over 8\pi}\,.$$
Taking the same limit in equation \eqns{\eCSFocalcul}, we obtain
$$X_0=-\ud M^2\Omega_2(M)+\frac{1}{4}\Omega_1(M)={\Lambda\over 16\pi}-{M\over8\pi}\,.$$
We thus have to add
$${\Lambda\over 48\pi}\,.$$
The complete expression reads
$$\eqalignno{X_0&=-\frac{1}{8}\left(k^2+4M^2\right){\cal B}_1(k)+\frac{1}{4}\Omega_1(M)+{\Lambda\over48\pi}\cr&\quad+{k_3^2\over k^2}
\left[\frac{1}{8}\left(k^2+4M^2\right){\cal B}_1(k)+\frac{1}{4}\Omega_1(M)-{\Lambda\over16\pi} \right]. &\eqnd{\eCSFocalculf} \cr}$$
%%%%%%%%%%%%%%%%%%%%%%%%%%%%%%%%%%%%%%%%%%%%%%%%%%%%%%%%%%%%%%%%%%%%%%%%%%%
\section{Other current components and WT identities}

We have only considered the $J_3$ component of the current. Here, we give a few details over the three-point vertex function
$$\tilde\Gamma_\mu(k;p-k/2,p+k/2)\equiv\left<\tilde J_\mu(k) \psi(p-k/2)\bar\psi(p+k/2) \right> $$  where $J_\mu$ is defined as in Eq.~\eqns{\eCSfermioncurrent}.
\par
First, we can write a WT identity that relation the conserved current to the fermion inverse two-point function. It reads
$$ k_\mu \tilde\Gamma_\mu(k;p-k/2,p+k/2)=\tilde\Gamma^{(2)}(p-k/2)-\tilde\Gamma^{(2)}(p+k/2).$$
In the large $N$ limit we know the right hand side exactly. From Eq.~\eqns{\eCSNvertexiipt},
$$\eqalign{\tilde\Gamma^{(2)}(p-k/2)-\tilde\Gamma^{(2)}(p+k/2)&=-U(p_3-k_3/2)\left[i(\sla{p}-\sla{k}/2)+M\right]U(p_3-k_3/2)\cr&\quad+U(p_3+k_3/2)\left[i(\sla{p}+\sla{k}/2)+M\right]U(p_3+k_3/2).\cr}$$
Taken into account the explicit form \eqns{\eCSUmatdef} of $U$ and after a simple commutation, we obtain
$$\eqalign{&\tilde\Gamma^{(2)}(p-k/2)-\tilde\Gamma^{(2)}(p+k/2) =ik_1\sigma_1+ik_2\sigma_2\cr&\  - U^2(p_3-k_3/2)\left[i(p_3-k_3/2)\sigma_3+M \right]+U^2(p_3+k_3/2)\left[i(p_3+k_3/2)\sigma_3+M \right].\cr}$$
One verifies then that $\tilde\Gamma_\mu $ has the simple form
$$\tilde\Gamma_\mu= i \sigma_\mu +E_\mu{\bf 1} +iF_\mu \sigma_3\,,$$
which is consistent with the WT identity.\par
Moreover, the quantities  $ \tilde\Gamma_\mu- i \sigma_\mu$, since they are only linear combinations of the matrices ${\bf 1}$ and $\sigma_3$, all satisfy an integral equation of the form \eqns{\eCSEFvertexinteq} but with different inhomogeneous terms.
\listrefs
\bye